\documentclass[journal,comsoc,twoside]{IEEEtran}
\usepackage[cmintegrals]{newtxmath}

\PassOptionsToPackage{draft}{hyperref}
\PassOptionsToPackage{obeyspaces}{url}
\usepackage{float, enumerate, xcolor, algorithm, algpseudocode, tabto, balance, amsmath, graphicx, epstopdf, hyperref, multirow, subfig, dblfloatfix, textcomp, listings, eurosym, lstautogobble, tikz} 
\usepackage[english]{babel}
\usepackage[utf8]{inputenc}
\usepackage[justification=centering]{caption}
\graphicspath{ {./images/} }
\usepackage[numbers,sort&compress]{natbib}
\usepackage{chngcntr}
\usepackage{mathtools}
\usepackage[overload]{empheq}

\newcommand{\ankit}[1]{\textcolor{black}{#1}}
\newcommand{\ma}{\textit{Module1}}
\newcommand{\mb}{\textit{Module2}}
\newcommand{\mc}{\textit{Module3}}
 
\usepackage{url}

\begin{document}
	\title{On the Economic Significance of Ransomware Campaigns: A Bitcoin Transactions Perspective}
	
	%
	

	\author{Mauro~Conti,
		~Ankit~Gangwal*,
		~Sushmita~Ruj
		\thanks{* Corresponding author}
		\thanks{M. Conti and A. Gangwal are with the Department of Mathematics, University of Padua, 35121, Padua, Italy (e-mail: conti@math.unipd.it; ankit.gangwal@phd.unipd.it).}
		\thanks{Sushmita Ruj is with Cryptology and Security Research Unit,
			Computer and Communication Sciences Division, Indian Statistical Institute, 700108, Kolkata,
			India (e-mail: sush@isical.ac.in).}
		}

	\maketitle	

	\begin{abstract}
		Bitcoin cryptocurrency system enables users to transact securely and pseudo-anonymously by using an arbitrary number of aliases (Bitcoin addresses). Cybercriminals exploit these characteristics to commit immutable and presumably untraceable monetary fraud, especially via ransomware; a type of malware that encrypts files of the infected system and demands ransom for decryption.
		\par
		In this paper, we present our comprehensive study on all recent ransomware and report the economic impact of such ransomware from the Bitcoin payment perspective. We also present a lightweight framework to identify, collect, and analyze Bitcoin addresses managed by the same user or group of users (cybercriminals, in this case), which includes a novel approach for classifying a payment as ransom. To verify the correctness of our framework, we compared our findings on CryptoLocker ransomware with the results presented in the literature. Our results align with the results found in the previous works except for the final valuation in USD. The reason for this discrepancy is that we used the average Bitcoin price on the day of each ransom payment whereas the authors of the previous studies used the Bitcoin price on the day of their evaluation. Furthermore, for each investigated ransomware, we provide a holistic view of its genesis, development, the process of infection and execution, and characteristic of ransom demands. Finally, we also release our dataset that contains a detailed transaction history of all the Bitcoin addresses we identified for each ransomware.


	\end{abstract} 
	
	
	\IEEEpeerreviewmaketitle

	\section{Introduction}
	\label{intro}
	Satoshi Nakamoto in 2008 proposed a decentralized cryptography-based electronic currency called Bitcoin~\cite{nakamoto2008bitcoin}. Such financial systems eliminate the control of centralized authority and provide ubiquity as well as fairness via (quasi) real-time transactions. Such digital currencies also guarantee a certain degree of anonymity, which raises novel and unique concerns, e.g., an inevitable-growth in illegal activities. 
	\par
	On another side, ransomware is a class of malware that restricts access to the system it infects until the victim pays the demanded ransom. Readily available toolkits such as~eda2\footnote{eda2 is an abandoned open-source ransomware kit that was distributed only for educational purposes.} and Ransomware-as-a-Service (RaaS) enable even a novice user to create and launch ransomware. Furthermore, the ransomware affiliate program lures users to spread ransomware in exchange for profit share. According to the annual threat report-2017 published by Symantec Inc.~\cite{symantec_report_2017}, ransomware continued to be the most dangerous cyber-crime threat to individual users and enterprises in 2016. Compared to the previous year, the number of detected ransomware infection increased by 36\% during 2016. Moreover, average ransomware detection rate reached over 1,500 incidents per day at the year-end. In particular, the average ransom amount rose 266\% from USD 294 in 2015 to USD 1,077. 
	
	\par
	The evolving class of ransomware has been exploiting privacy-preserving online services, e.g., the Tor hidden network~\cite{Dingledine2004} to remain anonymous. Moreover, the pseudo-anonymous nature of decentralized currencies such as Bitcoin makes it difficult to trace a payee. Hence, the cybercriminals have been misusing such payment systems to extort ransoms anonymously. In this paper, we present our comprehensive and~longitudinal study on recent ransomware and report the~economic impact of such ransomware from the Bitcoin~payment perspective.
	
	\par
	\textit{Contributions:}
	The major contributions of this paper are listed as follows:
	\begin{enumerate}
		\item We present a lightweight framework to identify, collect, and analyze addresses that belong to the same user. We~also propose a novel approach for classifying a payment as ransom.
		
		\item Using our framework, we analyzed the economic impact (in terms of ransoms extorted in Bitcoin) of all the recent ransomware: (i)~that used Bitcoin as at least one mode of ransom payment, and (ii)~for which at least one Bitcoin address is publicly known.
		
		
		\item We discuss the inception, evolution (where applicable), and functionality (including distribution, infection, and encryption procedure) of every analyzed ransomware along with the magnitude and timeline of their ransom demands.

		\item We also release our dataset\footnote{spritz.math.unipd.it/projects/btcransomware/} for future research endeavors. The dataset contains a detailed transaction history of all the addresses we identified for each ransomware. Hence, our results are fully reproducible. 
		
	\end{enumerate}
	
	\textit{Organization:}
	The remainder of this paper is organized as follows. In Section~\ref{related}, we explain the essential concepts related to ransomware infection and the Bitcoin currency system. Section~\ref{related_work} addresses the previous works on identification and assessment of cyber-crimes in the Bitcoin ecosystem. Section~\ref{evaluation} elucidates our framework for ransom identification. 
	In Section~\ref{ransomware}, we present our findings and enlighten the economic impact the ransomware that fulfilled our selection criteria. In Section~\ref{limitation}, we discuss the limitations of our proposed framework. Finally, Section~\ref{futurework} concludes the paper.
	
	
	\section{Preliminaries}
	\label{related}
	In this section, we describe the chronology of a typical ransomware infection and explain the fundamentals of the Bitcoin cryptocurrency system.
	
	\textit{Ransomware:} A typical ransomware infection includes the following events:
	\begin{enumerate}
		\item Infection: Similar to generic malware, ransomware are also distributed via various infection vectors. These vectors include, but not limited to, email spamming with malicious attachment (e.g., CryptoLocker) or link to the malicious payload (e.g., CryptoWall), exploit packs (e.g.,~Angler browser exploit in TeslaCrypt and Neutrino exploit kit in DMA Locker). Interestingly, recent ransomware incorporate self-propagation capabilities. For instance, NotPetya and WannaCry exploit vulnerabilities in the network protocols to infect local computers on the~same network.
		\item Encryption: After infiltration, ransomware silently encrypt files on the infected system. In particular, ransomware target those files that are valuable to the user, e.g.,~images, videos, documents. For the encryption process, ransomware use symmetric encryption algorithm, asymmetric encryption algorithm, or even combination of the both. The key for encryption is either generated locally or procured from a remote Command~and~Control~(C\&C). Generally, the backup files are also encrypted/deleted to prevent recovery. However, the files responsible for running the system are not affected, at least until the deadline for the ransom payment. 
		\item Extortion: After the encryption process, ransomware display a ransom note on the screen. The ransom note of recent ransomware includes a threat message, ransom amount specified in fiat currency such as US~dollar (for instance, USD~300 in NotPetya) or cryptocurrency such as Bitcoin (for instance, 1~BTC in CryptoLocker), a countdown timer that shows the time left before the deadline, and a payment address. The payment address can be a Bitcoin address or a website's address that shows this Bitcoin address. Typically, the ransom note also includes instructions on how and where to buy Bitcoin.
		\item Decryption: After confirmation of the ransom payment, the ransomware either automatically start the decryption process, or the victim is asked to download and run a~decryption tool.
	\end{enumerate}
	
	\textit{Bitcoin:} In 1993, researchers from Carnegie Mellon University~\cite{tygar1993cryptography} and University of Southern California~\cite{medvinsky1993netcash} discussed the need for a cryptography-based digital currency. On November~1,~2008, a person or a group of persons under a pseudonym Satoshi Nakamoto articulated the idea of peer-to-peer, decentralized, cryptography-based electronic currency system called Bitcoin~\cite{nakamoto2008bitcoin}. The basic terminology used in the Bitcoin protocol are as follows:
	\begin{itemize}
		\item Address: A Bitcoin address is a string identifier of a possible destination for a Bitcoin payment. It is 26~to~35~alphanumeric characters long and begins with the number~1~(Pay-to-Pub KeyHash or P2PKH type) or 3~(Pay to Script Hash or P2SH type). Bitcoin addresses are hashed public keys generated from the Elliptic Curve Digital Signature Algorithm (ECDSA). Hence, each Bitcoin is associated with the owner's public key.
		
		\item Wallet: A wallet is a file that stores Bitcoin addresses along with the corresponding private keys. It also maintains the Unspent Transaction Output (UTXO) corresponding to each address.
		
		\item Blockchain: The blockchain is a shared, public ledger on which the entire Bitcoin network relies. All confirmed transactions are included in the blockchain without any exception. This way, new transactions can be verified to be spending Bitcoin that are indeed owned by the spender. The integrity and the chronological order of the blockchain are enforced with cryptography.\footnote{bitcoin.org/en/how-it-works}
		
		\item Block: An individual unit of the blockchain is called a block. Each block includes the hash of the previous block to guarantee the integrity of the network, the nonce that assisted its mining, and a list of the transactions.
		
		\item Transaction: A transaction refers to a transfer of Bitcoin between Bitcoin addresses. To transfer Bitcoin, a payer creates a transaction message. In this message, the payer specifies the payee's Bitcoin address as well as an amount of Bitcoin to transfer. As shown in~\figurename{~\ref{figure_trx}}, the payer authenticates the transaction by digitally signing it with the private key of the corresponding address. Finally, Bitcoin network broadcasts and confirms (typically, in the following 10 minutes) the transaction through a process called mining. A confirmed transaction is irreversible.
		\enlargethispage{-\baselineskip}
		\begin{figure}[H]
			\centering
			\includegraphics[trim = 2mm 8mm 2mm 4mm, clip, width=.9\linewidth]{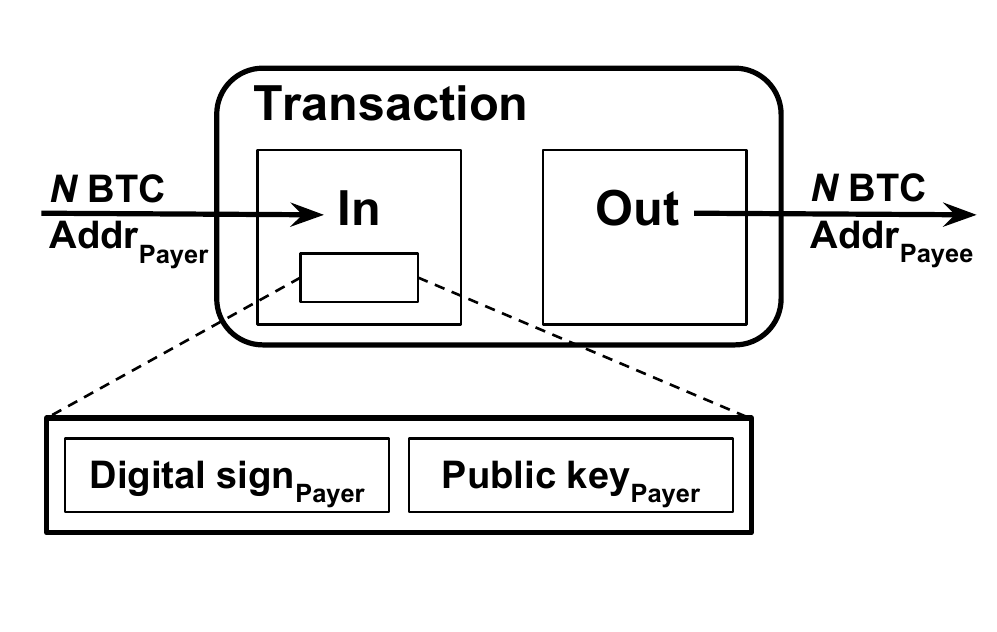}
			\caption[]{An example of a simple Bitcoin transaction}
			\label{figure_trx}
		\end{figure}

		\begin{figure*}[!htbp]
			\centering
			\includegraphics[trim = 2mm 2mm 2mm 2mm, clip, scale=.44]{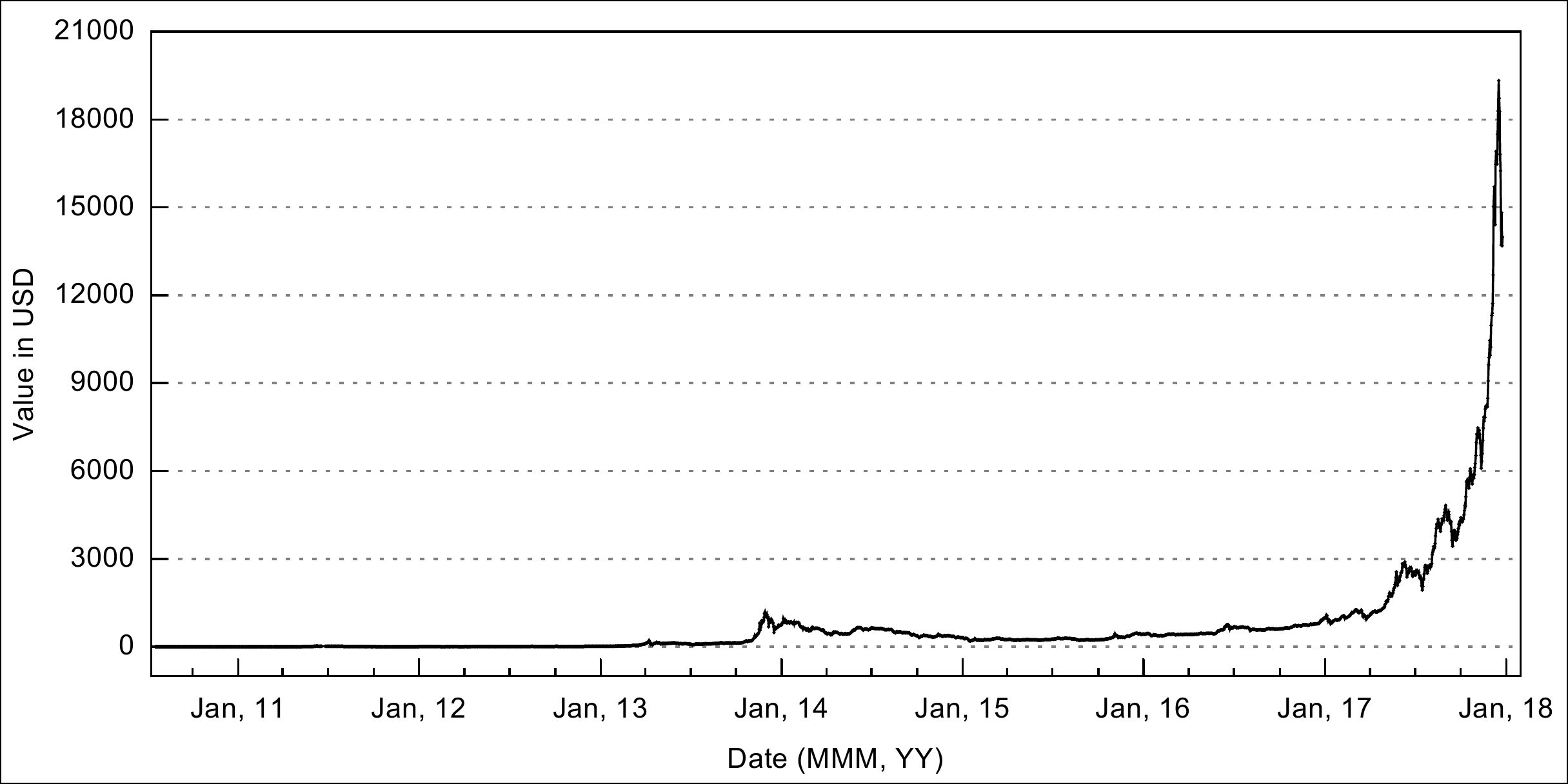}
			\caption[]{BTC-USD exchange rate trend}
			\label{figure1}
		\end{figure*}
	\end{itemize}
	\par
	A user can also purchase Bitcoin in exchange for other regulated currencies. The unit of the Bitcoin currency is ``Bitcoin,'' abbreviated as ``BTC.'' Like any other traded commodity, the price\footnote{We use the term ``price'' to refer BTC-USD exchange rate.} of Bitcoin varies. \figurename{~\ref{figure1}} depicts the BTC-USD exchange rate since July~18,~2010, the day when one of the world's first Bitcoin currency exchange market Mt.~Gox~was~established.
	
	\section{Related Work}
	\label{related_work}
	Law enforcement authorities as well as the research community have made several attempts to identify and measure cyber-crimes in the Bitcoin ecosystem. The authors in~\cite{bistarelli2017go, mcginn2016visualizing, di2015bitconeview, reid2011analysis} proposed tools to analyze transactions in the Bitcoin blockchain visually. Christin in~\cite{christin2013traveling} proposed a thorough analysis of the Silk Road anonymous marketplace and discussed the socio-economic implications of the findings. Ron and Shamir used the public blockchain data to estimate the wealth of the Silk Road marketplace's owner, known as Dread Pirate Robert~\cite{ron2014did}. Soska and Christin studied anonymous online marketplaces including Silk Road, Sheep Marketplace, etc. and examined how virtual marketplaces have evolved~\cite{soska2015measuring}. Meiklejohn~et~al.~\cite{meiklejohn2013fistful} proposed an approach to comprehend overall transaction patterns of the Bitcoin payments used for criminal or fraudulent purposes.
	\par		
	However, the literature on measuring the economic impact of ransomware that accepted ransoms via Bitcoin (hereinafter referred to as ``Bitcoin ransomware'') is rather limited. Huang~et~al.~\cite{huangtracking} discuss the ethical and technical issues of monitoring ransomware activities as well as the dynamics of ransom payments. Liao~et~al.~\cite{liao2016behind} analyzed the timestamps of ransom payments to CryptoLocker. The work~\cite{kharraz2015cutting} provides a holistic view of the general ransomware that appeared between 2006 and 2014. Additionally, the authors also estimated the financial intensives gained by CyptoLocker ransomware. Spagnuolo~et~al. proposed a framework called BitIodine \cite{spagnuolo2014bitiodine}. The authors used BitIodine to investigate Bitcoin addresses associated with CryptoLocker ransomware and Dread Pirate Roberts. The works~\cite{secureworks_cryptolocker,cryptolocker_intelligence_report} present a systematic analysis of CryptoLocker ransomware.
	\par
	It is noteworthy that previous works~\cite{spagnuolo2014bitiodine,kharraz2015cutting,liao2016behind,secureworks_cryptolocker,cryptolocker_intelligence_report} only considered either the daily average or highest Bitcoin price to classify ransom payments and do not take into account the variations that might occur due to the transaction fee. Furthermore, their estimation of the total worth of extorted ransoms is based on the Bitcoin price on the day of their evaluation, which exaggerates the results due to fluctuations (mostly, increase; see \figurename{~\ref{figure1}}) in the price of Bitcoin. Additionally, the systems proposed in the previous works~\cite{bistarelli2017go, mcginn2016visualizing, di2015bitconeview, reid2011analysis,spagnuolo2014bitiodine} demand high bandwidth, storage, and computational resources as they query the entire blockchain.
	
	\par
	To the best of our knowledge, our work is the first study that elaborates not only the characteristics and functionality of various Bitcoin ransomware, but it also gives more accurate insights on the economic impact of such ransomware. In particular, our work is different from the state-of-the-art on various dimensions: (i)~to identify a payment as ransom, we consider both the day-to-day lowest and highest Bitcoin price as well as the variations due to the transaction fee; (ii)~to accurately assess the worth (in USD) of extorted ransoms, we used the average Bitcoin price on the day of each ransom payment; and (iii)~our framework focuses only on the transactions belonging to the address(es) of interest rather than the entire blockchain.

	\section{Ransom Identification Framework}
	\label{evaluation}
	\ankit{To investigate the ransoms extorted by a ransomware, we first identify the Bitcoin addresses linked to the ransomware. Then, we obtain the transaction history of these addresses. Finally, we distinguish the transactions associated with the ransom payments. To this end,} we propose our framework, which consists of three stages/parts\ankit{/modules}: (i)~identifying the Bitcoin addresses belonging to the ransomware \ankit{(discussed in~Section~\ref{identify})}; (ii)~data \ankit{(transaction history)} collection and database generation from the blockchain \ankit{(presented in~Section~\ref{data_col})}; and (iii)~our considerations for classifying a payment as ransom \ankit{(elaborated in~Section~\ref{consideration})}.
	
	\subsection{\ankit{\ma:} Identification of ransomware addresses}
	\label{identify}
	Bitcoin offers privacy only through pseudonymity, and an increasing number of works~\cite{ron2013quantitative, ron2014did, soska2015measuring, reid2011analysis, biryukov2014deanonymisation, meiklejohn2013fistful} suggest that information available in public blockchain ledger can lead to de-anonymize (to a certain extent) Bitcoin transactions.
	\par
	To collect the addresses associated with a ransomware, we began by extensively searching various online resources:~ransomware knowledge base (e.g., ESET, Kaspersky Lab, Malwarebytes, Symantec); ransomware removal guides (e.g., BleepingComputer.com, MalwareTips.com, 2-spyware.com, ``How To'' videos on YouTube); reports from Counter Threat Units (CTU), Incident Responses (IR), and Security Operations Centers (SOC) (e.g., Dell SecureWorks, PhishMe.com); online fora (e.g., Reddit) where victims and researchers post Bitcoin addresses associated with the concerned ransomware; and screenshots of ransomware available in different image search engines (e.g., Google, Yahoo). \ankit{Considering the fact that not every address related to a ransomware is posted on the Internet, we used two clustering heuristics to identify the set of addresses controlled by the same user (cybercriminals, in our case).} 
	Our heuristics are based on the fundamental principles of the Bitcoin transaction protocol~\cite{nakamoto2008bitcoin} and are as follows:
	
	\subsubsection{Multi-input transactions} 
	A multi-input transaction usually\footnote{Nowadays, coin mixing services allow users to join their transactions to enhance anonymity and unlinkability. However, such services have many security and privacy concerns~\cite{conti_survey_2017}. Hence, for simplicity, we assume that the user commonly does not make use of Bitcoin mixers.} takes place when a user $U$ attempts to make a payment, and the payment amount $P$ cannot be sufficiently funded by any of the individual Bitcoin balance available in $U$'s wallet. In such a scenario, the Bitcoin protocol~allows grouping of a set of Bitcoin balances from $U$'s wallet to settle $P$ and make payment through a multi-input transaction. Hence, we can conclude that if a set of input addresses $S_{input}$ is used to disburse $P$, then $S_{input}$ is managed by the same user.

	\subsubsection{Shadow/change address} In the Bitcoin protocol, the whole input amount must be spent in the same transaction. To deliver the ``change'' back to the user $U$, a shadow address~$A_{shadow}$ is automatically generated and used to collect the unspent amount of the transaction.  If there are two addresses in the set of output addresses $S_{out}$, and one address has never been seen before in the whole blockchain while the other address has appeared before, then we can safely presume that the newly generated address is a shadow address~\cite{meiklejohn2013fistful}.
	
	\par
	Algorithm~\ref{algo1} explains our approach to identify the addresses managed by the same user, hereinafter referred to as ``Cluster''. Here, $S_{initial}$ represents the set of addresses collected from the online resources, $S_{input}$ is a set of input addresses in a transaction, and $A_{shadow}$ represents a shadow address generated (if any) in a transaction.
	\begin{algorithm}[H]
		\caption{Identifying addresses managed by the same user.}
		\label{algo1}
		\hspace*{\algorithmicindent} \textbf{Input:} $S_{initial}$ 
		\begin{algorithmic}[1]
			\State $Cluster := S_{initial}$
			\State $Cluster^\prime := \{\}$ \Comment{\{ \} is an empty set}
			\While{$Cluster \not = Cluster^\prime$}
			\State $Cluster^\prime := Cluster$
			\State $M := \{\}$ \Comment{$M$ stores $S_{input}$}
			\State $C := \{\}$ \Comment{$C$ stores $A_{shadow}$}
			
			\For{$i$ in $Cluster$}
			\State Get all transactions $Tx$ where $i$ is an input address
			\For{$t$ in $Tx$}
			\State $M \cup (S_{input}~in~t)$ \Comment{$\cup$ is set union}
			\State $C \cup (A_{shadow}~in~t)$
			\EndFor
			\EndFor			
			
			\State $Cluster := Cluster \cup M \cup C$
			\EndWhile
			\State \textbf{return} $Cluster$
		\end{algorithmic}
	\end{algorithm}
	\par
	Essentially, for a given list of addresses, our algorithm recursively finds all the addresses satisfying our heuristics.
	
	\begin{figure*}[!b]
		\centering
		\resizebox{2\columnwidth}{!}
		{
			\begin{tikzpicture}
			\draw (0,0) -- (24,0);
			
			\foreach \x in {0,1,2,3,4,5,6,7,8,9,10,11,12,13,14,15,16,17,18,19,20,21,22,23,24}
			\draw (\x cm,3pt) -- (\x cm,-3pt);
			
			\draw  (0,0)  node[above=3pt, rotate=45, xshift=-1.1cm,  yshift=-.5cm]  {Sep. 05, '13} node[below=3pt, rotate=45, xshift=1.2cm, yshift=0.5cm] {CryptoLocker};
			\draw  (1,0)  node[above=3pt, rotate=45, xshift=-0.85cm,  yshift=-.45cm]  {Feb. '14} node[below=3pt, rotate=45, xshift=1.22cm, yshift=0.5cm] {~CryptoDefense};
			\draw  (2,0)  node[above=3pt, rotate=45, xshift=-.75cm,  yshift=-.5cm]  {Q1 '14} node[below=3pt, rotate=45, xshift=0.95cm, yshift=0.5cm] {~CryptoWall};
			\draw  (3,0)  node[above=3pt, rotate=45, xshift=-1.1cm,  yshift=-.45cm]  {Mid-Jul. '14} node[below=3pt, rotate=45, xshift=1.1cm, yshift=0.5cm] {CTB-Locker};
			\draw  (4,0)  node[above=3pt, rotate=45, xshift=-1.1cm,  yshift=-.5cm]  {Feb. 05, '15} node[below=3pt, rotate=45, xshift=1.75cm,  yshift=0.5cm] {CryptoTorLocker2015};
			\draw  (5,0)  node[above=3pt, rotate=45, xshift=-1.2cm,  yshift=-.45cm]  {Mid-Feb. '15} node[below=3pt, rotate=45, xshift=1.0cm, yshift=0.5cm] {TeslaCrypt};
			\draw  (6,0)  node[above=3pt, rotate=45, xshift=-0.85cm,  yshift=-.45cm]  {Nov. '15} node[below=3pt, rotate=45, xshift=0.825cm, yshift=0.5cm] {Chimera};
			\draw  (7,0)  node[above=3pt, rotate=45, xshift=-0.85cm,  yshift=-.45cm]  {Dec. '15} node[below=3pt, rotate=45, xshift=1.22cm, yshift=0.5cm] {DMA Locker};
			\draw  (8,0)  node[above=3pt, rotate=45, xshift=-.75cm,  yshift=-.5cm]  {Q1 '16} node[below=3pt, rotate=45, xshift=0.9cm, yshift=0.5cm] {~Hi Buddy!};
			\draw  (9,0)  node[above=3pt, rotate=45, xshift=-0.85cm,  yshift=-.45cm]  {Mar. '16} node[below=3pt, rotate=45, xshift=0.6cm, yshift=0.5cm] {Petya};
			\draw  (10,0) node[above=3pt, rotate=45, xshift=-1.1cm,  yshift=-.5cm]  {Mar. 04, '16} node[below=3pt, rotate=45, xshift=0.9cm, yshift=0.5cm] {KeRanger};
			\draw  (11,0) node[above=3pt, rotate=45, xshift=-1.2cm,  yshift=-.45cm]  {Late  Mar. '16} node[below=3pt, rotate=45, xshift=0.7cm, yshift=0.5cm] {Jigsaw};       
			\draw  (12,0) node[above=3pt, rotate=45, xshift=-0.85cm,  yshift=-.5cm]  {May '16} node[below=3pt, rotate=45, xshift=0.72cm, yshift=0.5cm] {Mischa};
			\draw  (13,0) node[above=3pt, rotate=45, xshift=-1.15cm,  yshift=-.5cm]  {May 24, '16} node[below=3pt, rotate=45, xshift=0.85cm, yshift=0.5cm] {ZCryptor};
			\draw  (14,0) node[above=3pt, rotate=45, xshift=-0.85cm,  yshift=-.55cm]  {Aug. '16} node[below=3pt, rotate=45, xshift=1.05cm, yshift=0.5cm] {VenusLocker};
			\draw  (15,0) node[above=3pt, rotate=45, xshift=-0.85cm,  yshift=-.5cm]  {Dec. '16} node[below=3pt, rotate=45, xshift=0.95cm, yshift=0.5cm] {GoldenEye};
			\draw  (16,0) node[above=3pt, rotate=45, xshift=-0.85cm,  yshift=-.55cm]  {Dec. '16} node[below=3pt, rotate=45, xshift=.8cm, yshift=0.45cm] {KillDisk};                  
			
			\draw  (17,0) node[above=3pt, rotate=45, xshift=-0.85cm,  yshift=-.45cm]  {Feb. '17} node[below=3pt, rotate=45, xshift=1.55cm, yshift=0.5cm] {The Trump Locker};
			\draw  (18,0) node[above=3pt, rotate=45, xshift=-1.1cm,  yshift=-.5cm]  {Feb. 22, '17} node[below=3pt, rotate=45, xshift=0.78cm, yshift=0.5cm] {FindZip};              
			\draw  (19,0) node[above=3pt, rotate=45, xshift=-0.85cm,  yshift=-.45cm]  {Mar. '17} node[below=3pt, rotate=45, xshift=1.48cm, yshift=0.5cm] {The LLTP Locker};
			\draw  (20,0) node[above=3pt, rotate=45, xshift=-0.85cm,  yshift=-.5cm]  {May '17} node[below=3pt, rotate=45, xshift=1.18cm, yshift=0.5cm] {ThunderCrypt};
			\draw  (21,0) node[above=3pt, rotate=45, xshift=-1.1cm,  yshift=-.5cm]  {May 12, '17} node[below=3pt, rotate=45, xshift=0.9cm, yshift=0.5cm] {WannaCry};
			\draw  (22,0) node[above=3pt, rotate=45, xshift=-1.1cm,  yshift=-.5cm]  {Jun. 27, '17} node[below=3pt, rotate=45, xshift=0.9cm, yshift=0.5cm] {NotPetya};
			\draw  (23,0) node[above=3pt, rotate=45, xshift=-1.1cm,  yshift=-.5cm]  {Oct. 13, '17} node[below=3pt, rotate=45, xshift=1.22cm, yshift=0.5cm] {DoubleLocker};
			\draw  (24,0) node[above=3pt, rotate=45, xshift=-1.1cm,  yshift=-.5cm]  {Oct. 24, '17} node[below=3pt, rotate=45, xshift=1.05cm, yshift=0.5cm] {Bad Rabbit};
			\end{tikzpicture}
		}
		\caption[]{Occurrence of Bitcoin ransomware}
		\label{figure2}
	\end{figure*}
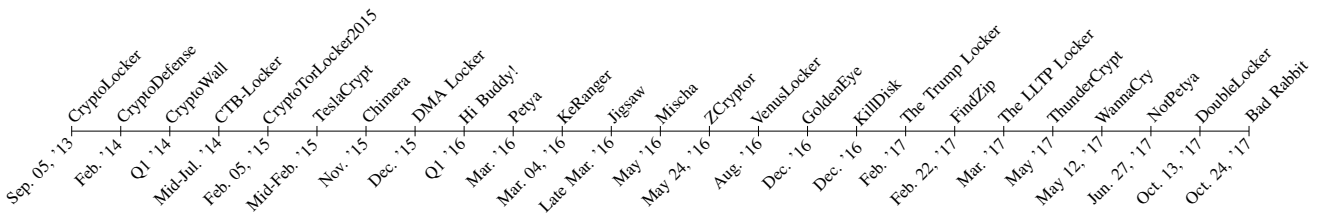

	\subsection{\ankit{\mb:} Data collection and database generation}
	\label{data_col}
	As explained in Section~\ref{related}, Bitcoin blockchain data is publicly available. At the time of writing (December~2017), block height of the blockchain was over 500,000 blocks, which means that downloading/querying the entire blockchain is very expensive in terms of bandwidth, storage, and computations. 
	To address these issues, we built a lightweight system that uses \textit{Blockchain Data API}\footnote{blockchain.info/api/blockchain\_api} to crawl and parse transactions associated only with the address(es) of interest. 
	\par
	For each transaction associated with an address of interest~(\textit{Address}), our system collects the hash of the transaction~(\textit{HASH}), remitted Bitcoin (\textit{BTC\_to\_Addr}), input address-es (\textit{Trx\_In\_Addrs}), output addresses (\textit{Trx\_Out\_Addrs}), GMT-based date (\textit{GMT\_Date}), and GMT-based~time (\textit{GMT\_Time}). Listing~\ref{schema} shows the SQL statement used to create our database.
	\begin{center}
		\begin{minipage}{.85\linewidth}
			\vspace{1em}
			\begin{lstlisting}[
			language=SQL,
			showspaces=false,
			basicstyle=\ttfamily,
			numbers=none,
			frame=single,
			commentstyle=\color{gray},
			autogobble=true,
			captionpos=b,
			caption=SQL statement for creating our database,
			label=schema
			]
			CREATE TABLE tx (
			HASH CHAR(64) NOT NULL PRIMARY KEY,
			BTC_to_Addr INT NOT NULL,
			Trx_In_Addrs TEXT,
			Trx_Out_Addrs TEXT,
			GMT_Date DATE,
			GMT_Time Time,
			Address CHAR(35) NOT NULL,
			Address_as_Input INT NOT NULL
			);
			\end{lstlisting}		
		\end{minipage}
	\end{center}
	\par
	The field \textit{HASH} serves as the \textit{Primary Key}, which implicitly discards any duplicate transactions reported for multiple participating/constituting addresses. \textit{Address\_as\_Input} denotes if the \textit{Address} was used as an input in the transaction. Our system also uses \textit{BitcoinAverage API}\footnote{apiv2.bitcoinaverage.com} to collect day-to-day highest, average, and lowest price of Bitcoin.
	
	\subsection{\ankit{\mc:} Considerations for classifying a payment as ransom}
	\label{consideration}
	\begin{table*}[!b]
		\centering
		\color{black}
		\begin{tabular}{|c|c|c|c|c|c|c|}
			\hline
			\multirow{2}{*}{\textbf{Ransomware}} & \multicolumn{3}{c|}{\textbf{Overall}}                 & \multicolumn{3}{c|}{\textbf{Ransom	}}                 \\ \cline{2-7} 
			& \textbf{Payments} & \textbf{BTC} & \textbf{USD Value} & \textbf{Payments} & \textbf{BTC} & \textbf{USD value} \\ \hline
			CryptoLocker                         & 51,766            & 133,045.9961 & 42,292,191.17      & 804               & 1403.7548    & 449,274.97         \\ \hline
			CryptoDefense                        & 128               & 138.3223     & 70,113.41          & 108               & 126.6960     & 63,859.49          \\ \hline
			CryptoWall                           & 51,278            & 87,897.8510  & 45,370,589.00      & 3,730             & 5,351.2329   & 2,220,909.12       \\ \hline
			DMA Locker                           & 298               & 1,433.3463   & 580,763.95         & 117               & 339.4591     & 178,162.77         \\ \hline
			NotPetya                             & 70                & 4.1787       & 10,284.42          & 33                & 4.0576       & 9,835.86           \\ \hline
			KeRanger                             & 13                & 10.0044      & 4,175.35           & 10                & 9.9990       & 4,173.12           \\ \hline
			WannaCry                             & 341               & 53.2906      & 99,549.05          & 238               & 47.1743      & 86,076.76          \\ \hline
		\end{tabular}
		\caption{\ankit{Summary of overall payments and ransom payments to the ransomware for which the observed payments align with their period of activity and ransom demands}}
		\label{table_compare}
	\end{table*}

	A Bitcoin transaction involves two varying factors: (i)~Bitcoin price, and (ii)~transaction fee. 
	The price of Bitcoin changes frequently. Therefore, considering only the daily average, highest, or lowest price of Bitcoin is not suitable, especially when the variation in the price is high. Furthermore, the transaction fee is paid on the top of the transaction amount. A victim may assume that the ransom amount to be paid includes (or excludes) the transaction fee, which leads to discrepancies in the payment-amount transferred to an address. Moreover, the transaction fee depends on the size of the transaction, i.e., a transaction that involves a larger number of addresses would incur more fee than a transaction with fewer addresses involved. Hence, to classify a payment as ransom, our framework considers both the day-to-day lowest and highest price of Bitcoin as well as the variation that might occur due to the transaction fee. 
	\par
	In general, the cybercriminals specify the ransom either in Bitcoin (e.g., 1~BTC) or USD equivalent BTC (e.g., Bitcoin equivalent to USD~300). Our framework classifies a payment $\rho$ to an address $\alpha$ in a transaction $\tau$ as ransom if it satisfies at least one condition in Eq.~(1a) or Eq.~(1b).


	\begin{subequations} 
		\label{equation1}
		\begin{align}[left={demand~in=\empheqlbrace}]
		BTC= {}& \begin{cases}
		r_b = d_b, &  \\
		r_b = d_b - f, &
		\end{cases} \\
		USD= {}& \begin{cases} 
		v_l \le d_u \le v_h, &  \\
		v_l \le d_u - f \le v_h, &
		\end{cases}
		\end{align}
	\end{subequations}
	where: 
	\begin{itemize}
		\item $f$ denotes the transaction fee, computed as the difference between the total amount being spent and the total amount being received in $\tau$.
		\item $d_b$ denotes the ransom asked in BTC.
		\item $d_u$ denotes the ransom asked in USD.
		\item $r_b$ denotes the BTC received by $\alpha$ in $\rho$.
		\item $v_l$ denotes the value of $r_b$ computed using the lowest BTC~price of the payment day.
		\item $v_h$ denotes the value of $r_b$ computed using the highest BTC~price of the payment day.
	\end{itemize}
	
	\par
	It is also important to mention that to evaluate the total ransom (in USD) received by a ransomware cluster, it would be unfair to use the Bitcoin price on the day of our evaluation as it would misrepresent the amount due to the variations in the price. Hence, unlike previous works, we used the average Bitcoin price on the day of each ransom payment.
	
	\vspace{-0.5em}
	\section{Economic Impact of Ransomware}
	\label{ransomware}
	
	We found twenty ransomware that fulfilled our selection criteria, i.e., those ransomware: (i) that used Bitcoin as at least one mode of ransom payment, and (ii) for which at least one Bitcoin address is publicly known. In this section, we discuss these twenty ransomware and their renamed/rebranded versions. Here, our main focus is to provide an insight into the economic impact of these ransomware from the Bitcoin payment perspective. \figurename{~\ref{figure2}} depicts the reported debut period of these ransomware as well as the occurrence of their renamed/rebranded versions.
	
	\par
	We performed the numerical assessment of the ransomware on December~7,~2017. Hence, all the data reported in this paper include the transactions until December~7,~2017. We begin with those ransomware for which the observed payments align with their period of activity and ransom demands.
	\ankit{Table~\ref{table_compare} presents a summary of overall payments received by the addresses of such ransomware. It also lists the payments classified as ransom by our framework. Furthermore, for each payment class, it includes equivalent BTC/USD value (using day-to-day average Bitcoin price). It is clear that CryptoLocker received the maximum number of payments, i.e., 51,766~payments that worth 133,045.9961~BTC, which is approximately USD~42,292,191.17. However, our framework classified 3,730~payments received by CryptoWall as ransom payments, which is the maximum number of ransom payments extorted by any ransomware. These payments worth 5,351.2329~BTC or USD~2,220,909.12. On another side, KeRanger received the~minimum number of overall payments as well as the ransom payments. Now, we discuss each ransomware in details.}

	\subsection{CryptoLocker}
	\ankit{\textit{Introduction:}} Appeared in September~2013, CryptoLocker targets computers running Windows operating system. It uses ``Microsoft Enhanced RSA and AES Cryptographic Provider (MS\_ENH\_RSA\_AES\_PROV)'' to create encryption keys and to encrypt users' files with the strong RSA~(CALG\_RSA\_KEYX) and AES~(CALG\_AES\_256) algorithms. Before beginning the encryption process, it establishes a connection with its C\&C to obtain an RSA~public key. It encrypts each file with a unique AES~key; after use, it encrypts each AES encryption key with the RSA~public key~\cite{secureworks_cryptolocker}.			 
	\par
	\ankit{\textit{Infection:}} CryptoLocker infection spread through two modes. In its initial release beginning from September~5,~2013, the cybercriminals especially targeted business professionals through spam emails. The messages of the emails were typical ``customer complaints" against recipients' firm. Attached to these emails was a ZIP archive that contained a single malicious Windows executable (exe) file. The names of both the ZIP file and malicious executable were identical (except for extensions) with 13 to 17 random alphabetical characters. Later versions of CryptoLocker, starting from October~7,~2013, were distributed by the peer-to-peer (P2P) Gameover ZeuS~\cite{secureworks_zeus}. In this case, Gameover Zeus used Cutwail spam botnet to send a huge number of spam emails miming popular online retailers and banking institutions. These emails often contained spoofed order confirmations, invoices, or urgent message for unpaid balances to entice victims to follow CryptoLocker exploit kits.
	\par
	\ankit{\textit{Ransom demand:}} The ransom note asks the victim to pay the ransom within 72~hours through any one of the various payment methods. It also threatens that not paying the ransom would lead to (allegedly) destruction of decryption keys. In the initial versions, the payment option included cashU\footnote{www.cashu.com}, Ukash\footnote{www.ukash.com}, paysafecard\footnote{www.paysafecard.com}, Bitcoin, or MoneyPak\footnote{www.attheregister.com/moneypak/}. However, later the ransoms were collected only via Bitcoin or MoneyPak. All these payments methods are anonymous (or at least pseudo-anonymous), which makes it difficult to track the payer and the payee. The amount of demanded ransom and their corresponding timelines (both the dates are included) are as~follow:
	
	\begin{itemize}
		\item 2 BTC between September~5,~2013 and November~11,~2013 allowing a three-day ransom period.
		\item 10 BTC between November~1,~2013 and November~11,~2013. The payment was the fee for using ``CryptoLocker Decryption Service'' that allowed victims, who failed to pay ransoms within the given time frame, to recover their files.
		\item 1 BTC between November~8,~2013 and November~13,~2013 to allowing a three-day ransom period.
		\item 0.5 BTC between November~10,~2013 and November~27,~2013 to allowing a three-day ransom period.
		\item 2 BTC between November~11,~2013 and January~31,~2014. In this case, the payment was the reduced fee for using  ``CryptoLocker Decryption Service''.
		\item 0.3 BTC between November~24,~2013 and December~31,~2013.
		\item 0.6 BTC between December~20,~2013 and January~31,~2014.
	\end{itemize}
	
	\par
	\ankit{\textit{Associated Bitcoin addresses and transactions:}} To evaluate the economic impact of CryptoLocker, we initially began with four Bitcoin addresses listed in Table~\ref{cryptolocker_address}. Using these addresses, \ankit{\ma~(Section~\ref{identify})} generated 956 addresses belonging to CryptoLocker cluster ($C_{CL}$). \ankit{We obtained the detailed transaction history of these addresses using \mb~(Section~\ref{data_col}).} Our analysis of transactions to $C_{CL}$ reveals that $C_{CL}$ received, in total, over 51,000 payments, which accounts for over 133,000~BTC (more than USD~42,000,000). Table~\ref{cryptolocker_inwards} presents a summary of the total payments credited to $C_{CL}$.
	
	\begin{table}[H]
		\centering
		\resizebox{\columnwidth}{!}
		{
			\begin{tabular}{|c|c|c|c|c|}
				\hline
				\textbf{\begin{tabular}[c]{@{}c@{}}Payments\end{tabular}} & \textbf{\begin{tabular}[c]{@{}c@{}}BTC\end{tabular}} & \textbf{\begin{tabular}[c]{@{}c@{}}USD valuez\\(daily highest\\BTC price)\end{tabular}} & \textbf{\begin{tabular}[c]{@{}c@{}}USD value\\(daily average\\BTC price)\end{tabular}} & \textbf{\begin{tabular}[c]{@{}c@{}}USD value\\(daily lowest\\BTC price)\end{tabular}} \\ \hline
				51,766                                                                         & 133,045.9961                                                                 & 42,722,858.15                                                                                          & 42,292,191.17                                                                                          & 41,734,959.83                                                                                         \\ \hline
			\end{tabular}
		}
		\caption{Total payments credited to $C_{CL}$ including all ransom and non-ransom payments}
		\label{cryptolocker_inwards}
	\end{table}
	
	\par
\begin{figure}[H]
		\centering
		\includegraphics[trim = 2mm 2mm 2mm 2mm, clip, width=\linewidth]{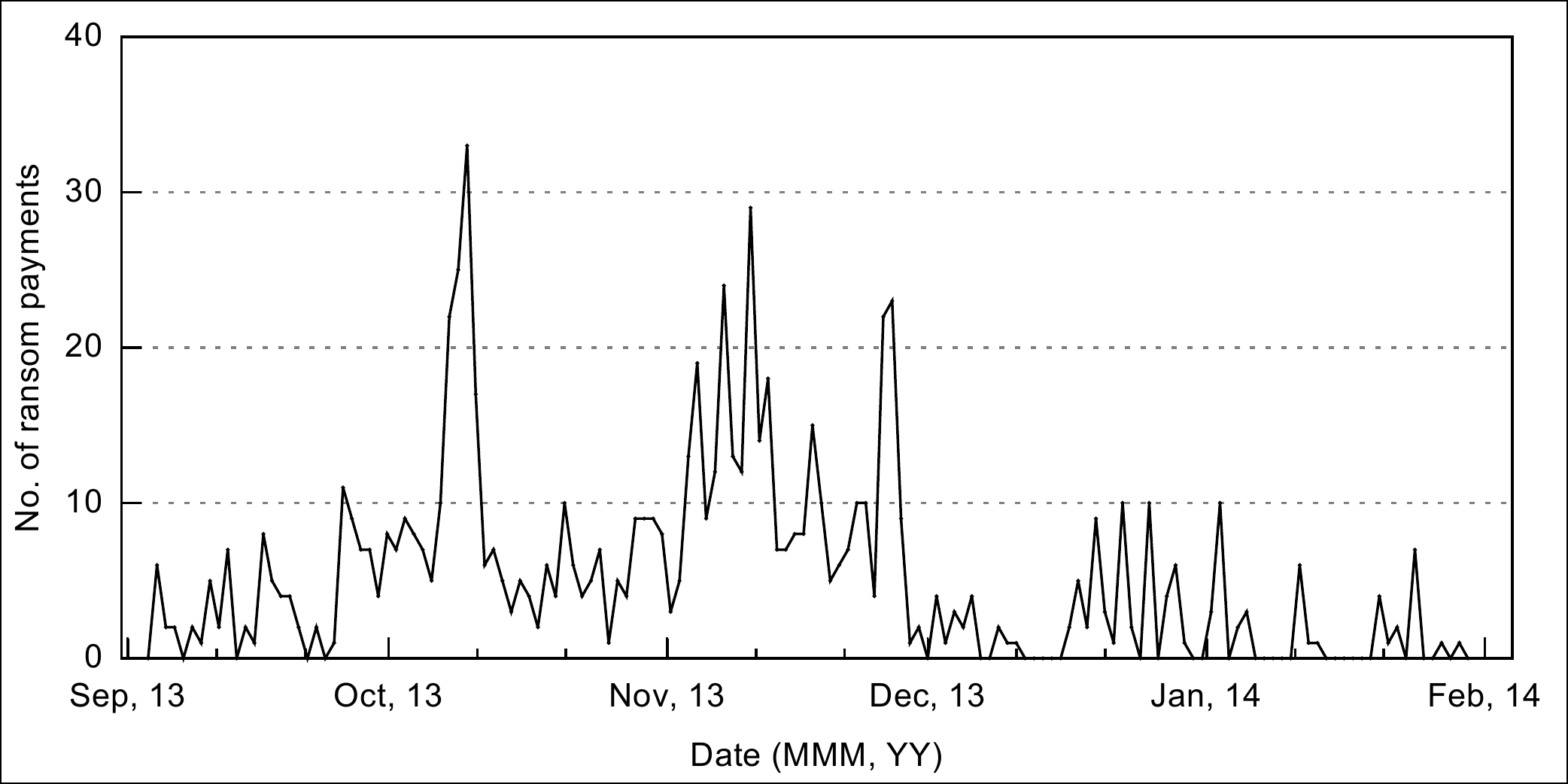}
		\caption[]{Number of ransoms paid to $C_{CL}$}
		\label{cryptolocker_ransom_payment_trend}
	\end{figure}
\begin{figure}[H]
		\centering
		\includegraphics[trim = 2mm 2mm 2mm 2mm, clip, width=\linewidth]{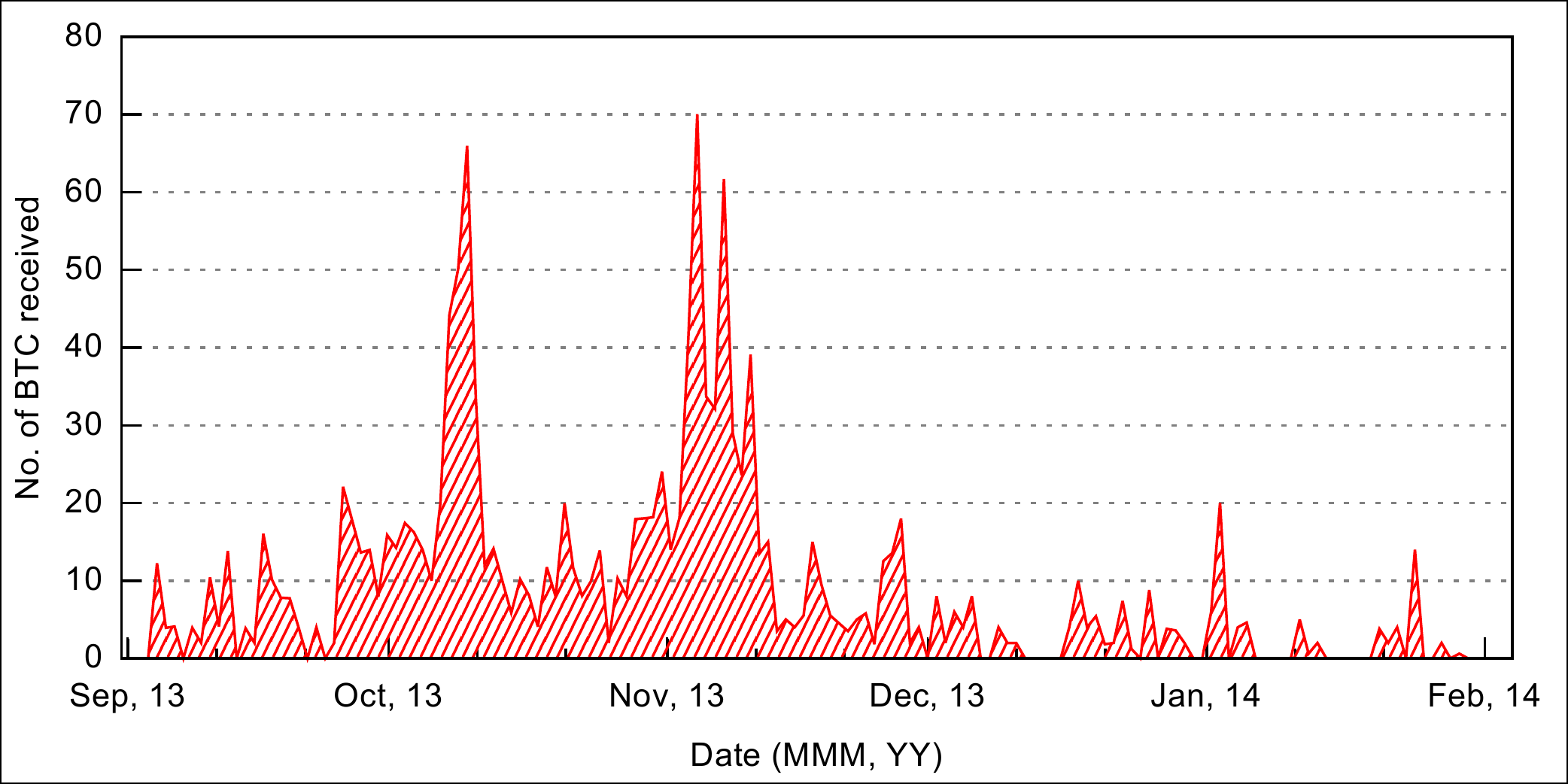}
		\caption[]{Number of Bitcoin received (in ransoms) by $C_{CL}$}
		\label{cryptolocker_BTC_USD1}
	\end{figure}
\begin{figure}[H]
		\centering
		\includegraphics[trim = 2mm 2mm 2mm 2mm, clip, width=\linewidth]{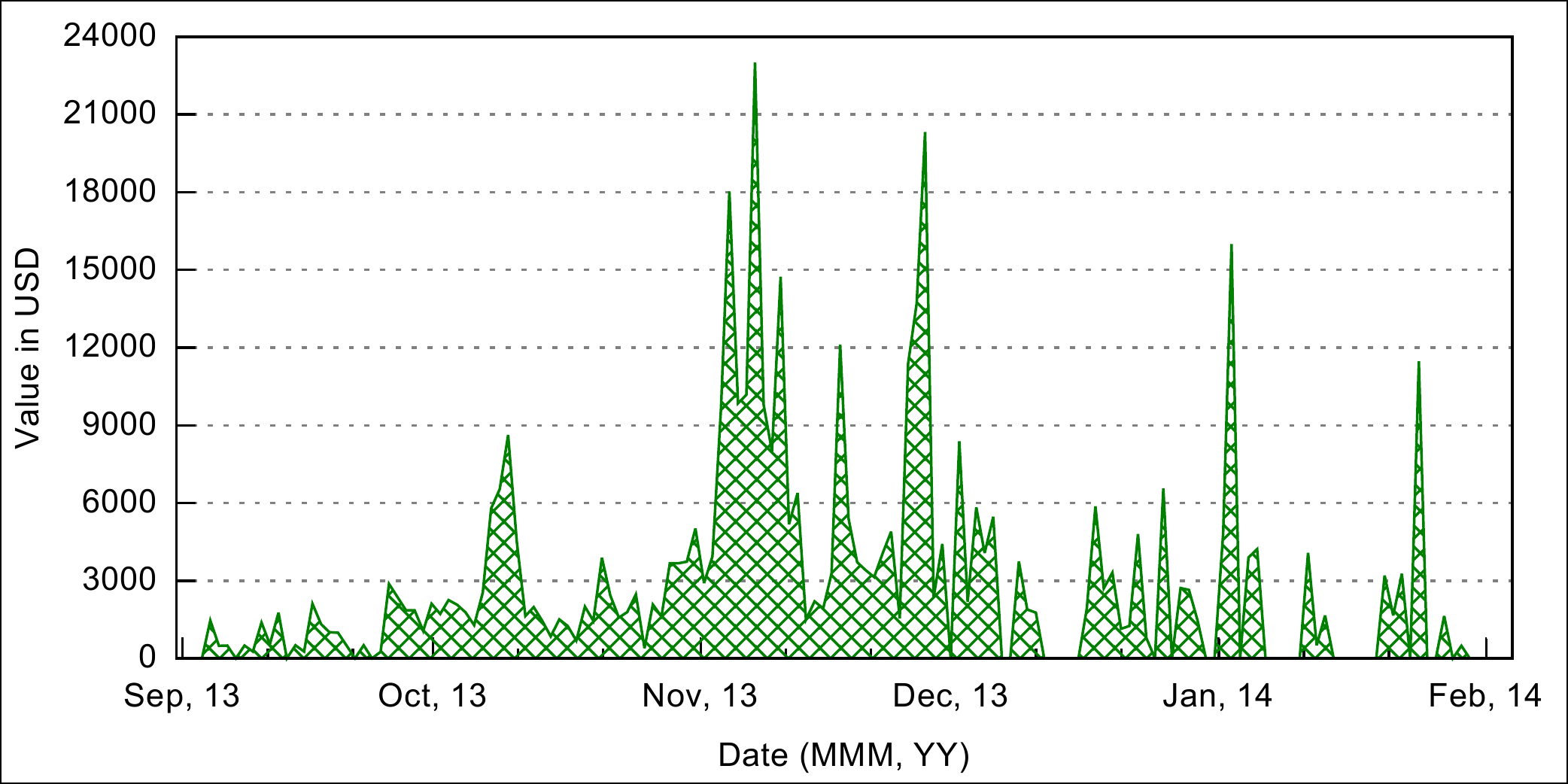}
		\caption[]{USD value of ransoms paid to $C_{CL}$}
		\label{cryptolocker_BTC_USD2}
	\end{figure}

	\ankit{\textit{Economy of ransom payments in Bitcoin:}}
	To evaluate the gross economic impact of only the ransom payments, we filtered the transactions using: (i) the ransom amounts and their timeline, (ii) our classification criteria \ankit{mentioned in \mc~(Section~\ref{consideration})}. \figurename{~\ref{cryptolocker_ransom_payment_trend}} shows the total number of ransoms paid by the victims by date. $C_{CL}$ received 33~payment on October~10,~2013, which is the maximum number of ransoms paid in a single day. However, as shown in \figurename{~\ref{cryptolocker_BTC_USD1}}, $C_{CL}$ received slightly more than 70 BTC on November~5,~2013, which is the maximum number of Bitcoin received in a single day. On another side, $C_{CL}$ received slightly above USD~23,000 on November~8,~2013, which is the maximum USD collected in a single day, see \figurename{~\ref{cryptolocker_BTC_USD2}}.
	
	\par
\begin{figure}[H]
		\centering
		\includegraphics[trim = 2mm 2mm 2mm 2mm, clip, width=\linewidth]{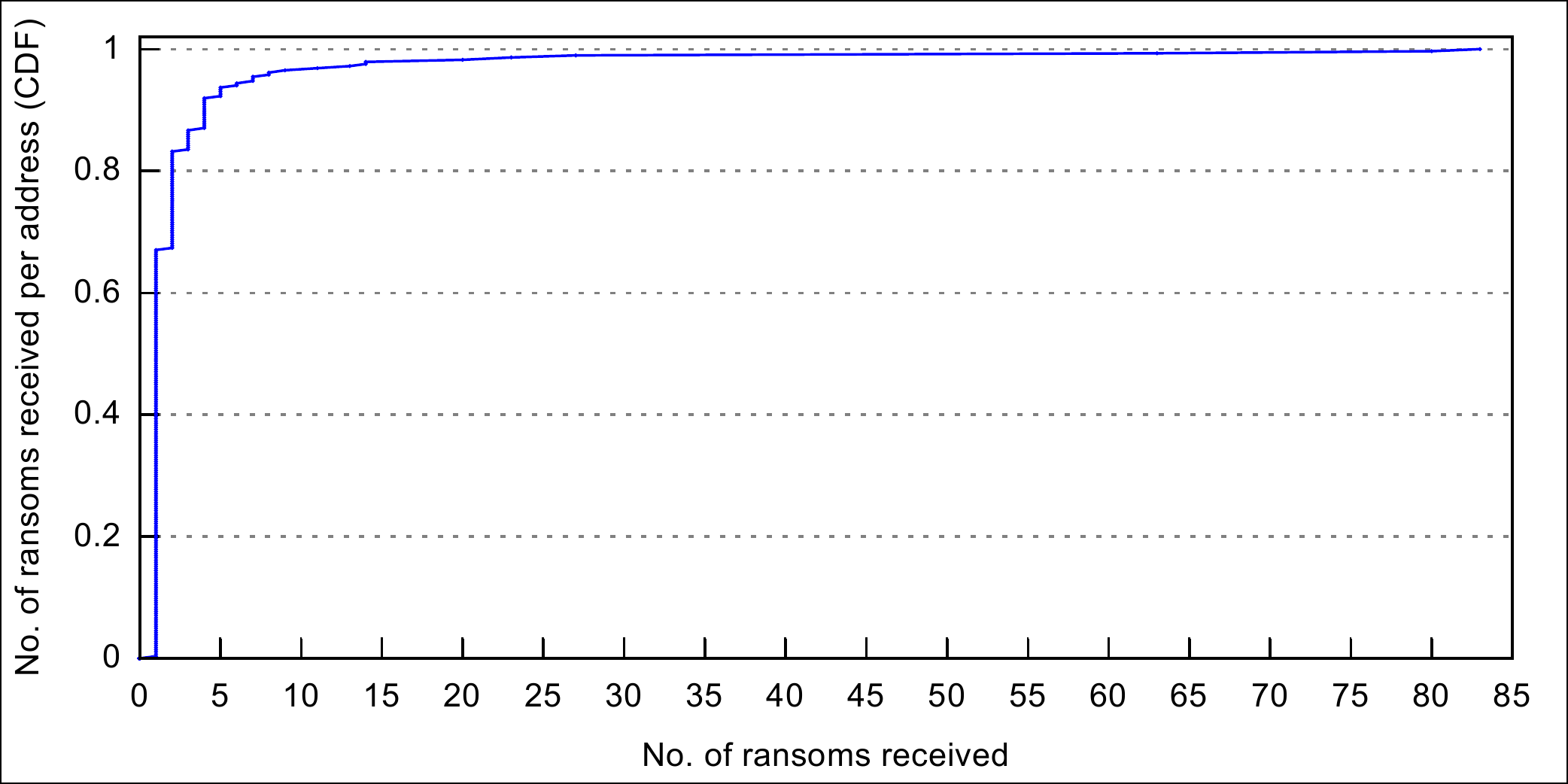}
		\caption[]{CDF of ransoms received per address in $C_{CL}$}
		\label{cryptolocker_ransom_Trx_In_CDF}
	\end{figure}
\begin{figure}[H]
		\centering
		\includegraphics[trim = 2mm 2mm 2mm 2mm, clip, width=\linewidth]{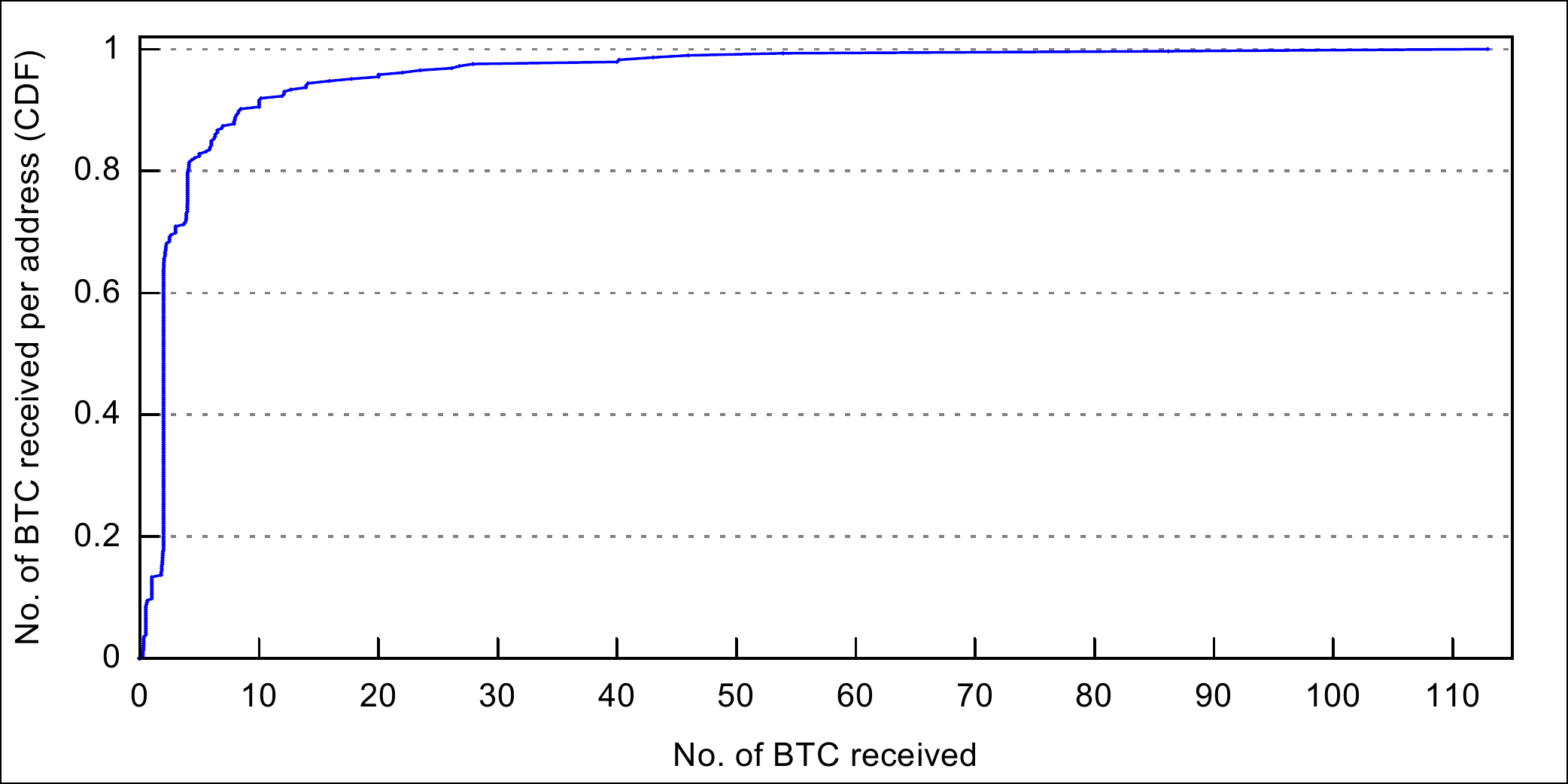}
		\caption[]{CDF of Bitcoin received (in ransoms) per address in $C_{CL}$}
		\label{cryptolocker_ransom_BTC_In_CDF}
	\end{figure}

	By further analyzing the addresses of $C_{CL}$, we discovered that approximately 83.16\% Bitcoin addresses received maximum two payments. Moreover, 13.33\% Bitcoin addresses received no more than one Bitcoin perhaps because victims were charged less due to a substantial increase in the Bitcoin value in late November~2013. Moreover, an address\footnote{16i7w5G2aoq8zqLDR3VJnawZ8VmYFZjVsd} collected 112.94~BTC while a different address\footnote{1HFLn7JP7FZrufvNKkQPEfAWGjKUdFZEmy} collected 83 ransom payments. These values correspond to the maximum number of Bitcoin and the maximum number of ransom collected by any address in $C_{CL}$. Figures~\ref{cryptolocker_ransom_Trx_In_CDF}~and~\ref{cryptolocker_ransom_BTC_In_CDF} depict Cumulative Distribution Function (CDF) of the number of ransoms and number of Bitcoin received (in ransoms) per address respectively.
	
	\par
	In total, we have identified 804 ransom payments to $C_{CL}$, which contribute to a total of 1,403.75 extorted BTC. Using day-to-day average Bitcoin price, we estimate that these ransoms convert to USD~449,274.97. Table~\ref{cryptolocker_ransom} summarizes the ransoms paid to CryptoLocker.
	
	\begin{table}[H]
		\centering
		\resizebox{\columnwidth}{!}
		{
			\begin{tabular}{|c|c|c|c|c|}
				\hline
				\textbf{Ransom} & \textbf{Time period}      & \textbf{Payments} & \textbf{BTC}     & \textbf{USD value} \\ \hline
				2 BTC           & Sep. 05, '13 - Nov. 11, '13          & 443                     & 884.9691           & 153,650.51            \\ \hline
				10 BTC (late)   & Nov. 01, '13 - Nov. 11, '13          & 17                      & 170.0000           & 47,549.90             \\ \hline
				1 BTC           & Nov. 08, '13 - Nov. 13, '13          & 38                      & 38.0000            & 14,302.26             \\ \hline
				0.5 BTC         & Nov. 10, '13 - Nov. 27, '13         & 118                     & 59.0000            & 37,108.27             \\ \hline
				2 BTC (late)    & Nov. 11, '13 - Jan. 31, '14         & 106                     & 212.0000           & 166,476.42            \\ \hline
				0.3 BTC         & Nov. 24, '13 - Dec. 31, '13         & 31                      & 9.1856             & 8,584.88              \\ \hline
				0.6 BTC         & Dec. 20, '13 - Jan. 30, '14         & 51                      & 30.6000            & 21,602.72             \\ \hline
				\textbf{Total}  & \textbf{Sep. 05, '13 - Jan. 30, '14} & \textbf{804}            & \textbf{1403.7548} & \textbf{449,274.97}   \\ \hline
			\end{tabular}
		}
		\caption{Summary of ransoms paid to CryptoLocker}
		\label{cryptolocker_ransom}
	\end{table}
	\par 
	Although we cannot be sure that the unaccounted transactions are not ransom payments, our results align with the findings presented in the works~\cite{spagnuolo2014bitiodine,liao2016behind,secureworks_cryptolocker,cryptolocker_intelligence_report} except for the final valuation in USD since the authors of these studies used the Bitcoin price on the day of their evaluation. More importantly, it implies that we can trust our methodology for evaluating other ransomware where a baseline for comparison is not~available.

	
	\subsection{CryptoDefense}
	\ankit{\textit{Introduction:}} With a sophisticated hybrid design, CryptoDefense first appeared in the last week of February~2014. It incorporates many powerful techniques that were used by previous ransomware. For example, use of Bitcoin and the Tor network for anonymity, RSA-2048 based public-key cryptography for strong encryption, and the typical pressure tactics such~as~a short deadline for payment with threats of increasing the~ransom after the deadline. It targets Windows systems. CryptoDefense encrypts files using the AES-256 algorithm. It generates the encryption key on the victim's computer using Windows CryptoAPI library. After the file encryption process completes, it encrypts the AES key using an RSA-2048 public~key.
	
	\par
	\ankit{\textit{Infection:}} Primarily, CryptoDefense ransomware infiltrated via spam emails that contained malicious payload disguised as a compressed PDF document. Upon successful infiltration, it attempts to contact its C\&C; and it sends the information about the infected system in the initial communication. Upon receiving the acknowledgment from the C\&C, it starts the encryption process.
	
	\par
	\ankit{\textit{Ransom demand:}} CryptoDefense asks USD/EUR~500 in Bitcoin within four~days to decrypt the files. The cost of decryption after four~days increases to USD/EUR~1,000. The attackers also provide a unique .onion page for each victim. Here, the victims could see a screenshot of their compromised system and decrypt one file as a proof of decryption.
	
	\par	
	\ankit{\textit{Associated Bitcoin addresses and transactions:}} We began with two publicly known Bitcoin payment addresses of CryptoDefense. These addresses are listed in Table~\ref{cryptodefense_address}. In our analysis, the CryptoDefense cluster ($C_{CD}$) had only two addresses as \ankit{\ma}~generates no new address from these addresses. Our analysis of transactions \ankit{(obtained using \mb)} to $C_{CD}$ indicates that $C_{CD}$ collected 128 payments. The total value of these payments is somewhat above 138~BTC (more than USD~70,000). Table~\ref{cryptodefense_inwards} presents a summary of the total payments credited to $C_{CD}$.
	
	\begin{table}[H]
		\centering
		\resizebox{\columnwidth}{!}
		{
			\begin{tabular}{|c|c|c|c|c|}
				\hline
				\textbf{\begin{tabular}[c]{@{}c@{}}Payments\end{tabular}} & \textbf{\begin{tabular}[c]{@{}c@{}}BTC\end{tabular}} & \textbf{\begin{tabular}[c]{@{}c@{}}USD value\\(daily highest\\BTC price)\end{tabular}} & \textbf{\begin{tabular}[c]{@{}c@{}}USD value\\(daily average\\BTC price)\end{tabular}} & \textbf{\begin{tabular}[c]{@{}c@{}}USD value\\(daily lowest\\BTC price)\end{tabular}} \\ \hline
				128                                                                           & 138.3223                                                                  & 72,342.26                                                                                              & 70,113.41                                                                                              & 67,715.88                                                                                             \\ \hline
			\end{tabular}
		}
		\caption{Total payments credited to $C_{CD}$ including all ransom and non-ransom payments}
		\label{cryptodefense_inwards}
	\end{table}
	
	\par
	\ankit{\textit{Economy of ransom payments in Bitcoin:}}
	Due to the limited number of transactions, we manually verified each payment to $C_{CD}$. As shown in Table~\ref{cryptodefense_ransom_BTC_address}, each Bitcoin address collected at minimum 35 ransom payments and a minimum of about 36.83 BTC.
	
	\begin{table}[H]
		\centering
		\resizebox{\columnwidth}{!}
		{
			\begin{tabular}{|l|c|c|}
				\hline
				\multicolumn{1}{|c|}{\textbf{Address}} & \textbf{Payments} & \textbf{BTC} \\ \hline
				19DyWHtgLgDKgEeoKjfpCJJ9WU8SQ3gr27     & 35                               & 36.8339                      \\ \hline
				1EmLLj8peW292zR2VvumYPPa9wLcK4CPK1     & 73                               & 89.8622                      \\ \hline
			\end{tabular}
		}
		\caption{Number of ransoms and Bitcoin received (in ransoms) per address in $C_{CD}$}
		\label{cryptodefense_ransom_BTC_address}
	\end{table}
	
	\par
\begin{figure}[H]
		\centering
		\includegraphics[trim = 2mm 2mm 1mm 2mm, clip, width=\linewidth]{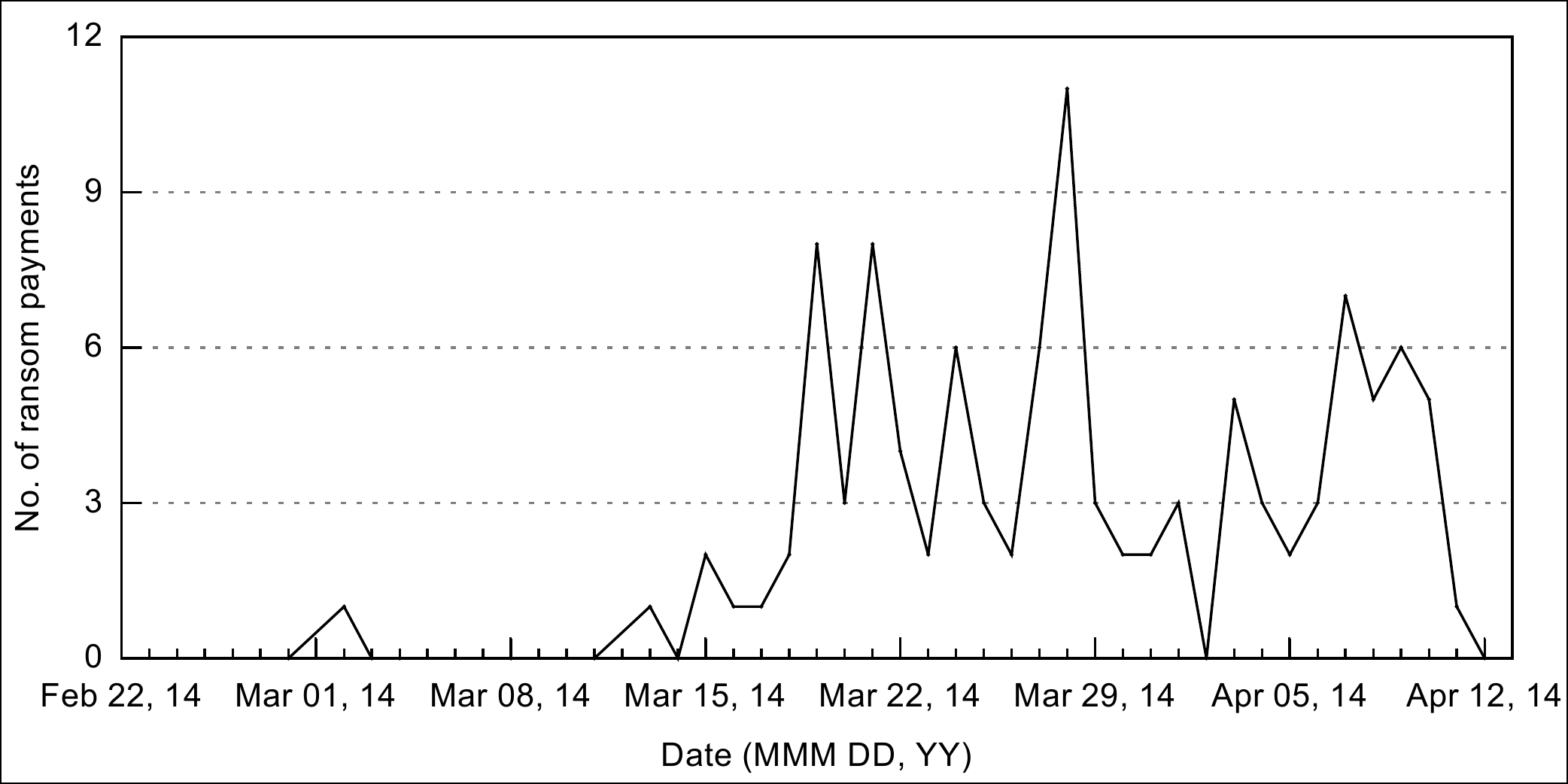}
		\caption[]{Number of ransoms paid to $C_{CD}$}
		\label{cryptodefense_ransom_payment_trend}
	\end{figure}
\begin{figure}[H]
		\centering
		\includegraphics[trim = 2mm 2mm 1mm 2mm, clip, width=\linewidth]{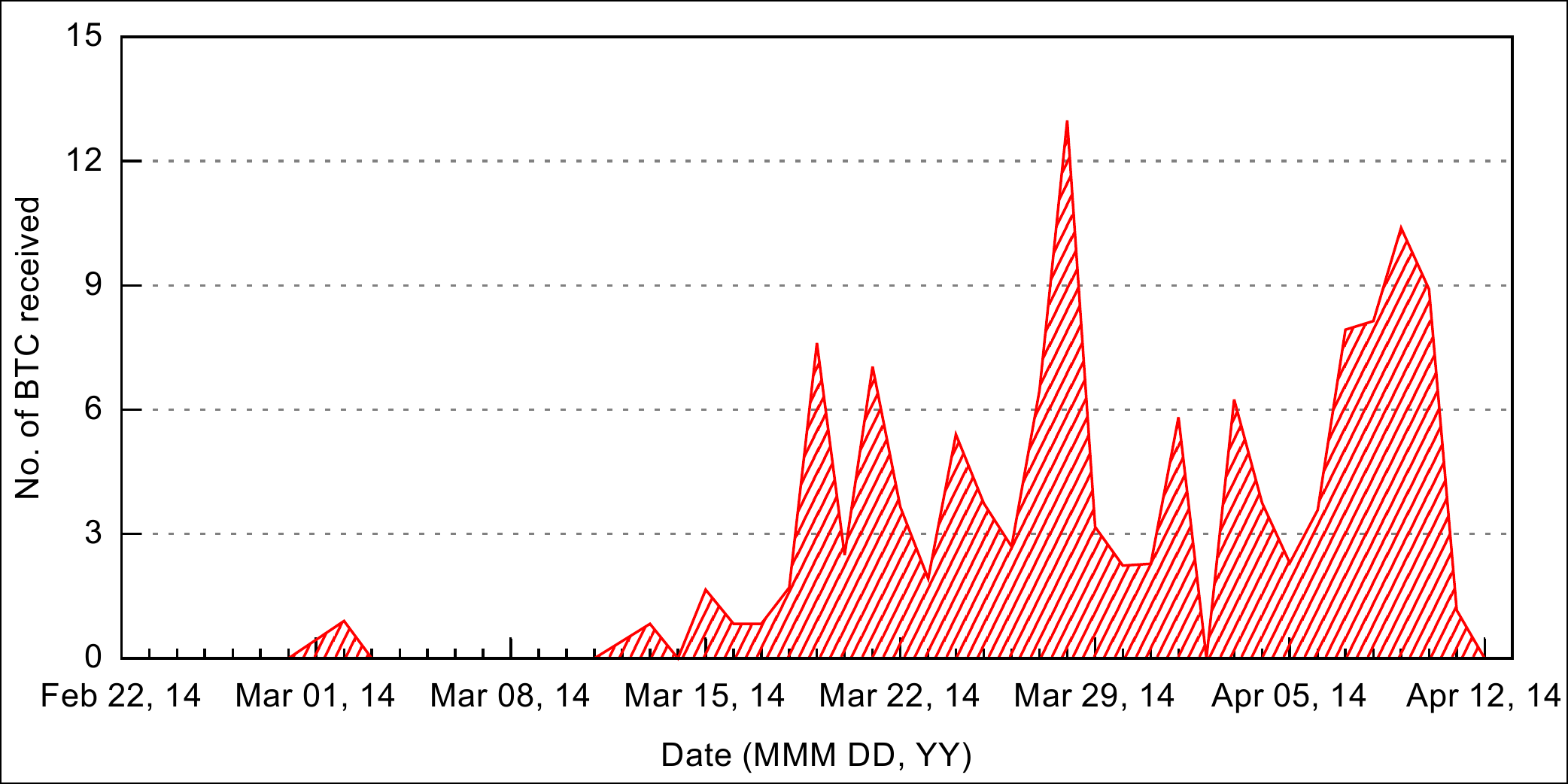}
		\caption[]{Number of Bitcoin received (in ransoms) by $C_{CD}$}
		\label{cryptodefense_BTC_USD1}
	\end{figure}
\begin{figure}[H]
		\centering
		\includegraphics[trim = 2mm 2mm 1mm 2mm, clip, width=\linewidth]{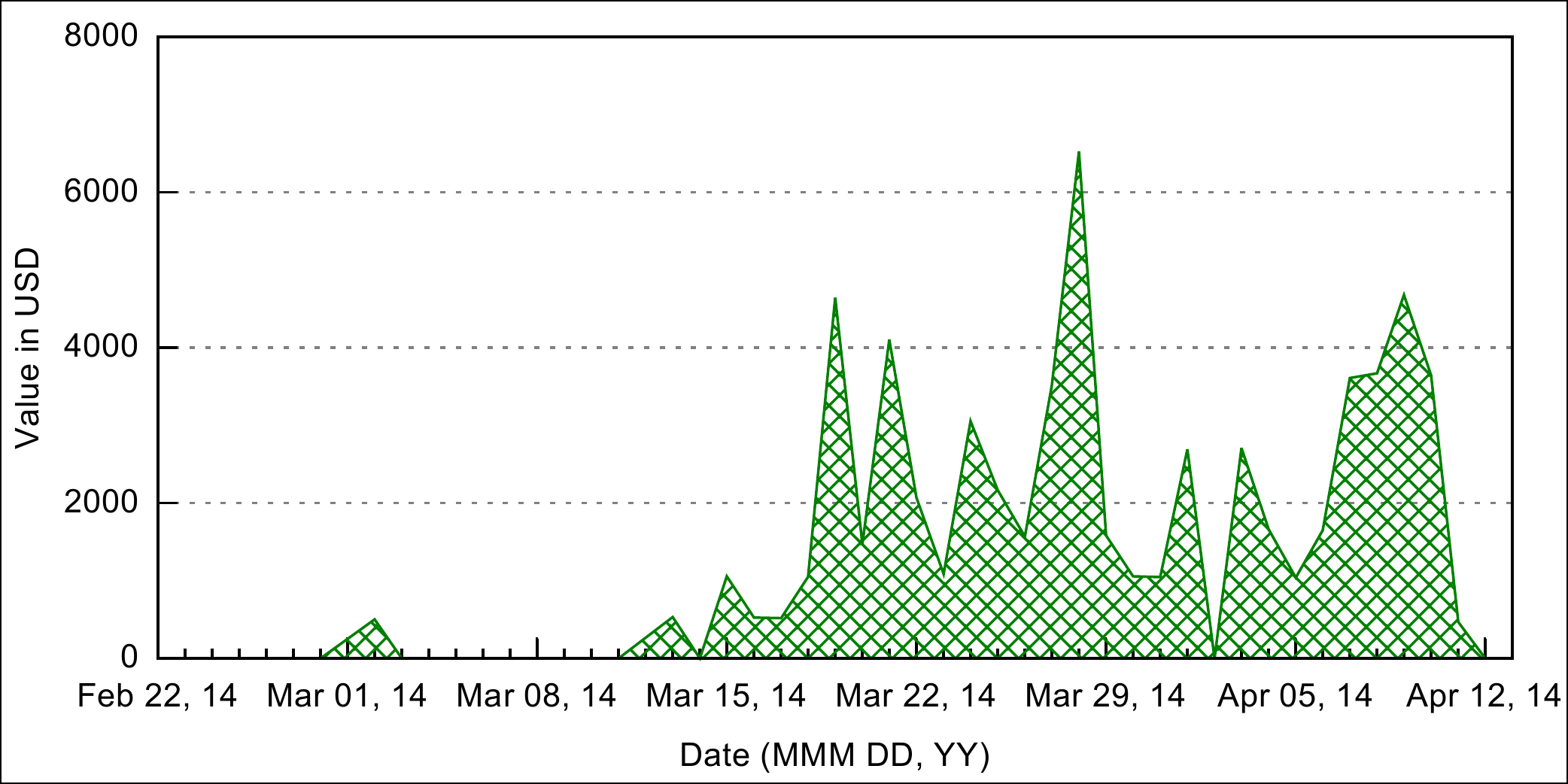}
		\caption[]{USD value of ransoms paid to $C_{CD}$}
		\label{cryptodefense_BTC_USD2}
	\end{figure}

	\figurename{~\ref{cryptodefense_ransom_payment_trend}} shows the total number of ransoms paid, and Figures~\ref{cryptodefense_BTC_USD1}~and~\ref{cryptodefense_BTC_USD2} depict the corresponding number of Bitcoin received and their value in USD. Figures~\ref{cryptodefense_ransom_payment_trend},~\ref{cryptodefense_BTC_USD1},~and~\ref{cryptodefense_BTC_USD2} also depict that on March~28,~2014, $C_{CD}$ collected around 13~BTC in 11~ransom payments, which amounts to approximately USD~6,500. It is the day when it received the maximum number of ransom payments/Bitcoin/USD in a~single~day.

	\par
	In total, we have distinguished 108 ransom payments to $C_{CD}$, which corresponds to 126.70 extorted BTC. Using day-to-day average Bitcoin price, we compute that the value of these ransom payments is equivalent to USD~63,859.49. Table~\ref{cryptodefense_ransom} summarizes the ransoms payments made to CryptoDefense.

	\begin{table}[H]
		\centering
		\resizebox{\columnwidth}{!}
		{
			\begin{tabular}{|c|c|c|c|c|}
				\hline
				\textbf{Ransom} & \textbf{Time period}               & \textbf{Payments} & \textbf{BTC}      & \textbf{USD value} \\ \hline
				\$/\euro500           & \multirow{3}{*}{Feb. 28, '14 - Apr. 11, '14} & 94                      & 96.1758           & 49,271.63              \\ \cline{1-1} \cline{3-5} 
				\$/\euro1,000          &                                    & 14                      & 30.5202           & 14,587.86              \\ \cline{1-1} \cline{3-5} 
				\textbf{Total}  &                                    & \textbf{108}            & \textbf{126.6960} & \textbf{63,859.49}     \\ \hline
			\end{tabular}
		}
		\caption{Summary of ransoms paid to CryptoDefense}
		\label{cryptodefense_ransom}
	\end{table}
	
	Unexpectedly, CryptoDefense has a built-in flaw.  It generates the asymmetric key pair on the victim's system. However, due to the poor implementation of the Microsoft's cryptographic infrastructure, it leaves a local copy of the keys. Anti-ransomware took advantage of this flaw to decrypt victim's computer. Such initiatives saved at least USD~175,000 worth ransoms~\cite{cryptodefense_2}.

	
	\subsection{CryptoWall}
	\ankit{\textit{Introduction:}} CryptoWall is recognized for its use of strong encryption algorithm, unique .CHM file infection mechanism, and strong C\&C activity over the anonymous Tor network. According to Dell SecureWorks Counter Threat Unit (CTU) research team~\cite{secureworks_cryptowall}, CryptoWall infection was spreading from the first half of November~2013. However, the attackers activated it in the first quarter of 2014. The earlier versions of CryptoWall closely impersonated both the appearance and the behavior of the CryptoLocker. CryptoWall affects Windows operating systems by encrypting files using the RSA-2048 (and the AES-256 encryption algorithm from version 3.0) encryption~algorithm.
	\par
	\ankit{\textit{Infection:}} Since its genesis, CryptoWall had spread through several infection vectors, which included drive-by downloads, browser exploit kits (e.g., Angler), and email attachments. Starting from late March~2014, it spread through download links sent via the Cutwail spam botnet and malicious email attachments. Interestingly, from June~2014, the malicious emails included links to popular cloud services such as Dropbox, MediaFire, and Cubby. The links pointed to a ZIP archive that contained the CryptoWall executable. Later these emails used a standard ``missed fax'' decoy and also mimicked message from government agencies or financial institutions that included links to malicious payload hosted over cloud services.
	\par
	\ankit{\textit{Evolution:}} Each version of CryptoWall lasted for a few months until a stealthier and enhanced version emerged.
	\begin{itemize}
		\item CryptoWall 1.0: Initial variants of CryptoWall lacked a unique name. It surfaced with its official name in the first quarter of 2014.
		\item CryptoWall 2.0: It appeared in November~2014. This version was almost identical to the previous version. However, unlike its predecessor, it creates a unique Bitcoin payment address for each victim and uses its own Web-2-Tor gateways.
		\item CryptoWall 3.0: The third version of CryptoWall emerged in January~2015. This version uses a local symmetric (AES-256) key for file encryption. The symmetric key is then encrypted using a unique public (RSA-2048) key generated by the C\&C server. Such process of encryption is much faster as compared to the previous versions.
		\item CryptoWall 4.0: Another updated version with improved communications and better code design to exploit more vulnerabilities appeared in November~2015.
	\end{itemize}
	
	\par
	\ankit{\textit{Ransom demand:}} The attackers originally accepted ransom payments through Litecoin~\cite{secureworks_cryptowall}. However, the only witnessed Litecoin address\footnote{LTv4m4y7NKHCXdw31dSEpTJmP6kXTinWDy} never collected any payment. Additionally, the victims could also pay the ransom via Bitcoin. The amount of ransom fluctuated frequently. Also, the time frame to pay the ransom varied up to seven~days. According to our observation, the demanded ransom and their corresponding timelines (both the dates are included) are as follow:
	\begin{itemize}
		\item \$200 worth BTC between March~2,~2014 and November~4,~2015.
		\item \$500 worth BTC between March~2,~2014 and December~22,~2015.
		\item Late payment of \$600 worth BTC between March~5,~2014 and November~5,~2015. This payment was three times the original ransom amount.
		\item Late payment of \$1,000 worth BTC between March~5,~2014 and December~2,~2015. This payment was two times the original ransom amount.
		\item \$700 worth BTC between March~10,~2014 and December~11,~2015.
		\item Late payment of \$1,400 worth BTC between March~11,~2014 and December~21,~2015. This payment was two times the original ransom amount.
	\end{itemize}
	
	\par
	\ankit{\textit{Associated Bitcoin addresses and transactions:}} We began with forty-two publicly known Bitcoin addresses of CryptoWall. These addresses are listed in Table~\ref{cryptowall_address}. Using these addresses, \ankit{\ma}~generated 2,944 addresses belonging to CryptoWall cluster ($C_{CW}$). Our analysis of transactions \ankit{(obtained using \mb)} to $C_{CW}$ shows that $C_{CW}$, in total, received over 51,000 payments. The total worth of these payments is nearly 88,000~BTC (more than USD~45,000,000). Table~\ref{cryptowall_inwards} presents a summary of the total payments credited to $C_{CW}$.
	
	\begin{table}[H]
		\centering
		\resizebox{\columnwidth}{!}
		{
			\begin{tabular}{|c|c|c|c|c|}
				\hline
				\textbf{\begin{tabular}[c]{@{}c@{}}Payments\end{tabular}} & \textbf{\begin{tabular}[c]{@{}c@{}}BTC\end{tabular}} & \textbf{\begin{tabular}[c]{@{}c@{}}USD value\\(daily highest\\BTC price)\end{tabular}} & \textbf{\begin{tabular}[c]{@{}c@{}}USD value\\(daily average\\BTC price)\end{tabular}} & \textbf{\begin{tabular}[c]{@{}c@{}}USD value\\(daily lowest\\BTC price)\end{tabular}} \\ \hline
				51,278                                                                        & 87,897.8510                                                               & 46,526,673.59                                                                                          & 45,370,589.00                                                                                          & 44,020,263.63                                                                                         \\ \hline
			\end{tabular}
		}
		\caption{Total payments credited to $C_{CW}$ including all ransom and non-ransom payments}
		\label{cryptowall_inwards}
	\end{table}
	
	\par
\begin{figure}[H]
		\centering
		\includegraphics[trim = 2mm 2mm 2mm 2mm, clip, width=\linewidth]{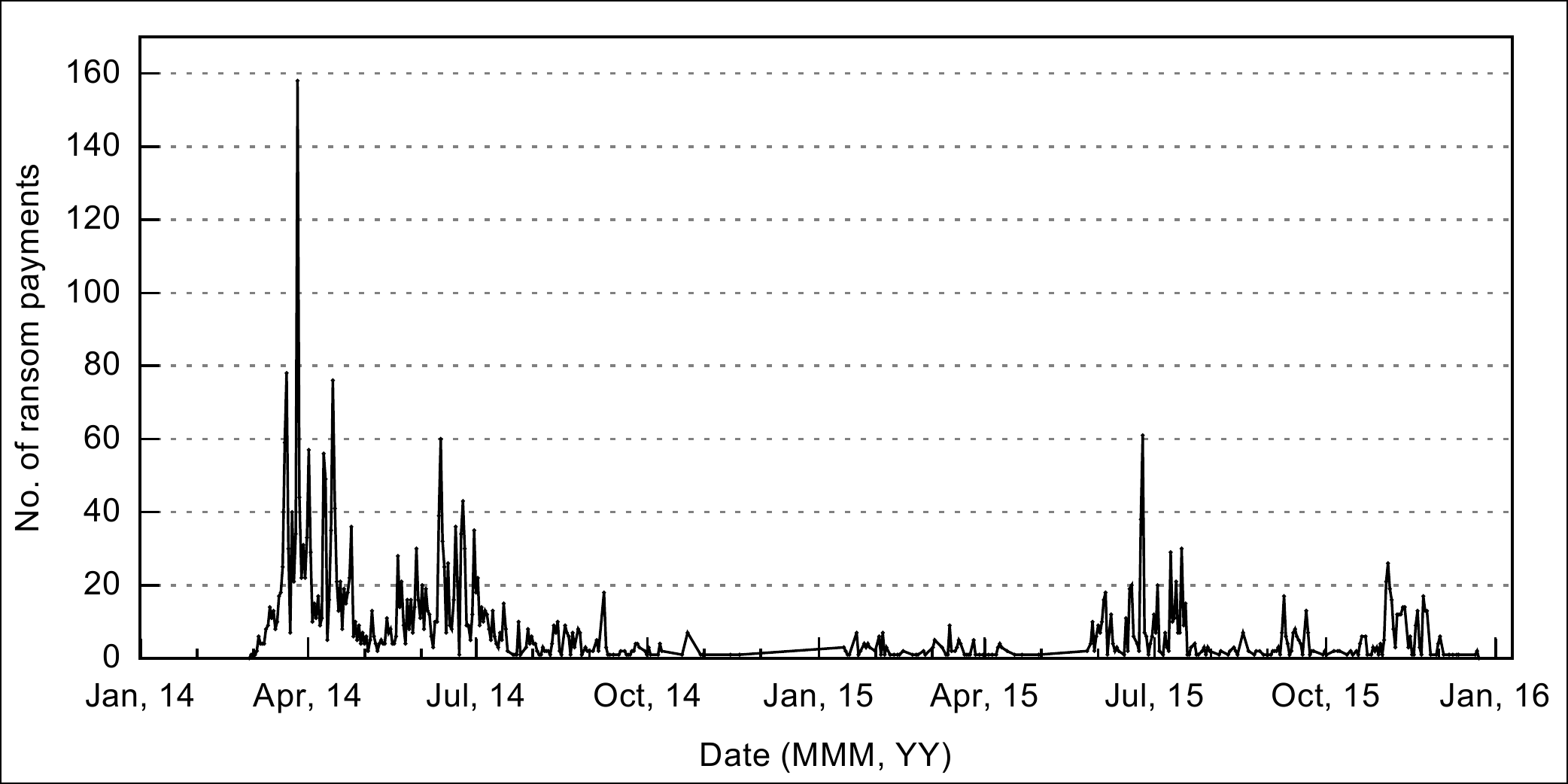}
		\caption[]{Number of ransoms paid to $C_{CW}$}
		\label{cryptowall_ransom_payment_trend}
	\end{figure}
\begin{figure}[H]
		\centering
		\includegraphics[trim = 2mm 2mm 2mm 2mm, clip, width=\linewidth]{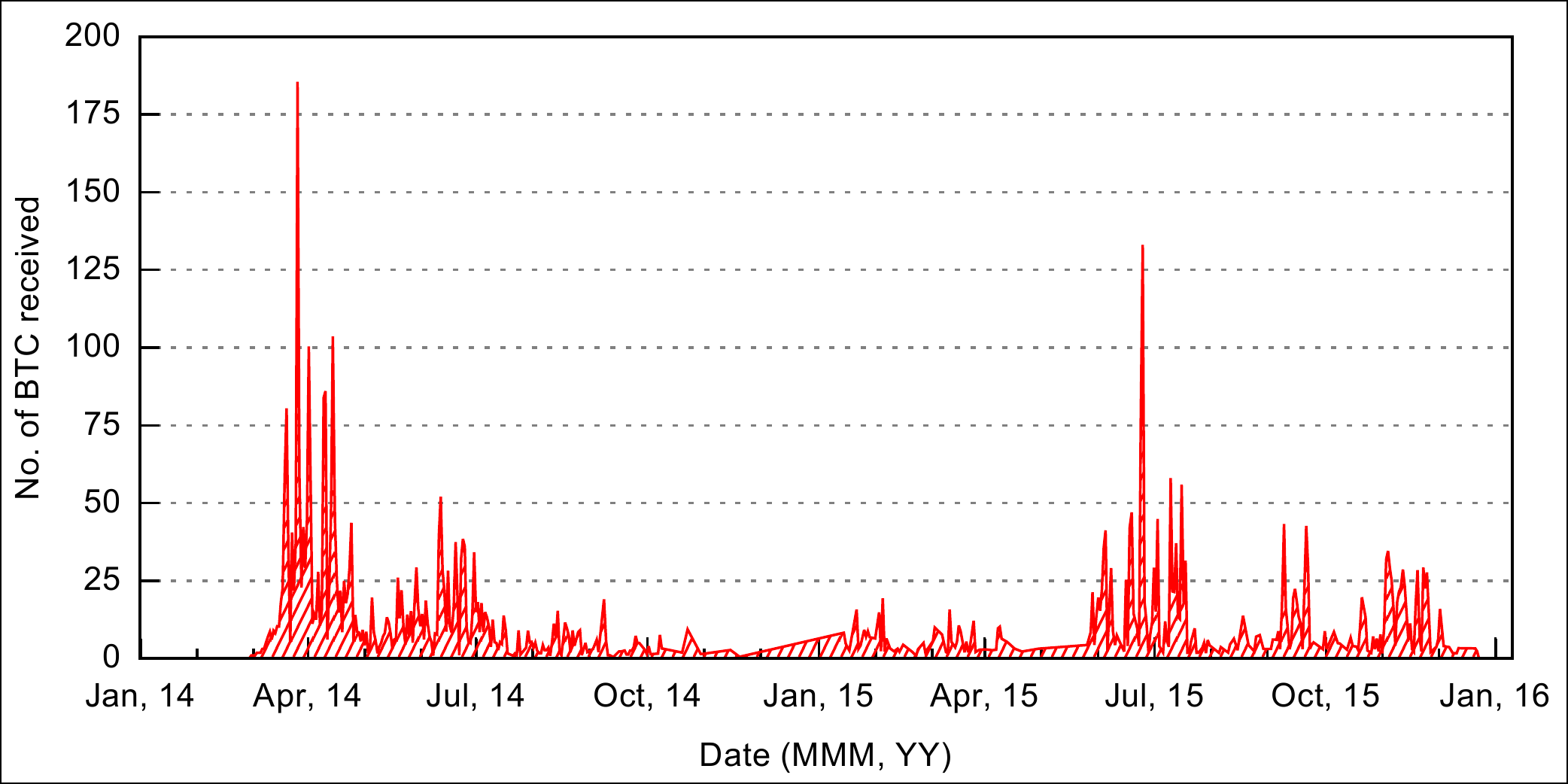}
		\caption[]{Number of Bitcoin received (in ransoms) by $C_{CW}$}
		\label{cryptowall_BTC_USD1}
	\end{figure}
\begin{figure}[H]
		\centering
		\includegraphics[trim = 2mm 2mm 2mm 2mm, clip, width=\linewidth]{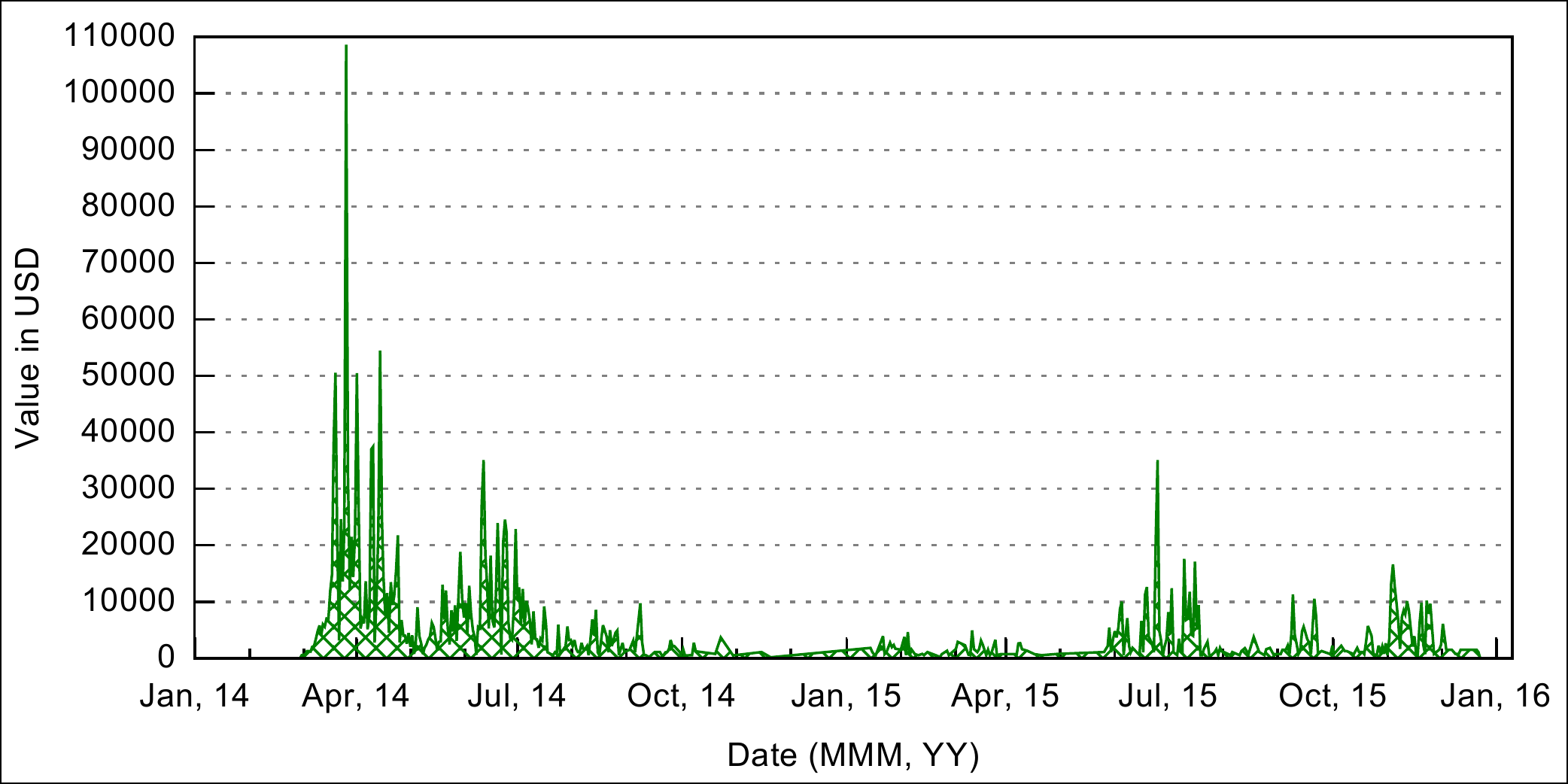}
		\caption[]{USD value of ransoms paid to $C_{CW}$}
		\label{cryptowall_BTC_USD2}
	\end{figure}

	\ankit{\textit{Economy of ransom payments in Bitcoin:}}
	Using the timeline of ransom demands, we carefully analyzed all the transactions \ankit{with \mc~}to distinguish ransom payments and evaluated the net worth generated by such payments. %
	As shown in Figures~\ref{cryptowall_ransom_payment_trend},~\ref{cryptowall_BTC_USD1},~and~\ref{cryptowall_BTC_USD2}, on March~27,~2014, $C_{CW}$ received slightly above 185~BTC in 158 payments. The total value of these payments is over USD~100,000. It is the day when it received the maximum number of ransom payments/Bitcoin/USD in a~single day.
	
	\par
\begin{figure}[H]
		\centering
		\includegraphics[trim = 2mm 2mm 2mm 2mm, clip, width=\linewidth]{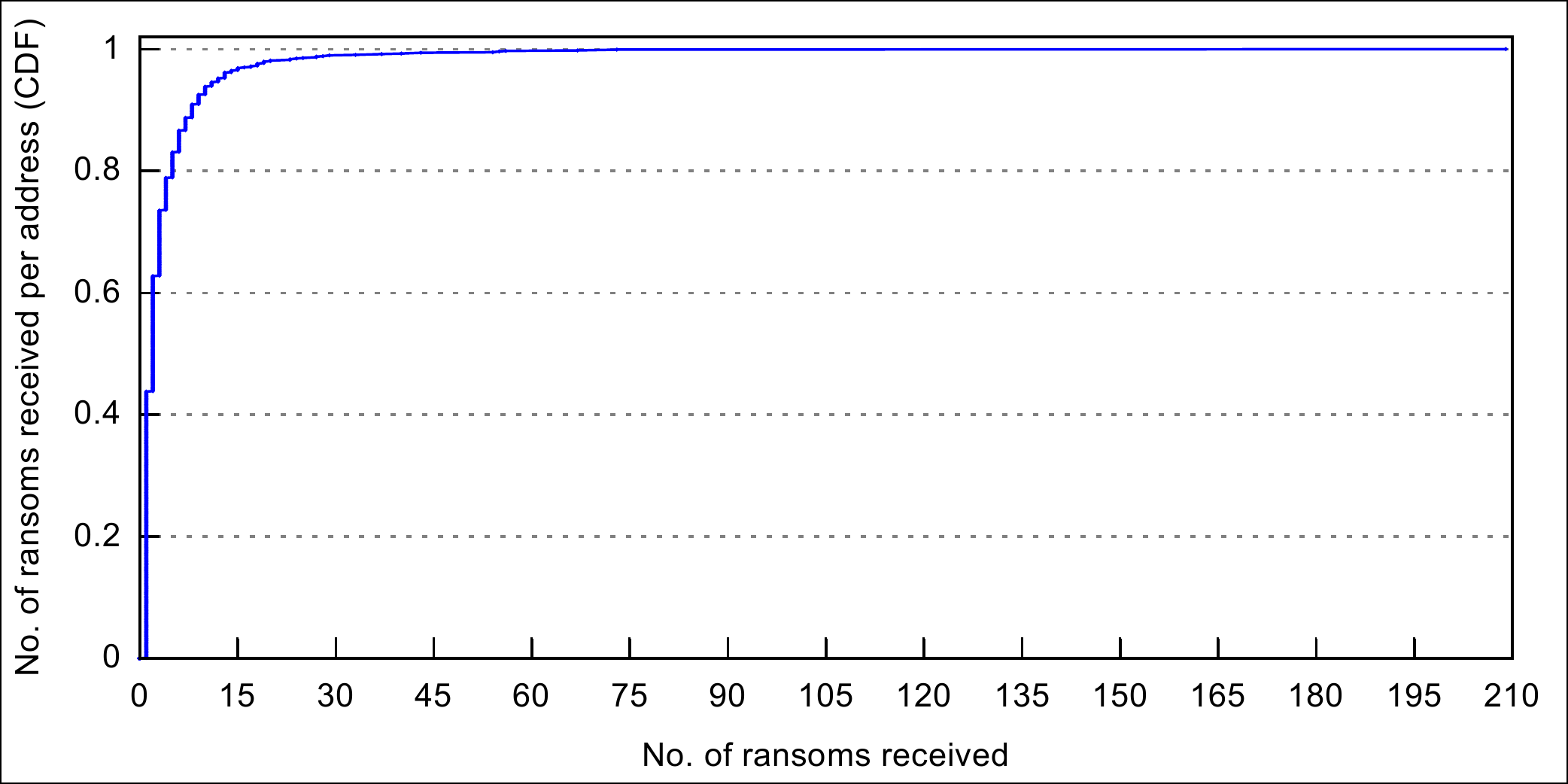}
		\caption[]{CDF of ransoms received per address in $C_{CW}$}
		\label{cryptowall_ransom_Trx_In_CDF}
	\end{figure}
\begin{figure}[H]
		\centering
		\includegraphics[trim = 2mm 2mm 2mm 2mm, clip, width=\linewidth]{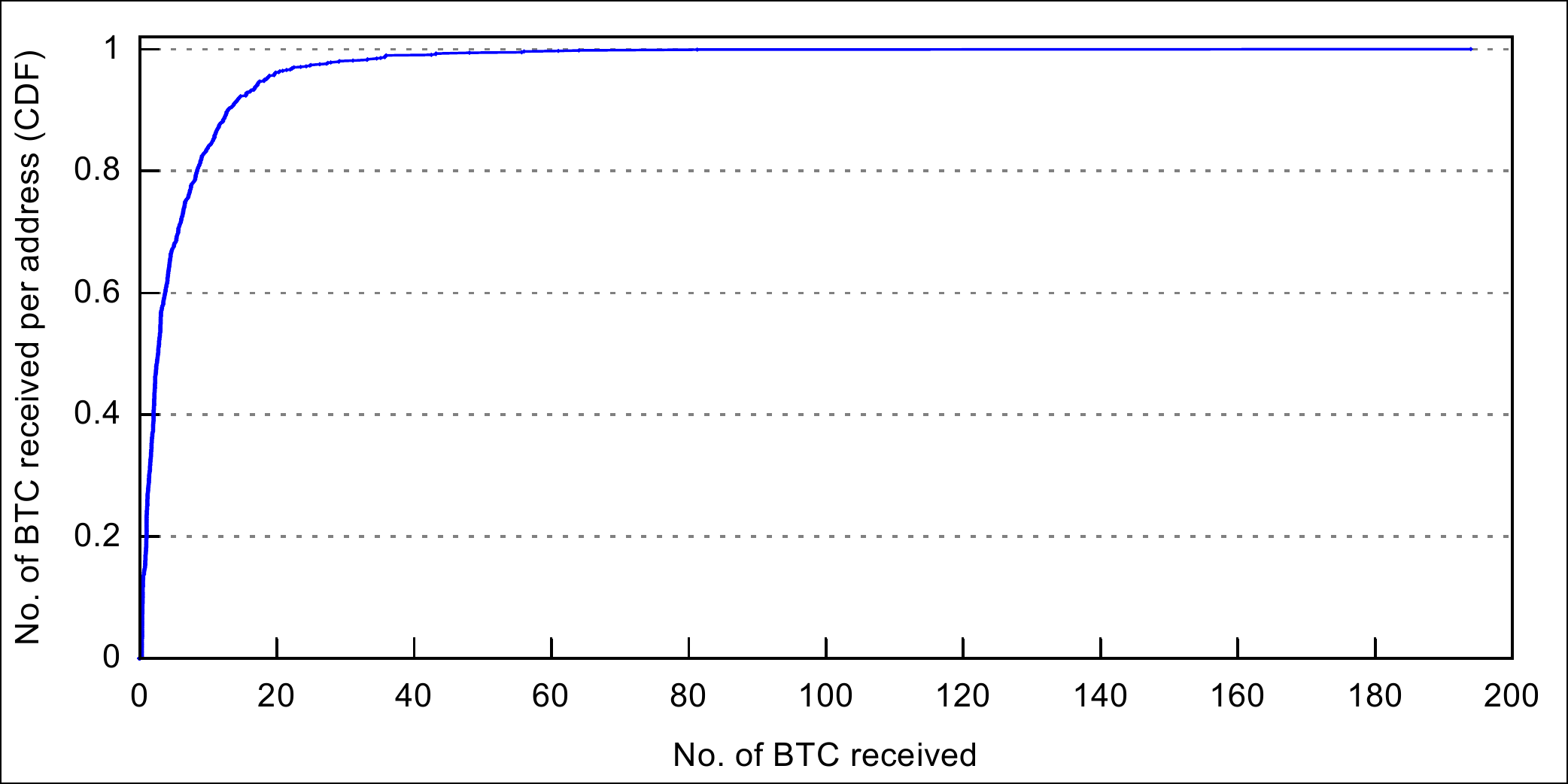}
		\caption[]{CDF of Bitcoin received (in ransoms) per address in $C_{CW}$}
		\label{cryptowall_ransom_BTC_In_CDF}
	\end{figure}

	By investigating the addresses of $C_{CW}$, we observed that approximately 43.77\% Bitcoin addresses received no more than one payment and 40.10\% Bitcoin addresses collected maximum two Bitcoin. On another side, an address\footnote{17AGazRCLStNguMDCxDoj7ZQHvaZBWTJZj} collected 193.94~BTC in 209 ransom payments. These values correspond to the maximum number of Bitcoin and the maximum number of ransom collected by any address in $C_{CW}$. Figures~\ref{cryptowall_ransom_Trx_In_CDF}~and~\ref{cryptowall_ransom_BTC_In_CDF} show the CDF of the number of ransoms and the number of Bitcoin received (in ransoms) per address respectively. 
	
	\par		
	We have identified 3,730 ransom payments to $C_{CW}$, which amount to 5,351.23 extorted BTC. Using day-to-day average Bitcoin price, we calculate that these ransom payments are equivalent to USD~2,220,909.12. Table~\ref{cryptowall_ransom} summarizes the ransoms paid to CryptoWall.
	
	\begin{table}[H]
		\centering
		\resizebox{\columnwidth}{!}
		{
			\begin{tabular}{|c|c|c|c|c|}
				\hline
				\textbf{Ransom} & \textbf{Time period}                 & \textbf{Payments} & \textbf{BTC}        & \textbf{USD value} \\ \hline
				\$200           & Mar. 02, '14 - Nov. 04, '15          & 614                     & 232.3343            & 121,849.84            \\ \hline
				
				\$500           & Mar. 02, '14 - Dec. 22, '15          & 1,631                   & 2220.9167           & 821,741.46            \\ \hline
				
				\$600 (late)    & Mar. 05, '14 - Nov. 05, '15          & 382                     & 444.5144            & 226,558.14            \\ \hline
				
				\$1,000 (late)   & Mar. 05, '14 - Dec. 02, '15          & 423                     & 836.5054            & 422,576.75            \\ \hline
				
				\$700           & Mar. 10, '14 - Dec. 11, '15          & 466                     & 966.7365            & 327,518.98            \\ \hline
				
				\$1,400 (late)   & Mar. 11, '14 - Dec. 21, '15          & 214                     & 650.2256            & 300,663.95            \\ \hline
				\textbf{Total}  & \textbf{Mar. 02, '14 - Dec. 22, '15} & \textbf{3,730}          & \textbf{5,351.2329} & \textbf{2,220,909.12} \\ \hline
			\end{tabular}
		}
		\caption{Summary of ransoms paid to CryptoWall}
		\label{cryptowall_ransom}
	\end{table}
	
	\par
	Moreover, according to the report by CTU researchers~\cite{secureworks_cryptowall}, CryptoWall attackers allowed the victims to decrypt their system by paying a further increased amount even after the expired deadline. 
	Although, we have not directly observed any sample of CryptoWall demanding such compensations. Nevertheless, the timing and the volume of such payments suggest that these payments pertain to ransoms. Table~\ref{cryptowall_weird_trx} summarizes such payments.
	\begin{table}[H]
		\centering
		\resizebox{\columnwidth}{!}
		{
			\begin{tabular}{|c|c|c|c|c|}
				\hline
				\textbf{Amount} & \textbf{Time period}                 & \textbf{Payments} & \textbf{BTC}       & \textbf{USD value} \\ \hline
				\$1,500         & Mar. 12, '14 - Dec. 12, '15          & 222                      & 678.7995           & 333,587.51            \\ \hline
				\$1,750         & Mar. 12, '14 - Nov. 04, '15          & 192                      & 647.5063           & 336,578.87            \\ \hline
				\$2,000         & Mar. 06, '14 - Jul. 06, '14          & 170                      & 650.7245           & 339,794.84            \\ \hline
				\$10,000        & Mar. 11, '14 - Jul. 11, '14          & 131                      & 2623.3381          & 1,316,778.41          \\ \hline
				\textbf{Total}  & \textbf{Mar. 06, '14 - Dec. 12, '15} & \textbf{715}             & \textbf{4600.3684} & \textbf{2,326,739.63} \\ \hline
			\end{tabular}
		}
		\caption{Summary of high value (possibly ransom) payments to CryptoWall}
		\label{cryptowall_weird_trx}
	\end{table}
	
	\par
	If we add these payments to the original ransom payments, then the revenue of CryptoWall reaches nearly 10,000 BTC, i.e., approximately USD~4,500,000.

	
	\subsection{DMA Locker}
	\ankit{\textit{Introduction:}} DMA Locker is one of the most actively developed and updated ransomware so far. From encryption algorithm to network communication, cybercrooks perpetually updated each component of DMA Locker. Initially, it used only the symmetric key cryptography for file encryption. However, later versions employ a stronger encryption approach by combining the AES-256 and the RSA-2048 encryption algorithms. It affects Windows operating system. 
	
	\par
	\ankit{\textit{Infection:}} The distribution mechanism of DMA Locker also evolved with the course of time. The malicious payload was hosted on compromised websites, and their links were distributed via email spamming. It also infiltrated by hacking Remote Desktops. The latest edition of the ransomware also spread via Neutrino exploit kit~\cite{dmalocker_2}.
	
	\par
	\ankit{\textit{Evolution:}} The development timeline of DMA Locker is discussed below:
	
	\begin{itemize}
		\item DMA Locker 1.0: The first version of DMA Locker was noticed in the last week of December~2015 with support for two languages: Polish and English. It performs file encryption by using the AES-256 algorithm in ECB mode. It uses a single AES key to encrypt target files, which is stored in the binary and deleted after use.
		
		\item DMA Locker 2.0: On February~3,~2016, DMA Locker was updated to use separate keys for each file. After encrypting a file, it encrypts the used AES key by RSA public key and stores the encrypted AES key in the encrypted file. The public key for RSA encryption comes hardcoded in the binary.
		
		\item DMA Locker 3.0: 
		Due to weak implementation of the random number generator, the AES key generated by the previous version can be guessed. In view to fix the flaw, the third edition was released on February~22,~2016. However, the entire campaign used the same RSA key-pair. Meaning that single private key can be reused for decrypting other infected systems.
		
		\item DMA Locker 4.0: The latest version of DMA Locker was released on May~19,~2016. This version generates a unique RSA key-pair on the server for each victim. Unlike previous versions, DMA Locker 4.0 can not work offline because it is designed to download the asymmetric public key from the server~\cite{dmalocker_1}.
	\end{itemize}
	
	\par
	\ankit{\textit{Ransom demand:}} The cybercrooks behind DMA Locker accepted ransom payments through Bitcoin. DMA Locker 4.0 gives payment instructions on a website. The website was a regularly (not Tor-based) hosted site.  Surprisingly, the payment site used the same IP address as the C\&C. Similar to other components, the ransom amount was also updated with time. Moreover, the first two versions stipulate a strict deadline of four~days to pay the ransom. Other versions allow an extension of three~days at the cost of an increased ransom. The demanded ransom and their corresponding timelines (both the dates are included) are as follow:
	
	\begin{itemize}
		\item 1 BTC between December~28,~2015 and July~22,~2016.
		
		\item 1.3 BTC between January~19,~2016 and May~30,~2016. 
		
		\item 2 BTC between January~28,~2016 and July~22,~2016 to allowing a three-day ransom period.
		
		\item 4 BTC between February~22,~2016 and June~5,~2016 to allowing a three-day ransom period.
		
		\item 8 BTC as late fee between February~22,~2016 and August~5,~2016. 
		
		\item 1.5 BTC as late fee between May~19,~2016 and July~11,~2016.
		
		\item 3 BTC between May~24,~2016 and August~25,~2016.
	\end{itemize}

	\par
	\ankit{\textit{Associated Bitcoin addresses and transactions:}} To understand the economic impact of DMA Locker, we began with eight Bitcoin addresses listed in Table~\ref{dmalocker_address}. Using these addresses, \ankit{\ma}~generated 28 addresses belonging to DMA Locker cluster ($C_{DL}$). Our scrutiny of transactions \ankit{(obtained using \mb)} to $C_{DL}$ shows that $C_{DL}$ received altogether 298 payments, i.e., more than 1,400~BTC (over USD~580,000). Table~\ref{dmalocker_inwards} presents a summary of the total payments credited to $C_{DL}$.
	
	\begin{table}[H]
		\centering
		\resizebox{\columnwidth}{!}
		{
			\begin{tabular}{|c|c|c|c|c|}
				\hline
				\textbf{\begin{tabular}[c]{@{}c@{}}Payments\end{tabular}} & \textbf{\begin{tabular}[c]{@{}c@{}}BTC\end{tabular}} & \textbf{\begin{tabular}[c]{@{}c@{}}USD value\\(daily highest\\BTC price)\end{tabular}} & \textbf{\begin{tabular}[c]{@{}c@{}}USD value\\(daily average\\BTC price)\end{tabular}} & \textbf{\begin{tabular}[c]{@{}c@{}}USD value\\(daily lowest\\BTC price)\end{tabular}} \\ \hline
				298                                                                           & 1,433.3463                                                                 & 593,498.26                                                                                              & 580,763.95                                                                                              & 567,543.86                                                                                             \\ \hline
			\end{tabular}
		}
		\caption{Total payments credited to $C_{DL}$ including all ransom and non-ransom payments}
		\label{dmalocker_inwards}
	\end{table}

	\par
\begin{figure}[H]
		\centering
		\includegraphics[trim = 2mm 2mm 2mm 2mm, clip, width=\linewidth]{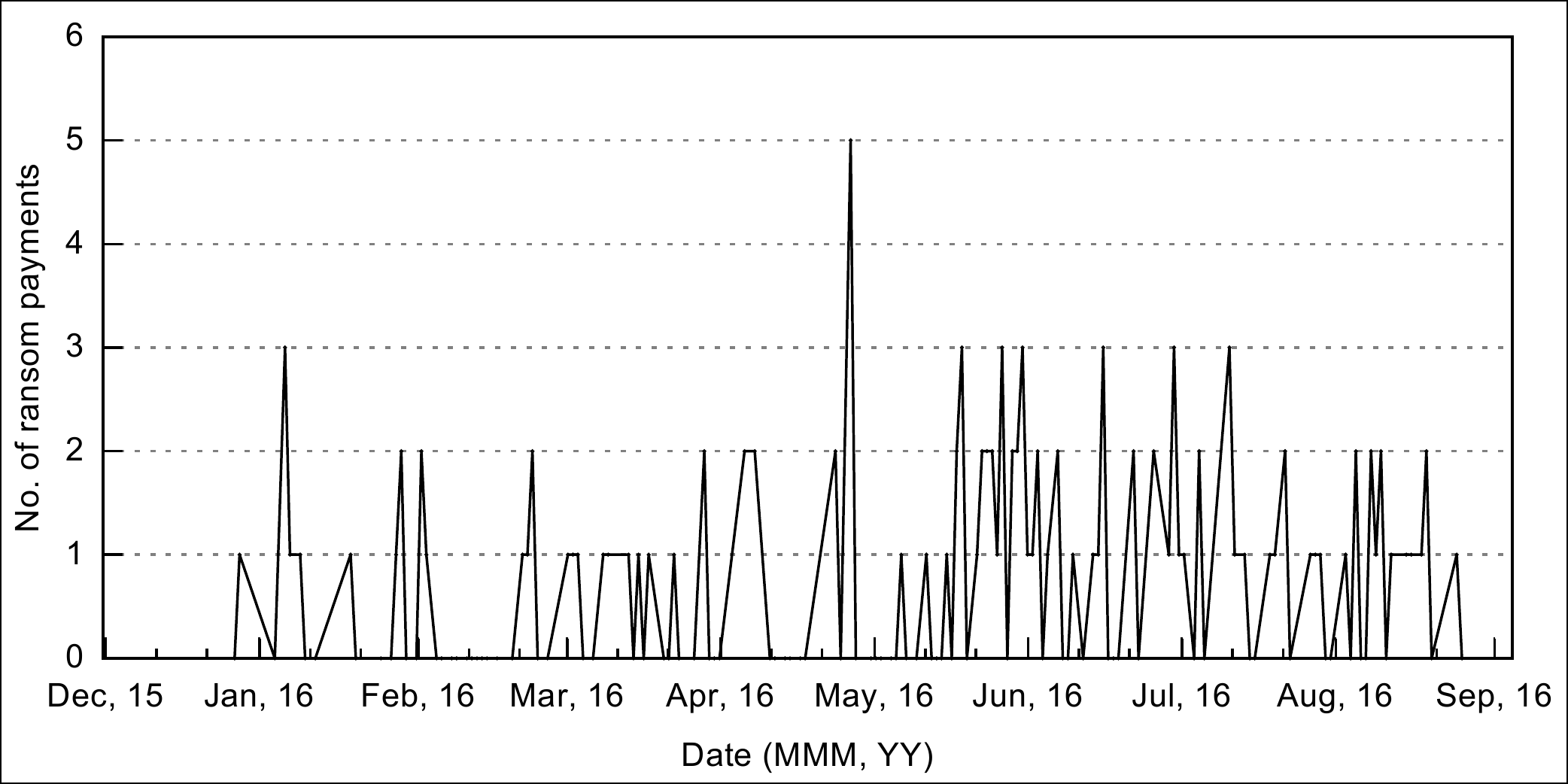}
		\caption[]{Number of ransoms paid to $C_{DL}$}
		\label{dmalocker_ransom_payment_trend}
	\end{figure}
\begin{figure}[H]
		\centering
		\includegraphics[trim = 2mm 2mm 2mm 2mm, clip, width=\linewidth]{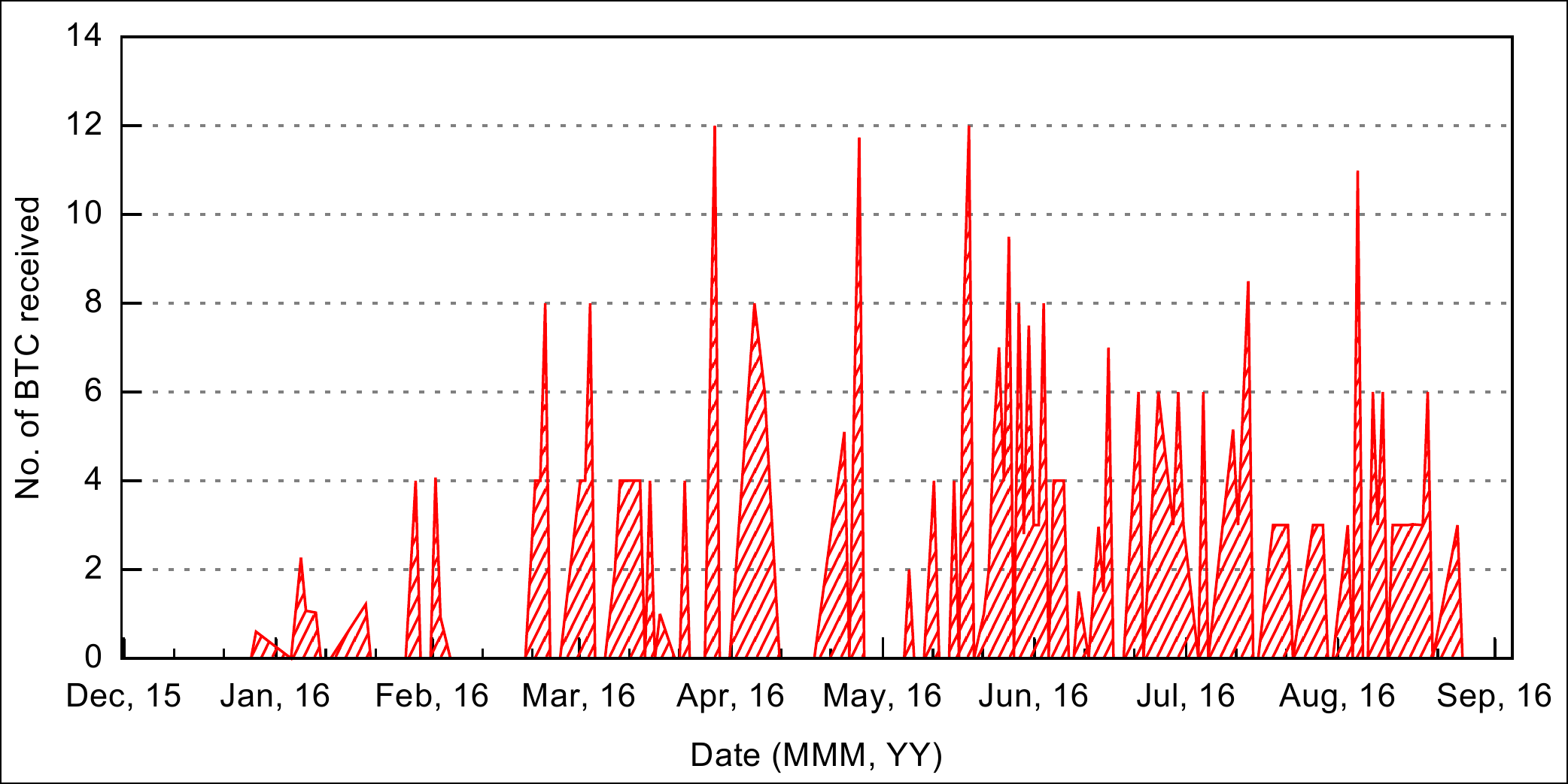}
		\caption[]{Number of Bitcoin received (in ransoms) by $C_{DL}$}
		\label{dmalocker_BTC_USD1}
	\end{figure}
\begin{figure}[H]
		\centering
		\includegraphics[trim = 2mm 2mm 2mm 2mm, clip, width=\linewidth]{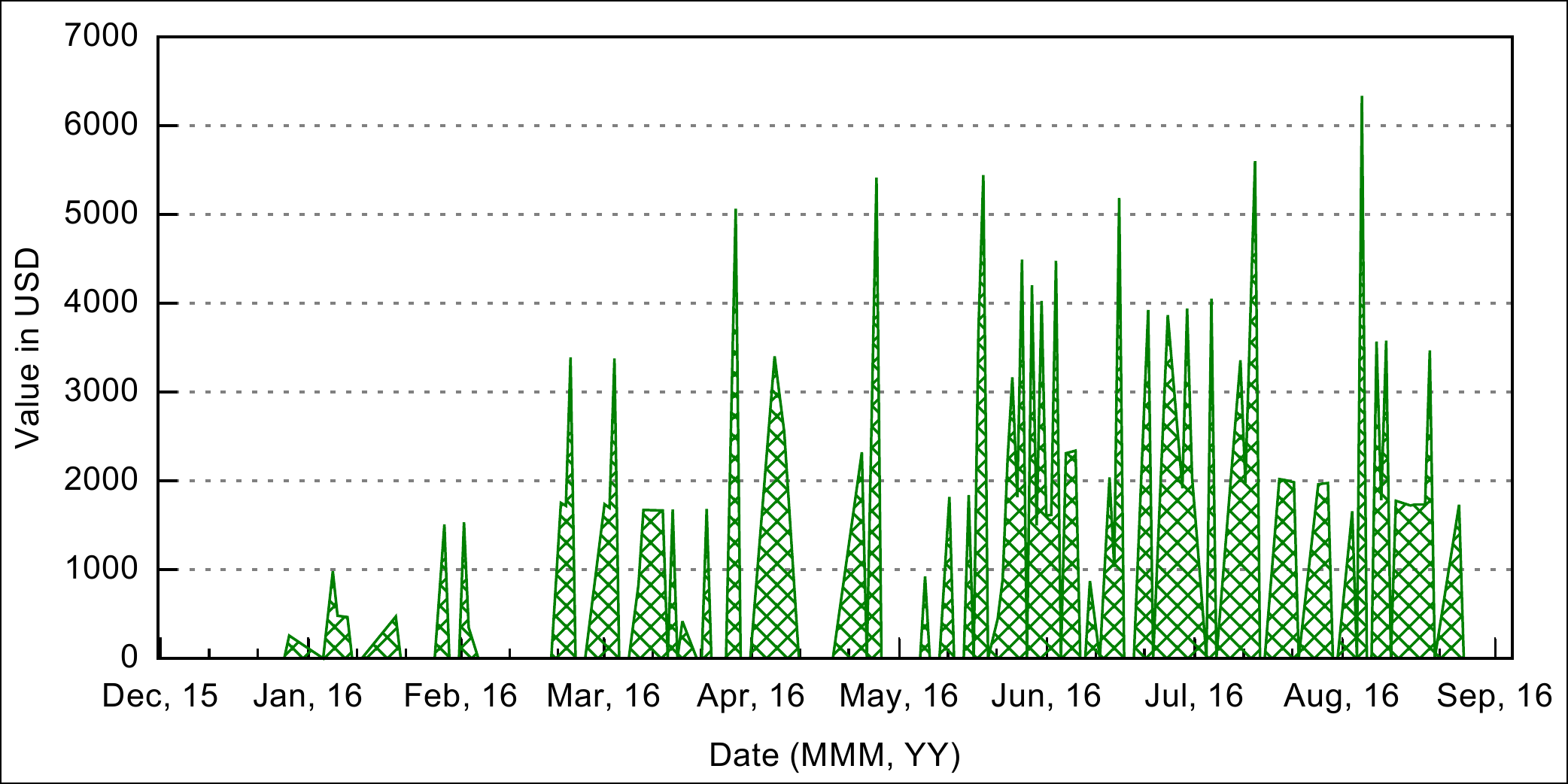}
		\caption[]{USD value of ransoms paid to $C_{DL}$}
		\label{dmalocker_BTC_USD2}
	\end{figure}

	\ankit{\textit{Economy of ransom payments in Bitcoin:}} \ankit{We used \mc, guided by the timeline of ransom demands, to separate ransom payments.} \figurename{~\ref{dmalocker_ransom_payment_trend}} depicts the total number of ransoms paid by date. $C_{DL}$ received 5~payment on April~27,~2016, which is the maximum number of ransoms paid in a single day. On another side, as shown in \figurename{~\ref{dmalocker_BTC_USD1}}, $C_{DL}$ collected 12~BTC on May~19,~2016, which corresponds to the maximum number of Bitcoin received in a single day. Furthermore, $C_{DL}$ received over USD~6,300 on August~5,~2016, which stands for the maximum USD received in a single day, see \figurename{~\ref{dmalocker_BTC_USD2}}.
	
	\par
	We further found that around 30\% addresses in $C_{DL}$ collected no more than one payment and nearly 20\% Bitcoin addresses received less than one Bitcoin. Furthermore, an address\footnote{1LPgKoErPUeM92SDY5axJzYCdQbeiRHD6i} collected 112.87~BTC in 38~ransom payments. These values correspond to the maximum number of Bitcoin and the maximum number of ransom collected by any address in $C_{DL}$. Table~\ref{dmalocker_ransom} summarizes the ransoms paid to DMA Locker.

	
	

	\begin{table}[H]
		\centering
		\resizebox{\columnwidth}{!}
		{
			\begin{tabular}{|c|c|c|c|c|}
				\hline
				\textbf{Ransom} & \textbf{Time period}       & \textbf{Payments} & \textbf{BTC}      & \textbf{USD value} \\ \hline
				1 BTC           & Dec. 28, '15 - Jul. 22, '16          & 16                      & 14.7526           & 7,052.37               \\ \hline
				1.3 BTC         & Jan. 19, '16 - May 30, '16           & 4                       & 5.2470            & 2,424.01               \\ \hline
				2 BTC           & Jan. 28, '16 - Jul. 22, '16          & 16                      & 32.0809           & 16,638.46              \\ \hline
				4 BTC           & Feb. 22, '16 - Jun. 05, '16          & 33                      & 131.9950          & 60,443.98              \\ \hline
				8 BTC (late)    & Feb. 22, '16 - Aug. 05, '16          & 4                       & 32.4892           & 16,960.59              \\ \hline
				1.5 BTC (late)  & May 19, '16 - Jul. 11, '16           & 6                       & 8.9147            & 5,136.87               \\ \hline
				3 BTC           & May 24, '16 - Aug. 25, '16           & 38                      & 113.9797          & 69,506.49              \\ \hline
				\textbf{Total}  & \textbf{Dec. 28, '15 - Aug. 25, '16} & \textbf{117}            & \textbf{339.4591} & \textbf{178,162.77}    \\ \hline
			\end{tabular}
		}
		\caption{Summary of ransoms paid to DMA Locker}
		\label{dmalocker_ransom}
	\end{table}
	
	\par
	We have identified 117 ransom payments to $C_{DL}$, which contribute to a total of 339.46 extorted BTC. Using day-to-day average Bitcoin price, we estimate that these ransom payments value USD~178,162.77.


	\subsection{Petya}
\begin{figure}[H]
		\centering
		\includegraphics[trim = 2mm 2mm 2mm 2mm, clip, scale=0.4]{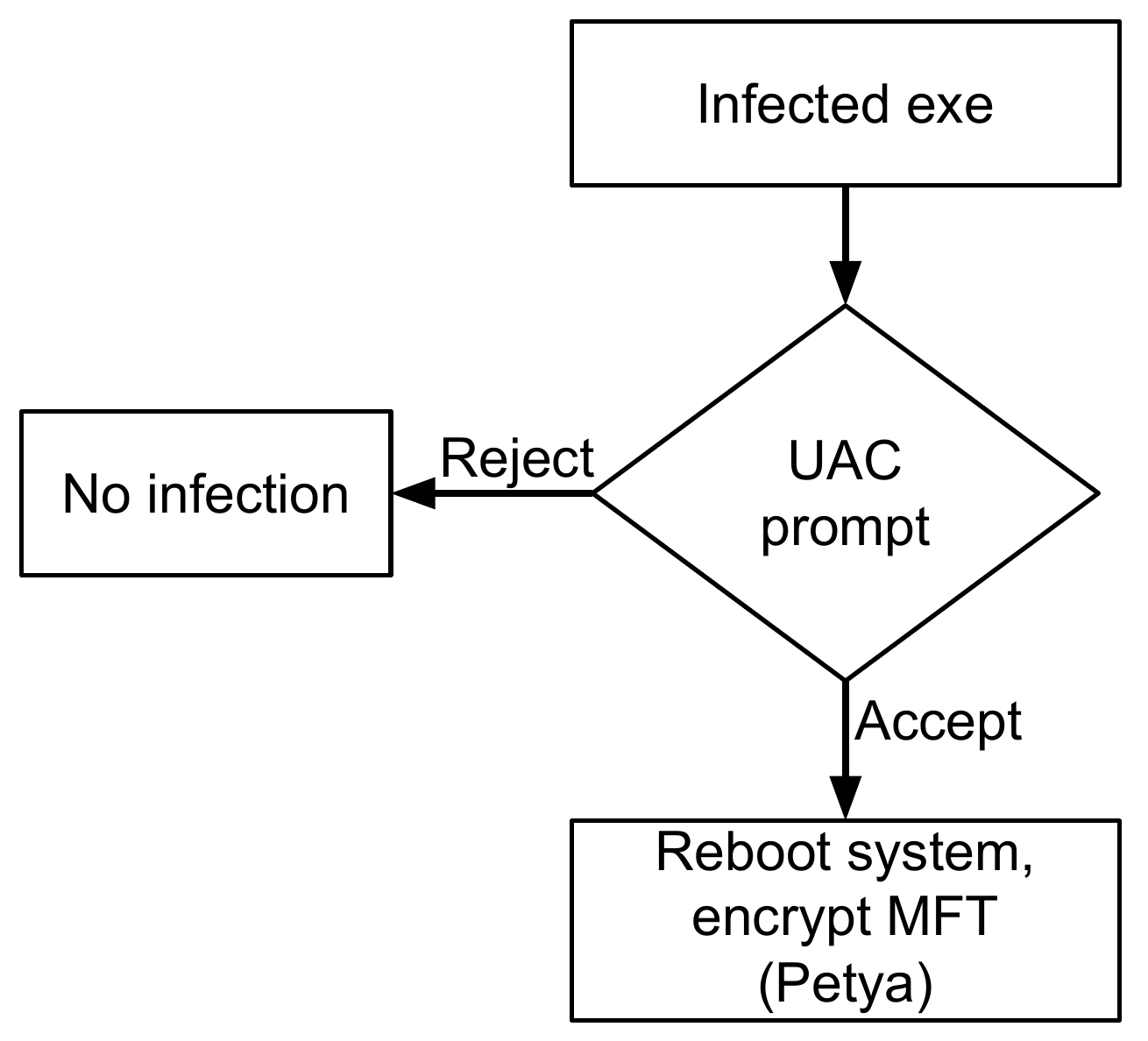}
		\caption[]{Workflow of Petya}
		\label{flowchart_petya}
	\end{figure}

	\ankit{\textit{Introduction:}} Initially seen in March~2016, this family of malware denies access to the full system by targeting the low-level structures on the disk. Petya spread via emails, and was delivered as Windows executable with an icon of a PDF document. Upon running, it opens a User Account Control (UAC) window. Accepting UAC allows Petya to run. In this case, it overwrites the Master Boot Record~(MBR) with a custom bootloader that loads a malicious kernel. Then, this kernel encrypts the Master File Table~(MFT) using Salsa20 stream cipher with a 32-byte long key, which leaves file system unreadable. \figurename{~\ref{flowchart_petya}} depicts the full process of Petya.
	
	\par
\begin{figure}[!htbp]
		\centering
		\includegraphics[trim = 2mm 2mm 2mm 2mm, clip, scale=0.4]{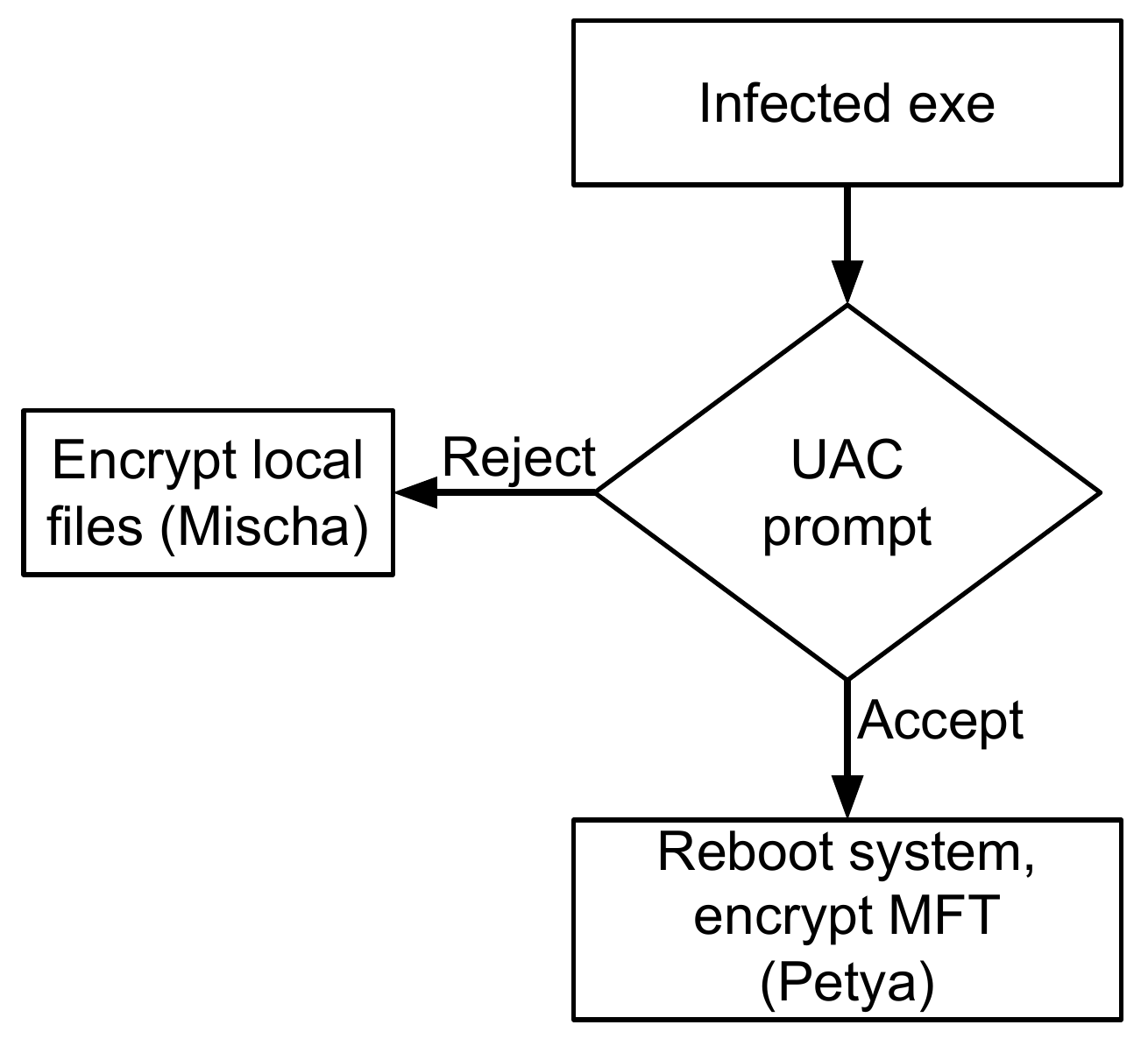}
		\caption[]{Workflow of Mischa}
		\label{flowchart_mischa}
	\end{figure}

	\ankit{\textit{Mischa:}} In May~2016, the malware was modified to integrate another malicious payload known as Mischa. Mischa was designed as a backup strategy to Petya. Altogether, they target different (both high-level and low-level) layers of a system. In this version, denying the UAC prompt directs Mischa to encrypt local files on the victim computer; otherwise, Petya proceeds. \figurename{~\ref{flowchart_mischa}} depicts the full process of Mischa. Both Petya and Mischa can work offline without communicating with their C\&C. The payload from the dropper\footnote{The file that launches a malware.} uses CryptGenRandom function from the Windows CryptoAPI library to generate a random encryption key. Mischa uses a CBC-style file encryption utilizing a randomly generated key along with the previously generated master key. Interestingly, Mischa can encrypt documents as well as executables~\cite{petya_avast_mischa}. The cybercriminals also offered RaaS through their own affiliate~program. 
	\par
\begin{figure}[H]
		\centering
		\includegraphics[trim = 0mm 0mm 0mm 1mm, clip, scale=0.41]{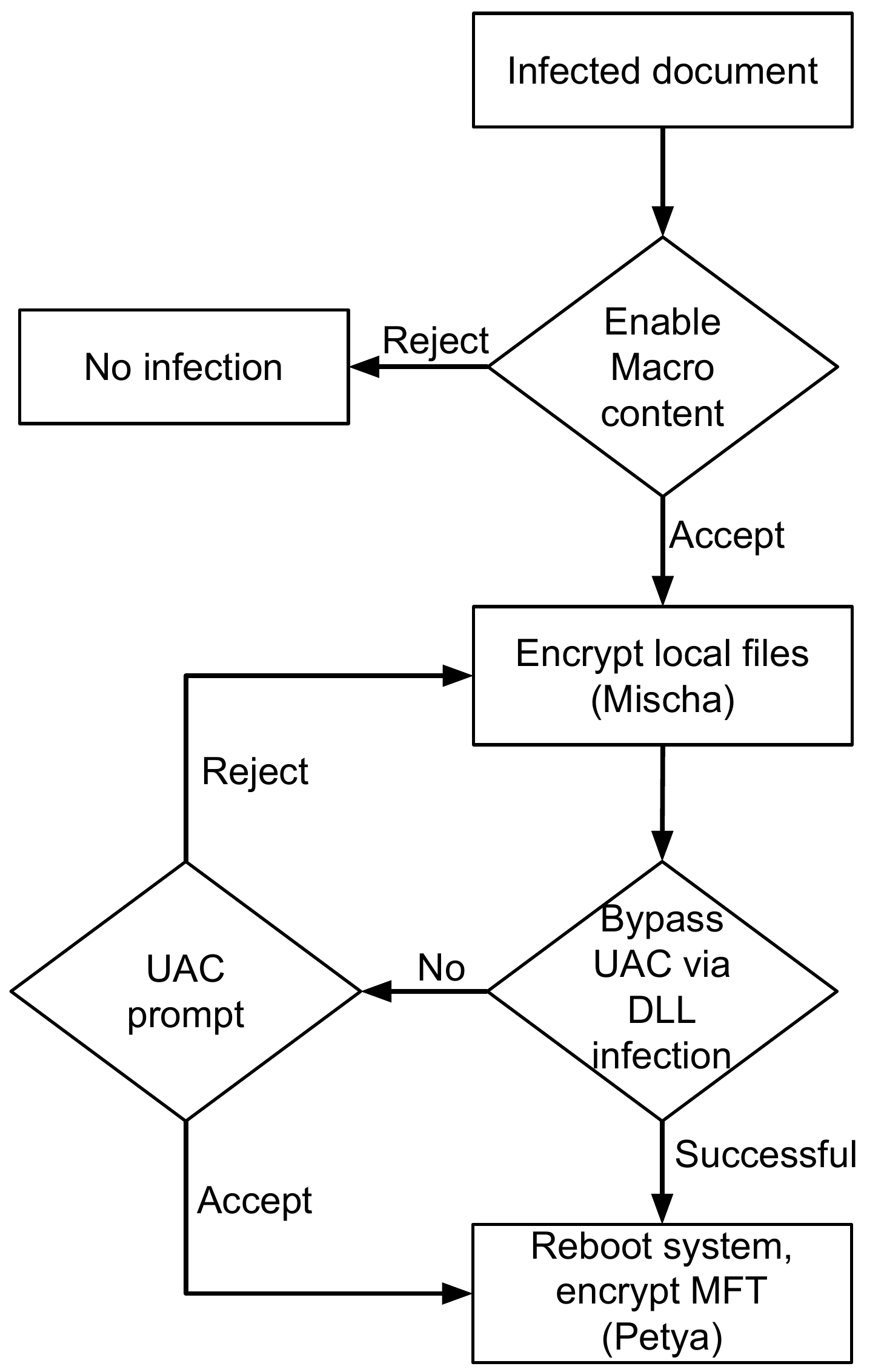}
		\caption[]{Workflow of GoldenEye}
		\label{flowchart_goldeneye}
	\end{figure}

	\ankit{\textit{GoldenEye:}} The malware was again rebranded as GoldenEye in early December~2016. In contrast with the previous versions, GoldenEye executes both payloads, where possible. Similar to its predecessors, it was also distributed via email. But, the payload was attached to an MS Excel document. The document prompts the user to enable Macro content. Enabling Macro content executes a malicious Visual Basic Script, which runs the Mischa payload to encrypt documents on the system. After Mischa finishes, it attempts to gain system privileges via DLL injection (Windows 7 - 8.1), or a UAC prompt is shown (Windows 10). If DLL infection succeeds or the UAC prompt is accepted, Petya payload encrypts the MFT. \figurename{~\ref{flowchart_goldeneye}} depicts the full process of GoldenEye.
	
	\par
	\ankit{\textit{NotPetya:}} The latest variant of Petya surfaced on June~27,~2017. Kaspersky unofficially named\footnote{www.kaspersky.com/blog/new-ransomware-epidemics/17314/} it NotPetya/ExPetr due to significant differences in its operations compared to the earlier versions. Initially, NotPetya was distributed as an update to MeDoc\footnote{www.medoc.ua/uk} accounting software prevalent in Ukraine. After infiltration, it self-propagates via two methods. One of the methods is the EternalBlue exploit, which is an exploit of Windows'~Server Message Block (SMB) protocol. The same exploit is also used by WannaCry ransomware, which was released only a month before NotPetya. It can also spread across network shares by Windows Management Instrumentation Command-line (WMIC), for which it uses credentials acquired from the local machine. In contrast with other ransomware, it focuses on the local network to spread rather than the Internet. NotPetya works as a destructive data wiper tool due to its inability to restore the encrypted sectors of the physical~disk~\cite{petya_symantec}.
	
	\par			
	\ankit{\textit{Associated Bitcoin addresses and transactions:}} We discuss the financial transactions associated with only NotPetya because the payments received by the address clusters generated for Mischa and GoldenEye (using addresses listed in Tables~\ref{mischa_address}~and~\ref{goldeneye_address} respectively) were significantly less~(no more than USD~3) than the demanded ransoms (roughly USD~1,000). For NotPetya, cybercriminals used a single Bitcoin payment address to collect a fixed ransom of USD~300. The address is listed in Table~\ref{NotPetya_address}. NotPetya cluster ($C_{NP}$) generated by \ankit{\ma}~also had only one Bitcoin addresses. \ankit{We acquired the detailed transaction history of this address using \mb.} $C_{NP}$ received exactly 70 payments. These payments worth slightly above 4~BTC (over USD~10,000). Table~\ref{NotPetya_inwards} summarizes the payments credited in $C_{NP}$.
	
	\begin{table}[H]
		\centering
		\resizebox{\columnwidth}{!}
		{
			\begin{tabular}{|c|c|c|c|c|}
				\hline
				\textbf{\begin{tabular}[c]{@{}c@{}}Payments\end{tabular}} & \textbf{\begin{tabular}[c]{@{}c@{}}BTC\end{tabular}} & \textbf{\begin{tabular}[c]{@{}c@{}}USD value\\(daily highest\\BTC price)\end{tabular}} & \textbf{\begin{tabular}[c]{@{}c@{}}USD value\\(daily average\\BTC price)\end{tabular}} & \textbf{\begin{tabular}[c]{@{}c@{}}USD value\\(daily lowest\\BTC price)\end{tabular}} \\ \hline
				70                                                                            & 4.1787                                                                      & 10,717.74                                                                                              & 10,284.42                                                                                              & 9,958.33                                                                                              \\ \hline
			\end{tabular}
		}
		\caption{Total payments credited to $C_{NP}$ including all ransom and non-ransom payments}
		\label{NotPetya_inwards}
	\end{table}
	
	\par
\begin{figure}[H]
		\centering
		\includegraphics[trim = 2mm 2mm 2mm 2mm, clip, width=\linewidth]{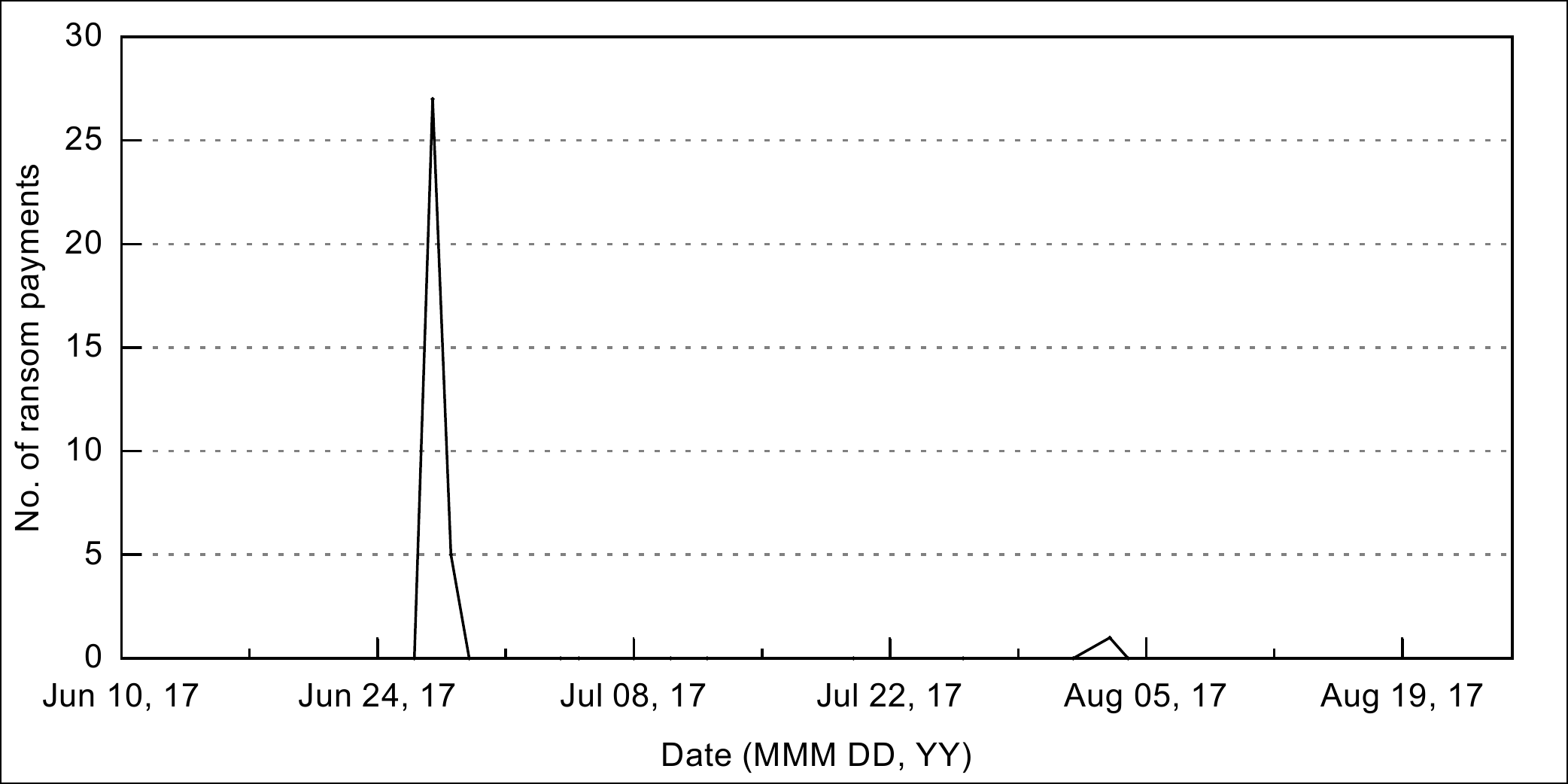}
		\caption[]{Number of ransoms paid to $C_{NP}$}
		\label{NotPetya_ransom_payment_trend}
	\end{figure}
\begin{figure}[H]
		\centering
		\includegraphics[trim = 2mm 2mm 2mm 2mm, clip, width=\linewidth]{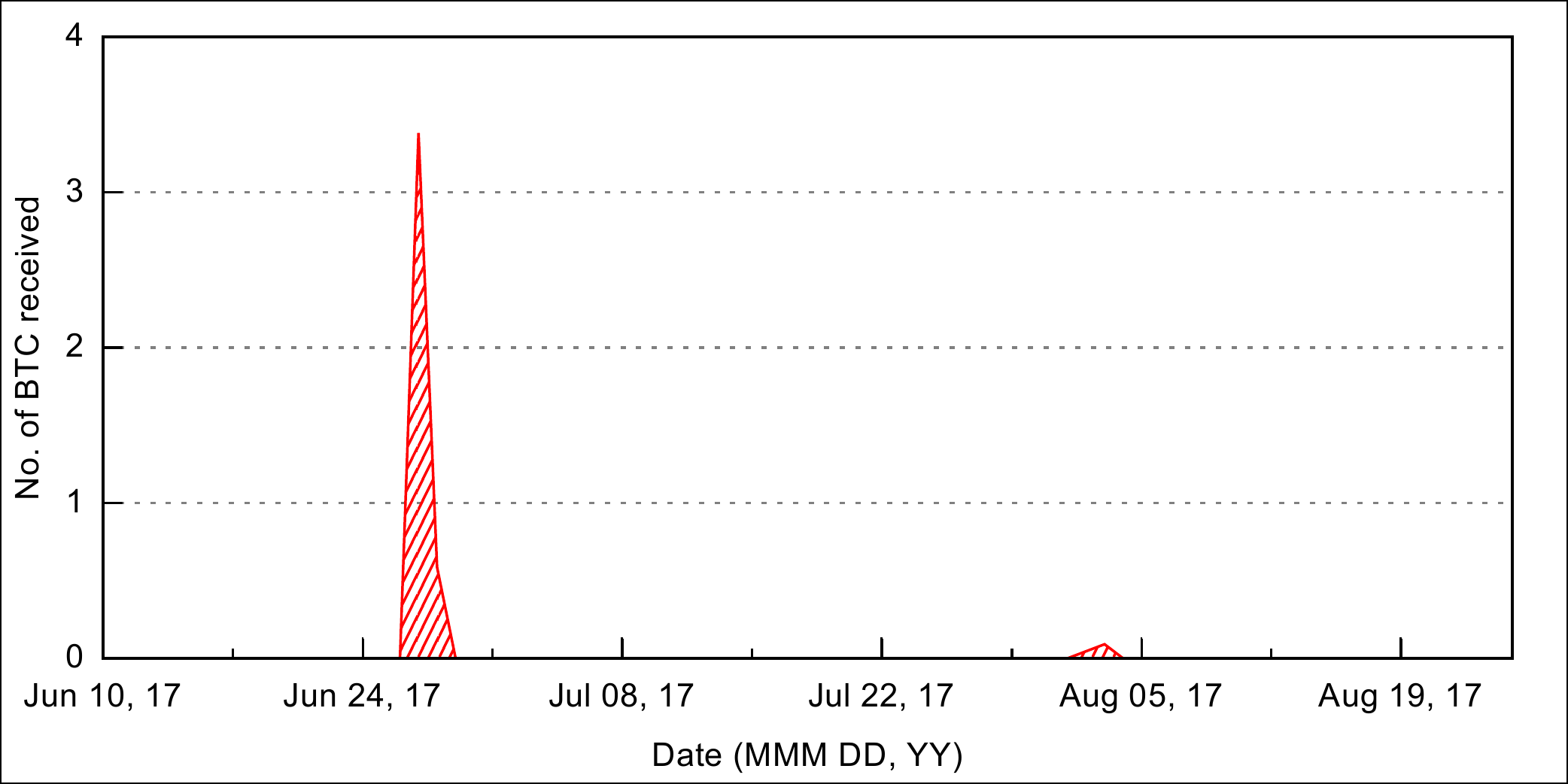}
		\caption[]{Number of Bitcoin received (in ransoms) by $C_{NP}$}
		\label{NotPetya_BTC_USD1}
	\end{figure}
\begin{figure}[H]
		\centering
		\includegraphics[trim = 2mm 2mm 2mm 2mm, clip, width=\linewidth]{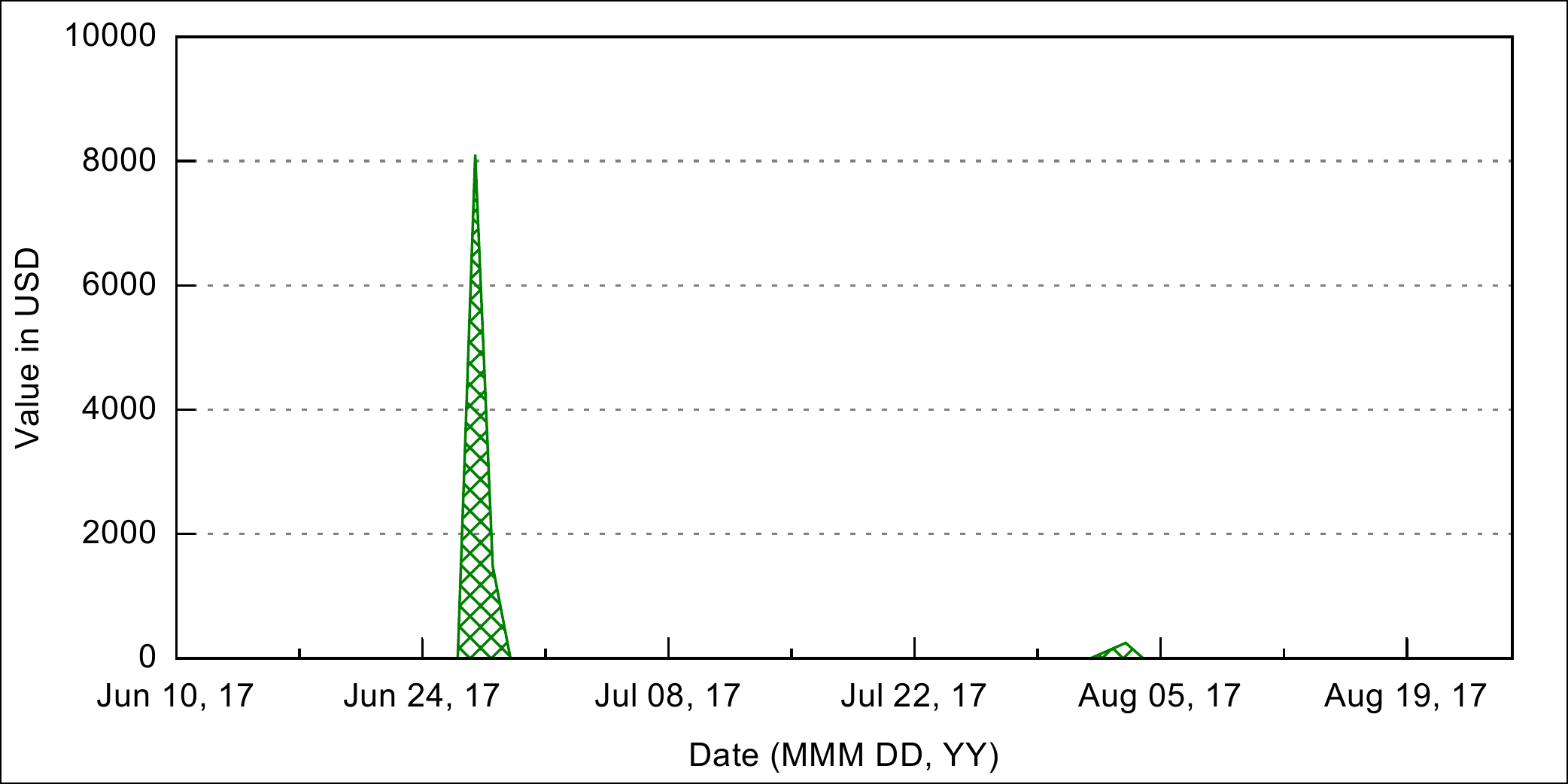}
		\caption[]{USD value of ransoms paid to $C_{NP}$}
		\label{NotPetya_BTC_USD2}
	\end{figure}

	\ankit{\textit{Economy of ransom payments in Bitcoin:}}
	\ankit{We segregated ransom payments using \mc.} As shown in Figures~\ref{NotPetya_ransom_payment_trend},~\ref{NotPetya_BTC_USD1},~and~\ref{NotPetya_BTC_USD2}, on the day of its outbreak, i.e., on June~27,~2017 $C_{NP}$ received somewhat above 3~BTC in total 27 payments that amount approximately USD~8,000.
	It collected the maximum number of ransom payments/Bitcoin/USD on this day.
	
	\par
	In total, we have identified 33 ransom payments to $C_{NP}$, which add up to roughly 4.06 extorted BTC. Using day-to-day average Bitcoin price, we calculate that these ransom payments worth equivalent to USD~9,835.86. Table~\ref{NotPetya_ransom} summarizes the ransoms paid to NotPetya.
	
	\begin{table}[H]
		\centering
		\resizebox{\columnwidth}{!}
		{
			\begin{tabular}{|c|c|c|c|c|}
				\hline
				\textbf{Ransom} & \textbf{Time period} & \textbf{Payments} & \textbf{BTC} & \textbf{USD value} \\ \hline
				\$300           & Jun. 27, '17 - Aug. 03, '17    & 33                      & 4.0576         & 9,835.86               \\ \hline
			\end{tabular}
		}
		\caption{Summary of ransoms paid to NotPetya}
		\label{NotPetya_ransom}
	\end{table}
	
	\par
	Given the irreversible destructive nature and the targeted-software of NotPetya, many researchers suggested that the primary aim of NotPetya was not money. Other researchers speculated that it was probably a second level attack to wipe traces of an early intrusion~\cite{petya_secureworks,petya_logrhythm}.


	\subsection{KeRanger}
	\ankit{\textit{Introduction:}} KeRanger emerged as the first fully functional ransomware that targets macOS operating system. It was discovered on March~4,~2016, by Palo~Alto~Networks. By nature it is a trojan horse, it uploads infected system's information (e.g., model name, UUID) to its C\&C over the Tor network to obtain an RSA public key. Along with the key it also receives victim-specific information that it is writes to a file named ``README\_FOR\_DECRYPT.txt.'' KeRanger encrypts each file $F$ as follows:
	\begin{enumerate}
		\item Generate a random number ($R$).
		\item Generate an Initialization Vector ($I$) using F's content.
		\item Encrypt $R$ with the RSA key (obtained from C\&C), and store it at the beginning of $F.encrypted$ file.
		\item Store $I$ inside the $F.encrypted$ file.
		\item Mix $R$ and $I$ to generate an AES key.
		\item Encrypt data of the original file with the AES key and write the encrypted data to $F.encrypted$ file~\cite{keranger_palo}.
	\end{enumerate}
	
	\par			
	\ankit{\textit{Infection:}} KeRanger was disseminated via two infected installers for the open source BitTorrent client project Transmission version~2.90, which were available for download on the official website. Moreover, these installers were signed with a valid Mac app development certificate; hence, they bypassed OS X’s Gatekeeper security feature.
	
	\par
	\ankit{\textit{Ransom demand:}} To decrypt the encrypted files, the cybercrooks asked the victims to pay exactly one Bitcoin (around USD~400) through a website hosted on the Tor network. 
	
	\par
	\ankit{\textit{Associated Bitcoin addresses and transactions:}} We began with six identified Bitcoin address of KeRanger. These addresses are listed in Table~\ref{keranger_address}. \ankit{\ma}~identified ten new addresses from these six addresses. Therefore, KeRanger cluster ($C_{KR}$) had a total of 16 addresses in our analysis. The transactions \ankit{(obtained using \mb)} to $C_{KR}$ show that $C_{KR}$, in total, received only 13 payments. These transactions worth around 10~BTC (nearly USD~4,200). Table~\ref{keranger_inwards} presents a summary of the total payments credited to $C_{KR}$
	
	\begin{table}[H]
		\centering
		\resizebox{\columnwidth}{!}
		{
			\begin{tabular}{|c|c|c|c|c|}
				\hline
				\textbf{\begin{tabular}[c]{@{}c@{}}Payments\end{tabular}} & \textbf{\begin{tabular}[c]{@{}c@{}}BTC\end{tabular}} & \textbf{\begin{tabular}[c]{@{}c@{}}USD value\\(daily highest\\BTC price)\end{tabular}} & \textbf{\begin{tabular}[c]{@{}c@{}}USD value\\(daily average\\BTC price)\end{tabular}} & \textbf{\begin{tabular}[c]{@{}c@{}}USD value\\(daily lowest\\BTC price)\end{tabular}} \\ \hline
				13                                                                            & 10.0044                                                                     & 4,204.54                                                                                               & 4,175.35                                                                                               & 4,147.01                                                                                              \\ \hline
			\end{tabular}
		}
		\caption{Total payments credited to $C_{KR}$ including all ransom and non-ransom payments}
		\label{keranger_inwards}
	\end{table}

\begin{figure}[H]
		\centering
		\includegraphics[trim = 2mm 2mm 2mm 2mm, clip, width=\linewidth]{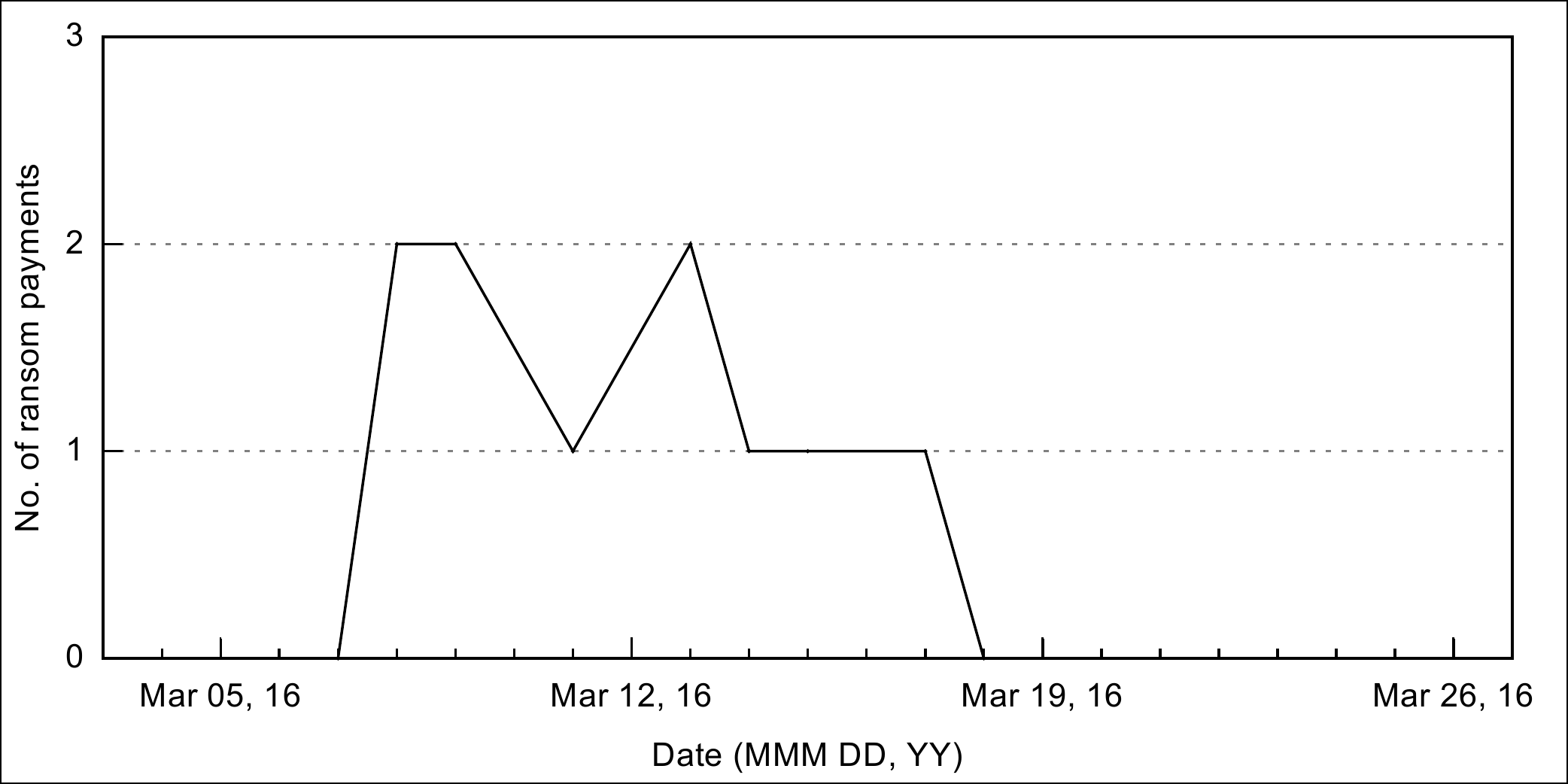}
		\caption[]{Number of ransoms paid to $C_{KR}$}
		\label{keranger_ransom_payment_trend}
	\end{figure}

\begin{figure}[H]
		\centering
		\includegraphics[trim = 2mm 2mm 2mm 2mm, clip, width=\linewidth]{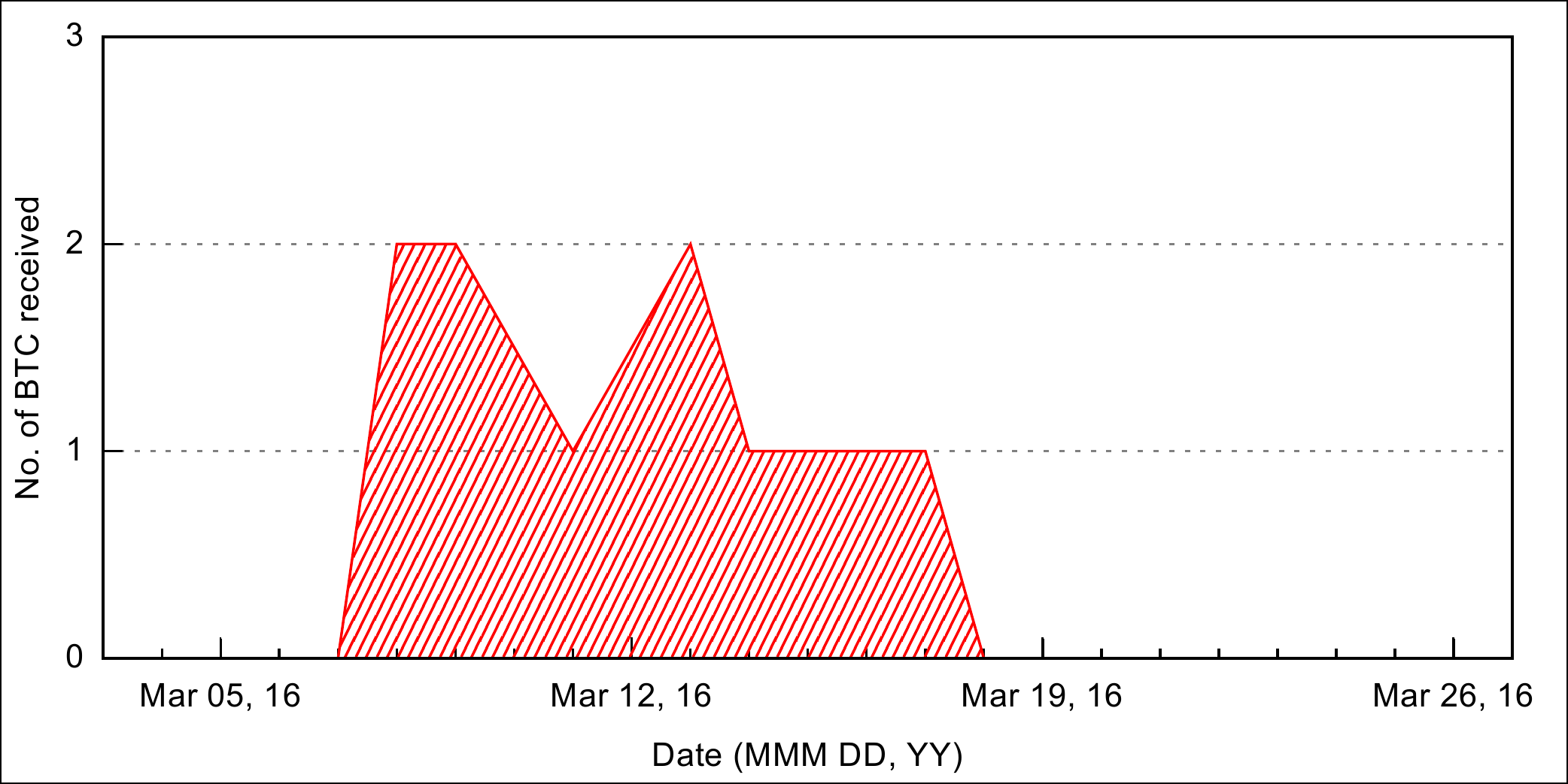}
		\caption[]{Number of Bitcoin received (in ransoms) by $C_{KR}$}
		\label{keranger_BTC_USD1}
	\end{figure}

\begin{figure}[H]
		\centering
		\includegraphics[trim = 2mm 2mm 2mm 2mm, clip, width=\linewidth]{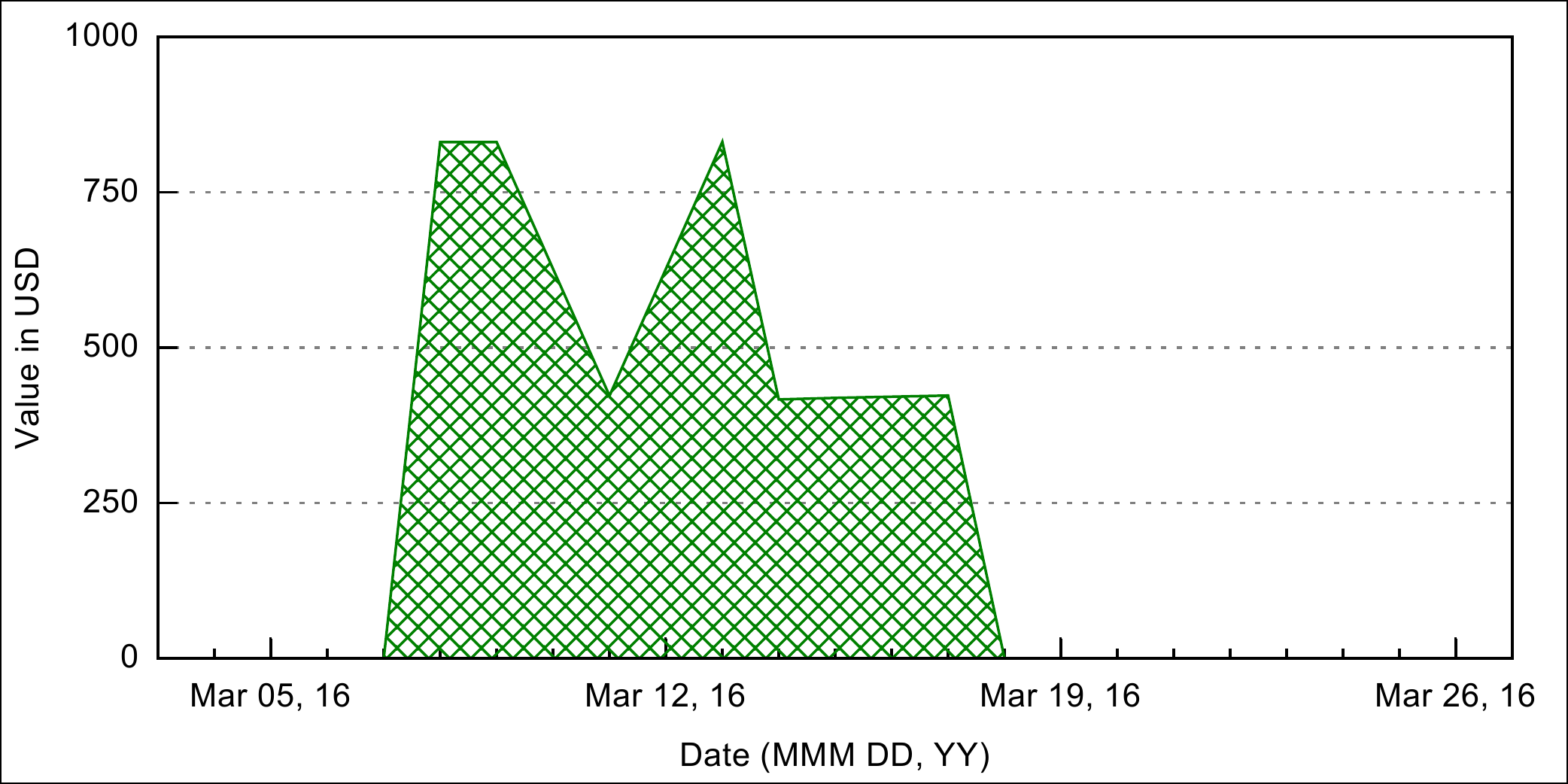}
		\caption[]{USD value of ransoms paid to $C_{KR}$}
		\label{keranger_BTC_USD2}
	\end{figure}

	\ankit{\textit{Economy of ransom payments in Bitcoin:}} 
	\ankit{We isolated ransom payments using \mc.} \figurename{~\ref{keranger_ransom_payment_trend}} shows the total number of ransoms paid to $C_{KR}$. $C_{KR}$ received the last ransom payment on March~17,~2016. Figures~\ref{keranger_BTC_USD1}~and~\ref{keranger_BTC_USD2} depict the total number of Bitcoin received (in ransom) and their corresponding value in USD. Moreover, we found that none of the address received more than one Bitcoin (more than one ransom, in other words).
	
	According to our analysis, $C_{KR}$ received only 10 ransom payments, which contribute to roughly 9.99 extorted BTC. Using day-to-day average Bitcoin price, we estimate that these ransoms convert to USD~4,173.12. Table~\ref{keranger_ransom} summarizes the ransoms paid to KeRanger.
	\begin{table}[H]
		\centering
		\resizebox{\columnwidth}{!}
		{
			\begin{tabular}{|c|c|c|c|c|}
				\hline
				\textbf{Ransom} & \textbf{Time period} & \textbf{Payments} & \textbf{BTC} & \textbf{USD value} \\ \hline
				1 BTC           & Mar. 04, '16 - Mar. 17, '16    & 10                      & 9.9990        & 4,173.12               \\ \hline
			\end{tabular}
		}
		\caption{Summary of ransoms paid to KeRanger}
		\label{keranger_ransom}
	\end{table}
	\par
	One of the possible reasons for such low number of ransom payments could be that by March~5,~2016, Transmission project removed the infected installers from the website, and Apple revoked the abused certificate that allowed Gatekeeper to block the infected installers.

	
	\subsection{WannaCry}
	\ankit{\textit{Introduction:}} WannaCry (also known as WCry, WanaCrypt0r, Wana~Decrypt0r 2.0) is blended threat with characteristics of both a worm and a ransomware. It was first seen on May~12,~2017. It affects Windows system by encrypting files using a combination of the RSA and the AES algorithms. Interestingly, it encrypts each file with a separate 128-bit AES encryption key in CBC mode. Furthermore, it encrypts each AES key individually using the RSA-2048 encryption algorithm~\cite{secureworks_wannacry}.
	\par
	\ankit{\textit{Infection:}} WannaCry scans explicitly for the presence of the DoublePulsar backdoor on a target. If the DoublePulsar backdoor is not present, then it tries to compromise the system using the EternalBlue exploit~\cite{wannacry_talosintelligence}. The EternalBlue exploit was exposed merely a few months before the WannaCry attack by a hacker group known as The Shadow Brokers.
	\par
	\ankit{\textit{Kill switch and kill mutex:}}
	A kill switch is usually employed to terminate a program's execution. In case of WannaCry, the kill switch was a domain name\footnote{www.iuqerfsodp9ifjaposdfjhgosurijfaewrwergwea.com}. Upon initialization, WannaCry tries to connect to the domain over HTTP. If the connection is successful, then it stops and exits. Possibly, it was designed to evade a sandbox testing. The kill switch domain was hardcoded in the source code and was discovered by Marcus Hutchins\footnote{en.wikipedia.org/wiki/MalwareTech}. On another side, before beginning the encryption process, WannaCry attempts to create a mutex named ``MsWinZonesCacheCounterMutexA'' and exits if the mutex is already present.
	\par
	\ankit{\textit{Ransom demand:}} The ransom note asks the victims to pay USD~300 ransom in Bitcoin within three~days. The ransom note also states that the ransom amount would become double (i.e., USD~600) after three~days, and if the ransom is not paid within seven~days from the day of infection, all the encrypted files would~be~deleted.
	\par
	\ankit{\textit{Associated Bitcoin addresses and transactions:}} Cybercriminals intended to create a unique Bitcoin payment address for each victim. But a race condition bug prevents the correct execution of the code. In this situation, it presents one of three hard-coded Bitcoin addresses to collect the ransom~\cite{wannacry_symantec}. These addresses are listed in Table~\ref{wannacry_address}. Moreover, using these addresses, \ankit{\ma}~generated no new address. Hence, WannaCry cluster ($C_{WC}$) generated by our framework had only three Bitcoin addresses during our analysis. \ankit{We procured the detailed transaction history of these three addresses using \mb.} $C_{WC}$ received 341 payments. These payments worth over 50~BTC (approximately USD~100,000). Table~\ref{wannacry_inwards} summarizes the payments credited in $C_{WC}$.
	
	\begin{table}[H]
		\centering
		\resizebox{\columnwidth}{!}
		{
			\begin{tabular}{|c|c|c|c|c|}
				\hline
				\textbf{\begin{tabular}[c]{@{}c@{}}Payments\end{tabular}} & \textbf{\begin{tabular}[c]{@{}c@{}}BTC\end{tabular}} & \textbf{\begin{tabular}[c]{@{}c@{}}USD value\\(daily highest\\BTC price)\end{tabular}} & \textbf{\begin{tabular}[c]{@{}c@{}}USD value\\(daily average\\BTC price)\end{tabular}} & \textbf{\begin{tabular}[c]{@{}c@{}}USD value\\(daily lowest\\BTC price)\end{tabular}} \\ \hline
				341                                                                           & 53.2906                                                                     & 102,141.19                                                                                             & 99,549.05                                                                                              & 96,497.20                                                                                             \\ \hline
			\end{tabular}
		}
		\caption{Total payments credited to $C_{WC}$ including all ransom and non-ransom payments}
		\label{wannacry_inwards}
	\end{table}
	
	\par
	\ankit{\textit{Economy of ransom payments in Bitcoin:}}
	Due to comparatively a smaller number of transactions, we manually verified each payment to $C_{WC}$. As shown in Table~\ref{wannacry_ransom_BTC_address}, each Bitcoin address collected at minimum 69 ransom payments and a minimum of nearly 13.52 BTC.
	
	\begin{table}[H]
		\centering
		\resizebox{\columnwidth}{!}
		{
			\begin{tabular}{|l|c|c|}
				\hline
				\multicolumn{1}{|c|}{\textbf{Address}} & \textbf{Payments} & \textbf{BTC} \\ \hline
				12t9YDPgwueZ9NyMgw519p7AA8isjr6SMw     & 77                      & 15.1129        \\ \hline
				13AM4VW2dhxYgXeQepoHkHSQuy6NgaEb94     & 92                      & 18.5431        \\ \hline
				115p7UMMngoj1pMvkpHijcRdfJNXj6LrLn     & 69                      & 13.5183        \\ \hline
			\end{tabular}
		}
		\caption{Number of ransoms and Bitcoin received (in ransoms) per address in $C_{WC}$}
		\label{wannacry_ransom_BTC_address}
	\end{table}
	
	\par
\begin{figure}[H]
		\centering
		\includegraphics[trim = 2mm 2mm 2mm 2mm, clip, width=\linewidth]{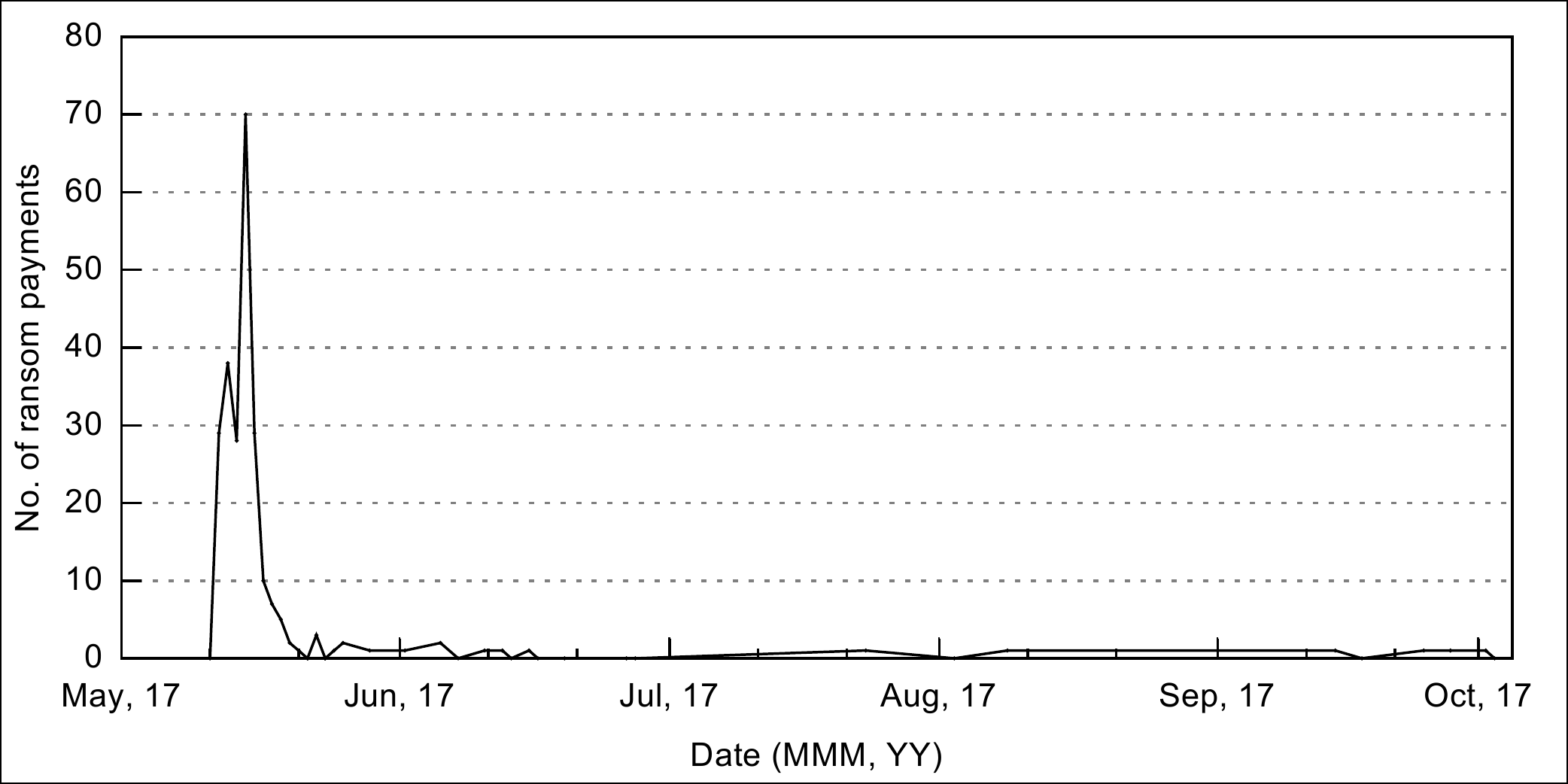}
		\caption[]{Number of ransoms paid to $C_{WC}$}
		\label{wannacry_ransom_payment_trend}
	\end{figure}
\begin{figure}[H]
		\centering
		\includegraphics[trim = 2mm 2mm 2mm 2mm, clip, width=\linewidth]{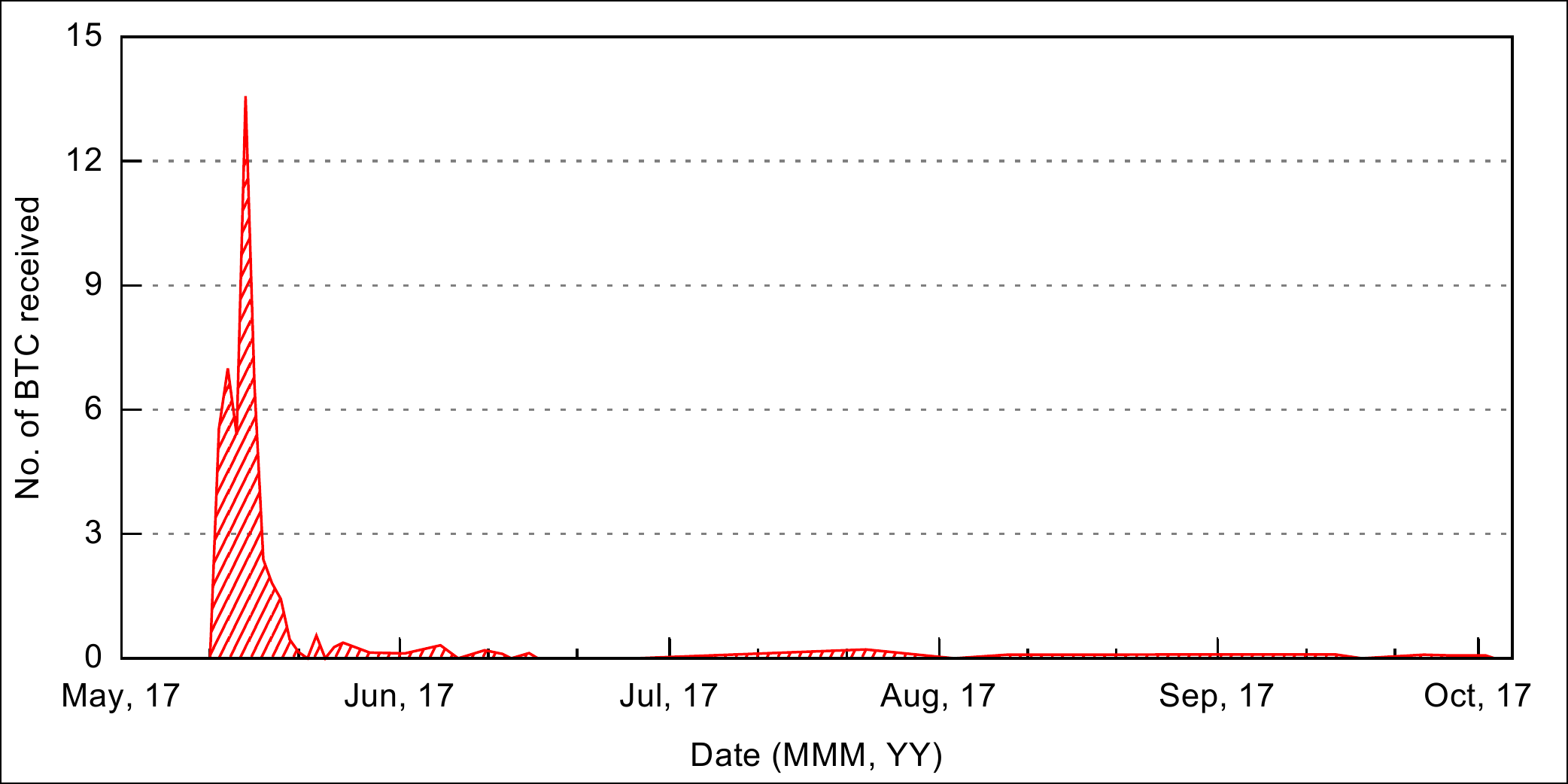}
		\caption[]{Number of Bitcoin received (in ransoms) by $C_{WC}$}
		\label{wannacry_BTC_USD1}
	\end{figure}
\begin{figure}[H]
		\centering
		\includegraphics[trim = 2mm 2mm 2mm 2mm, clip, width=\linewidth]{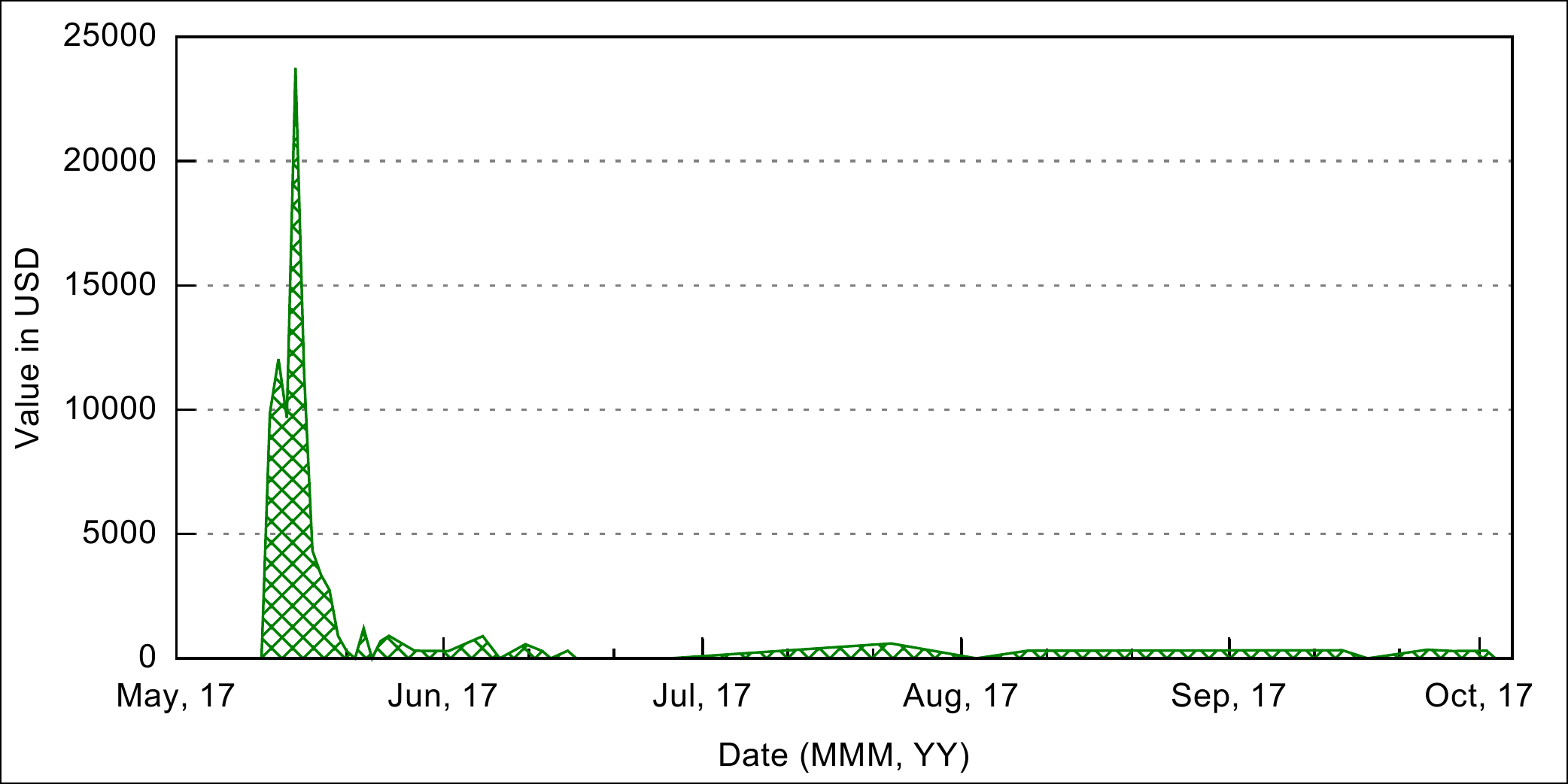}
		\caption[]{USD value of ransoms paid to $C_{WC}$}
		\label{wannacry_BTC_USD2}
	\end{figure}

	Figures~\ref{wannacry_ransom_payment_trend},~\ref{wannacry_BTC_USD1},~and~\ref{wannacry_BTC_USD2} indicate that on May~15,~2017, $C_{WC}$ received 70 payments that amount to nearly 14 BTC, which is approximately USD~24,000. It is the day when it received the maximum number of ransom payments/Bitcoin/USD in a~single~day.

	
	
	\par
	In total, we have identified 238 ransom payments to $C_{WC}$, which add up to 47.17 extorted BTC. Using day-to-day average Bitcoin price, we calculate that these ransom payments worth equivalent to USD~86,076.76. Table~\ref{wannacry_ransom} summarizes the ransoms payments made to WannaCry.
	
	\begin{table}[H]
		\centering
		\resizebox{\columnwidth}{!}
		{
			\begin{tabular}{|c|c|c|c|c|}
				\hline
				\textbf{Ransom} & \textbf{Time period}       & \textbf{Payments} & \textbf{BTC}   & \textbf{USD value} \\ \hline
				\$300           & \multirow{3}{*}{May 12, '17 - Oct. 02, '17}          & 192                     & 32.3430          & 58,416.62              \\ \cline{1-1} \cline{3-5}
				\$600           &           & 46                      & 14.8313          & 27,660.14              \\ \cline{1-1} \cline{3-5}
				\textbf{Total}  &  & \textbf{238}            & \textbf{47.1743} & \textbf{86,076.76}     \\ \hline
			\end{tabular}
		}
		\caption{Summary of ransoms paid to WannaCry}
		\label{wannacry_ransom}
	\end{table}
	
	\par
	The overall impact (including financial losses) due to WannaCry infection could have been worse. But, thanks to the early detection of the kill switch, which prevented the infected computers from spreading WannaCry further.
	

	\textit{Other Bitcoin ransomware:}
	We now briefly discuss all those Bitcoin ransomware, for which the observed payments either are entirely different (merely a few dollars) against the demanded ransom or the date of transactions do not match with the activity of the ransomware. Nevertheless, we make no solid claims without further evidence. However, we make their addresses and corresponding dataset available in Appendix~\ref{ransomware_addresses} and our repository  (mentioned before) respectively for future efforts in this direction of research.
	\subsection{CTB-Locker}
	\ankit{\textit{Introduction:}} CTB-Locker (Curve-Tor-Bitcoin-Locker) first appeared in mid-July~2014 as Critroni. Initially, it targeted individual Windows users. But, soon its focus shifted to vulnerable WordPress websites. The latter version encrypts the homepage of a website and replaces the original homepage with a new page containing the ransom note. It is also infamous for using Elliptic Curve Cryptography (ECC), the Tor network to hide the C\&C, Bitcoin for ransom payment, and its availability in multiple (seven) languages. As a pioneer, it uses ECC for encryption, which enables it to obtain the equivalent level of security as the RSA with much smaller key sizes. E.g., a 256-bit ECC offers equivalent security as 3072-bit RSA. It obtains a secret key by applying a SHA256 hash function to a 52~bytes long random sequence, and Curve25519 generates the corresponding public key. In fact, it uses a combination of symmetric and asymmetric encryption algorithm where the AES algorithm encrypts the user's files~\cite{ctblocker_encryption}.
	\par
	\ankit{\textit{Infection:}} Researchers\footnote{malware.dontneedcoffee.com/2014/07/ctb-locker.html} and allegedly participants\footnote{www.reddit.com/r/Malware/comments/2uffwc/ctb\_locker\_ama/} disclosed that the attackers used an affiliate program to spread the infection in return of profits. Generally, in an affiliate program, the participants attempt to spread the infection via several possible vectors. CTB-Locker was primarily distributed through exploit kits (e.g., Rig and Nuclear) and malicious email spam (e.g., overdue phone invoices, missed fax, bank statements) campaigns that exploit Dalexis or Elenoocka downloader component. 
	\par
	\ankit{\textit{Ransom demand:}} In the beginning, the ransom was set at 0.5~BTC (about USD~300) for US, Europe, and Canada while 0.25~BTC for other countries. Later, the ransom was changed to 0.4~BTC~(about USD~150) that doubles after four~days. In addition, the victims could decrypt five files for free and could also do test transaction of 0.0001~BTC on one of the two dedicated Bitcoin addresses.
	\par
	\ankit{\textit{Associated Bitcoin addresses and transactions:}} The addresses belonging to CTB-Locker that we found are listed in Table~\ref{ctblocker_address}. The last two addresses listed in the table are those Bitcoin addresses where the victims could do a test transaction. However, the cluster generated \ankit{(using \ma)} from these addresses did not receive any payment except for two test transactions. \ankit{One of the possible reasons could be the nature of the target audience. Most of the web hosting plans facilitate periodic backups. If a web page becomes inaccessible/encrypted, the webmaster can restore a relatively fresh version without paying the ransom.}

	
	\subsection{CryptoTorLocker2015}
	On February~5,~2015, Symantec discovered CryptoTorLocker2015 as a very low-level threat for Window operating system. It utilizes only public key cryptography for file encryption. In particular, it uses the RSA-2048 encryption algorithm, for which it downloads the RSA public key from an attacker-controlled C\&C. 
	Being a trojan, it spread via classical infection mechanisms such as drive-by download. It asks the victims to pay 0.5~BTC (equivalent to USD/EUR~100) within five~days of infection to decrypt the files.
	\par
	\ankit{\ma~}of our framework generated six new addresses belonging to CryptoTorLocker2015 from the single address listed in Table~\ref{cryptotorlocker2015_address}. These seven addresses received almost USD~1100 worth 5~BTC in 136 payments. But, only one transaction\footnote{blockchain.info/tx/36f2bbc56e7ce7bea59265ce1b7f9ac42040dc5491f01a 4b338f619293515820} that happened on February~11,~2015 satisfies the criteria of the ransom demand specified by the attackers.


	\subsection{TeslaCrypt}
	\ankit{\textit{Introduction:}} TeslaCrypt or AlphaCrypt began to spread in mid February~2015. It searches explicitly for game-related user content (e.g., custom maps and progress/save files) along with other personal documents and pictures. TeslaCrypt ignores audio files, video files, and removable (e.g., USB) storage. It does not scan connected networks as well. It uses the AES algorithm to encrypt files, but with an aim to mislead the victims, it appends ``ecc'' extension to the encrypted files while the ransom note message claims that it has used the RSA-2048 encryption algorithm. Its C\&C hid in the Tor anonymity network and required an SSL encrypted connection from a victim machine for communication. Preventing TeslaCrypt from interacting with the C\&C does not prevent the encryption process because it generates the encryption keys locally. 
	\par
	\ankit{\textit{Infection and ransom demand:}} TeslaCrypt was distributed exclusively through Angler and Nuclear browser exploit kits~\cite{secureworks_teslacrypt}. The attackers accepted the ransoms via various payment methods. The ransom amount in Bitcoin was 1.5~BTC within seven~days, 2.5~BTC otherwise. The victims from North American region could also choose to pay USD~1000 with PayPal My Cash cards while the European victims could pay EUR~600 with Ukash or paysafecard.
	\par
	\ankit{\textit{Associated Bitcoin addresses and transactions:}} The payments collected by the address cluster (generated from the addresses listed in Table~\ref{teslacrypt_address}) do not match with the ransom amount demanded by the attackers. However, FireEye research team in their study~\cite{fireeye_teslacrypt} describes that the attackers negotiated with the victims and gave ``discounts'' on the ransom amount. In this case, the attackers accumulated around 254.6~BTC, which converts to about USD~57,272. \ankit{Later, the attackers publicly released\footnote{www.bleepingcomputer.com/news/security/teslacrypt-shuts-down-and-releases-master-decryption-key/} the master decryption key.}


	\subsection{Chimera}
	In the November~2015, cybercrooks began to target English- and German-speaking Windows users with Chimera ransomware. The cybercrooks distributed Chimera via targeted e-mails to small businesses and companies. Unlike other ransomware, it does not use a Tor website to handle payment instructions or to hide the C\&C. Instead, it is the first ransomware that uses Bitmessage\footnote{bitmessage.org/wiki/Main\_Page} P2P protocol to interact with the C\&C and to obtain the RSA public/private keys. In such scenario, it is difficult, if not impossible, to take down all peers in the network that are assisting ransomware's operations. 
	\par
	It is also the first ransomware to use doxing as a pressure tactic. It threatens the victims to post personal data including pictures and videos on the internet if the ransom is not paid. However, several ransomware analysts showed that it was a bogus threat~\cite{malwarebyte_chimera}. Chimera asks the victims to pay between 0.9~BTC to 2.5~BTC as ransom. The addresses belonging to Chimera that we found are listed in Table~\ref{chimera_address}. The payments received by the cluster generated \ankit{(using \ma)} from these addresses do not fall in the range of the ransom asked by the cybercrooks.
	\ankit{Dramatically, rival ransomware developers leaked the RSA private keys of Chimera\footnote{twitter.com/JanusSecretary/status/757951375561072640}.}


	\subsection{Hi Buddy!}
	Hi Buddy! was active in the first quarter of 2016. Like many other ransomware, it encrypts only user's files and leaves the system files responsible for running Windows operating system. Upon execution, it attempts to connect with its C\&C over Tor2Web service and sends information such as the version of operating system, location (country) of the victim machine. The response from the C\&C includes a string variable whose value (``FIRST'' or ``RECEIVED'') depends on whether it should encrypt or decrypt the files. The values ``FIRST'' and ``RECEIVED'' correspond to encryption and decryption respectively. It is important to note that the encryption/decryption process does not execute until it receives the response from the C\&C. It uses the AES-256~algorithm to encrypt user's files. The encryption key is generated by hashing (SHA-256) a string variable named ``password,'' which is obtained from the C\&C. It spread via general spamming techniques. After encryption, it shows a ransom note asking about 0.8~BTC for decryption. The address cluster generated from the address listed in Table~\ref{hibuddy_address} did not receive any~ransom.


	\subsection{Jigsaw}
	Jigsaw was released in late March~2016 to affect systems running Windows operating system. It is considered to be the most dramatic ransomware so far. It was released in several different languages, while each variant was hard-coded to execute only after a specific date. As a representative example, the English version was set to execute after March~23,~2016 while the Portuguese version was written to run after April~6,~2016.	Moreover, It employs an unprecedented extortion strategy. During the first 24~hours it deletes a few files every hour; after 24~hours, hundreds of files every hour; and after 48~hours, thousands of files every hour. And, if the ransom is not paid within 72~hours, it deletes all the remaining files. If the victim shuts-down or restarts the computer, it destroys 1000 files as ``punishment.'' Furthermore, each variant demands a distinct amount of ransom ranging from USD~23 to USD~5000 to be paid through Bitcoin.
	\par
	The cybercriminals hosted the payload on free cloud storage services such as 1fichier.com and distributed the links to the malicious payload through email spamming. Jigsaw works offline and uses the AES-128 encryption algorithm in CBC mode to encrypt user's files. Using 19 addresses (listed in Table~\ref{jigsaw_address}), our framework generated 24 addresses belonging to Jigsaw ransomware. Altogether these addresses collected approximately 2.5~BTC (USD~1,200) in 58~payments. However, all these payments occurred from March~2016 to August~2016, i.e., during the period when Jigsaw was active. Hence, we may argue that perhaps all these transactions were ransom payments.


	\subsection{ZCryptor} 
	A security researcher called Jack first reported\footnote{malwarefor.me/zcrypt-ransomware/} ZCryptor on May~24,~2016. It targets computers running Windows operating system. After obtaining victim-specific encryption key from its C\&C, it uses the RSA encryption algorithm to encrypt user's files. ZCryptor exhibits worm-like behavior. It is one of the few ransomware that can self-propagate to other connected computers and network devices even without using an exploit kit or spamming. For initial infection, the cybercrooks used conventional distribution techniques such as email spamming, fake software (e.g., Adobe Flash) updater, and macro malware in Microsoft Office suite. It also attempts to distract the victim by showing benign pop-ups while performing the file encryption. Once the encryption process completes, ZCryptor displays its ransom message in which it asks for 1.2~BTC (about USD~500) to be paid within four~days. Nevertheless, it permits additional three~days for the payment at the cost of 5~BTC (about USD~2,100). The address cluster generated from the address listed in Table~\ref{zcryptor_address} did not receive any ransom. \ankit{It is noteworthy that on May~26,~2016, i.e., within two~days of its discovery, Microsoft issued an alert\footnote{blogs.technet.microsoft.com/mmpc/2016/05/26/link-lnk-to-ransom/} to its users about ZCryptor and also updated the definition base of Windows Defender to protect against ZCryptor.}


	\subsection{VenusLocker}
	\ankit{\textit{Introduction, infection, and ransom demand:}} At the beginning of August~2016, VenusLocker, a new eda2-based ransomware began to target Windows based systems. 
	Similar to most ransomware, VenusLocker encrypts data files using the AES-256 algorithm. It generates the AES encryption key on the victim's system from a cryptographically-strong random number generator and encrypts it with an embedded RSA-2048 public key before sending to the C\&C. It also creates and conveys a unique ID to C\&C to identify the infected system. It spread primarily via drive-by download. It allows only three~days (with no extension) to pay the ransom in Bitcoin. At first, it demanded USD~100 as ransom. But, soon it asked USD~500. However, the ransom amount settled on one BTC with an update in December~2016.
	
	\par
	\ankit{\textit{Associated Bitcoin addresses and transactions:}} Initially, we identified three addresses belonging to VenusLocker. These addresses are listed in Table~\ref{venuslocker_address}. \ankit{\ma}~of our framework identified three new addresses from the addresses. Therefore, VenusLocker cluster ($C_{VL}$) had a total of six addresses in our analysis. The transactions \ankit{(obtained using \mb)} to $C_{VL}$ reveal that $C_{VL}$, in total, received 11~payments. The total worth of these transactions is almost 7~BTC (more than USD~6,500). Table~\ref{venuslocker_inwards} presents a summary of the total payments credited~to~$C_{VL}$.

	\begin{table}[H]
		\centering
		\resizebox{\columnwidth}{!}
		{
			\begin{tabular}{|c|c|c|c|c|}
				\hline
				\textbf{\begin{tabular}[c]{@{}c@{}}Payments\end{tabular}} & \textbf{\begin{tabular}[c]{@{}c@{}}BTC\end{tabular}} & \textbf{\begin{tabular}[c]{@{}c@{}}USD value\\(daily highest\\BTC price)\end{tabular}} & \textbf{\begin{tabular}[c]{@{}c@{}}USD value\\(daily average\\BTC price)\end{tabular}} & \textbf{\begin{tabular}[c]{@{}c@{}}USD value\\(daily lowest\\BTC price)\end{tabular}} \\ \hline
				11                                                                            & 6.8155                                                                    & 6,861.73                                                                                                & 6,753.81                                                                                                & 6,637.06                                                                                               \\ \hline
			\end{tabular}
		}
		\caption{Total payments credited to $C_{VL}$ including all ransom and non-ransom payments}
		\label{venuslocker_inwards}
	\end{table}
	
	%
	
	\par
	The campaign was launched again as The~Trump~Locker in February~2017 and rebranded in March~2017 as The~LLTP~Locker. The ransom amount in The~Trump~Locker varied from USD~50 to USD~150 (in Bitcoin), while The~LLTP~Locker targeted specifically Spanish users asking USD~200 (in Bitcoin) as ransom. The address clusters generated for The~Trump~Locker and The~LLTP~Locker (using the addresses listed in Tables~\ref{trumplocker_address}~and~\ref{lltp_address} respectively) never received any payment.


	\subsection{KillDisk}
	KillDisk debuted as a data wiper malware. It affected energy industry, finance sector, sea transport, and news agencies in 2015 and 2016. In early December 2016, the malware was updated to integrate a ransomware component. The KillDisk ransomware targets not only Windows operating system but also Linux workstations and servers, which magnifies its damage potential. It targets every drive (local and network) that the victim can access. 
	Both the Windows and Linux variant work differently. 
	\par			
	The Windows variant, detected by CyberX~\cite{killdisk_cyberx}, encrypts each file with a separate AES-256 encryption key (generated using CryptGenRandom function from the Windows CryptoAPI library). After use, it encrypts the AES keys using a public RSA-1028 key. It obtains the RSA public key via the Telegram API from the C\&C.  It uses the whitelisting technique to avoid sandbox analysis. 
	\par The Linux variant, detected by ESET~\cite{killdisk}, performs encryption using Triple-DES applied to 4096-byte blocks where each file is encrypted using a different set of 64-bit encryption keys generated locally. However, the encryption keys are neither stored locally nor sent to the C\&C, which means that the decryption is virtually impossible. Furthermore, it also makes the machine unbootable as it rewrites the boot sector and uses the GRUB bootloader to show the ransom note.
	\par
	Both the variants show the same ransom note, asking an enormous ransom of 222~BTC to be paid on the same Bitcoin address. The address is listed in Table~\ref{killdisk_address}. The address cluster generated by our framework also had only one Bitcoin address, which did not receive any ransom.


	\subsection{FindZip}
	FindZip ransomware, also known as Filecoder, was discovered and reported by ESET researchers~\cite{findzip} on February~22,~2017. It is written in Swift programming language to infect systems with macOS operating system. It encrypts all mounted external and network storage. Upon execution, it locally generates a 25-characters long random string, which it uses to create a separate encrypted .zip file for each user file using the ``zip'' shell command. Next, it deletes all the original files by the ``rm'' command and sets the encrypted file's time to February~13,~2010 using the ``touch'' command. 
	It was distributed as a ``Patcher'' application from torrent distribution sites. The torrent file downloads a single ZIP archive file that contains fake patching applications for premium software such as Adobe Premiere Pro and Microsoft Office for Mac. However, the applications are not signed with an Apple-recognized key.
	\par
	The ransom message is hardcoded inside the ransomware. Hence, it uses the same Bitcoin address for each victim. It demands 0.25 BTC for decrypting the files and instructs the victims to wait for 24 hours after paying the ransom. But it promises to start the decryption in 10 minutes if the victim pays 0.45 BTC. The address cluster generated from the addresses listed in Table~\ref{findzip_address} did not collect any payment.
	

	\subsection{ThunderCrypt}
	ThunderCrypt emerged in the first week of May~2017. It targeted primarily Taiwanese Windows users for ransom extortion. It carefully encrypts the user's data by a hybrid RSA-2048 public key encryption algorithm. It does not encrypt the essential files of the operating system so that the system keeps on working and has an active internet connection. To distribute the ransomware, the cybercriminals injected a malicious script into a Taiwanese forum ``ENVY.'' The script triggers a pop-up, which requests permission to run a fake Adobe Flash Player installer. The bogus installer was designed to drop the ThunderCrypt payload on the victim machine.
	\par
	ThunderCrypt demands exactly 0.345 Bitcoin (roughly USD~500) from the victims. And like most ransomware, it threatens the victim to erase the key from the server if the ransom is not paid. Additionally, it allows the victim to decrypt one file to prove that the decryption is possible. The addresses belonging to ThunderCrypt that we found are listed in Table~\ref{thundercrypt_address}. However, the cluster generated from these addresses did not receive any payment. \ankit{It is also worth mentioning that in communication\footnote{wccftech.com/thundercrypt-ransomware-taiwanese-man/} with a victim, cybercriminals admitted that their campaign failed.}


	\subsection{DoubleLocker}		
	On October~13,~2017, ESET researchers~\cite{doublelocker_addr_1} reported the first-ever ransomware that targets Android operating system. DoubleLocker is rewritten from an Android banking Trojan named ``Android.BankBot.211.origin.'' It abuses Android's accessibility services to elevate privileges on the victim system. Unlike its banking parent, it does not steal victims' banking credentials. It rather changes the device's PIN code and encrypts device's primary storage using the AES-256 encryption algorithm, which leaves the device inaccessible to the user. Similar to its banking parent, the attackers distributed it as a fake Adobe Flash Player app over compromised websites. To release the decryption key, DoubleLocker asks the victims to pay 0.0130~BTC (about USD~50) within 24~hours and waits for three~confirmations of the payment. The address cluster (generated from the addresses listed in Table~\ref{doublelocker_address}) received only one payment, which is far too less than the asked ransom.


	\subsection{Bad Rabbit}
	Bad Rabbit started to spread from October~24,~2017. Similar to NotPetya, it encrypts files as well as the MFT on the Windows machine, and it also replaces the MBR with a custom bootloader. For file encryption, it uses the AES-128 encryption algorithm in CBC mode, and the RSA-2048 encryption secures the keys. It uses DiskCryptor driver in AES-XTS mode to encrypt disk partitions on the infected system. Nevertheless, it is not a wiper like NotPetya. The attackers distributed it via drive-by attack as a dropper-file named ``install\_flash\_player.exe,'' which prompts a standard UAC to elevate administrative privileges. Additionally, it exploits the EternalBlue exploit to infect machines in the local network. The ransom note asks the victims to use a Tor website to make a payment of 0.05~BTC within 42~hours, after which the price of decryption goes up~\cite{securelist_badrabbit}. The payments collected by the address cluster (generated from the addresses listed in Table~\ref{badrabbit_address}) are significantly lesser than the asked ransom.				
	

	\section{Limitations}
	\label{limitation}
	\ankit{
		One of the most important and decisive elements for the quality of the outcomes of our framework is the address identification module, presented in Section~\ref{identify}. It relies on the Bitcoin addresses collected from the public sources; the quality of data collected from the public sources could be a concern. One of the promising alternatives is to collect binaries of the ransomware and execute them several times in a virtual environment to witness/obtain Bitcoin addresses. However, the question of integrity and authenticity of the binaries remains the same. Given the nature of the problem, we followed the approach used in the previous studies~\cite{liao2016behind,spagnuolo2014bitiodine} and took extreme precaution while collecting addresses from the public sources.
	}
	\par
	The fundamental principles of the Bitcoin protocol implicitly impart two types of flaws in our address identification module: overestimation and underestimation. Our methodology would overestimate when multiple users pool their transactions into a single transaction; as in the case of mixers. On another side, it would underestimate when there exists no evidence (in the blockchain) of an address owned by a user being used in conjunction with any other address of the same user. However, in a given scenario, it would report more accurate results as compared to the existing approaches due to its attributes of ransom classifications.

	\section{Conclusion and Future Work}
	\label{futurework}
	Pseudo-anonymity and irreversibility of Bitcoin transaction protocol have made Bitcoin a dexterous utility among cybercriminals. Unlike genuine users, who seek to transact securely and efficiently; cybercrooks exploit these characteristics to commit immutable and presumably untraceable monetary fraud. In this paper, we have presented our comprehensive and longitudinal study on twenty recent Bitcoin ransomware along with their renamed/rebranded versions. We have also introduced our framework to identify, collect, and analyze Bitcoin addresses that belong to the cybercriminals behind the ransomware. Moreover, we elaborated the characteristics and the functionality of the ransomware as well as reported the economic impact of such ransomware from the Bitcoin payment perspective.

	\par
	In the future, we will extend our identification framework to other cryptography-based currencies. We will also investigate the ransoms extorted via other payment options; we~hope~to~present a comprehensive report that will include ransom payments from all payment option endorsed by the ransomware. Finally, we will attempt to trace how the received ransoms were used and by whom.

	\section*{Acknowledgment}
	Ankit Gangwal is pursuing his Ph.D. with a fellowship for international students funded by Fondazione Cassa di Risparmio di Padova e Rovigo~(CARIPARO). This work is partially supported by the EU TagItSmart! Project (agreement H2020-ICT30-2015-688061), the EU-India REACH Project (agreement ICI+/2014/342-896), the grant n. 2017-166478 (3696) from Cisco University Research Program Fund and Silicon Valley Community Foundation, and by the grant ``Scalable IoT Management and Key security aspects in 5G systems'' from~Intel.

	%

	\appendix
	\setcounter{table}{0}
	\counterwithin{table}{section}
	\subsection{Ransomware' Bitcoin addresses identified in our initial investigation}
	\label{ransomware_addresses}
	
	\begin{table}[H]
		\centering
		\begin{tabular}{|c|l|c|}
			\hline
			\textbf{\#} & \multicolumn{1}{c|}{\textbf{Address}} & \textbf{Source(s)} \\ \hline
			1           & 135N2nfAkextd6E25quXpM98qLSi2BccCb    & \multirow{2}{*}{\cite{cryptolocker_addr_1}} \\ \cline{1-2}
			2           & 1AEoiHY23fbBn8QiJ5y6oAjrhRY1Fb85uc    &                    \\ \hline
			3           & 18iEz617DoDp8CNQUyyrjCcC7XCGDf5SVb    & \multirow{2}{*}{\cite{cryptolocker_addr_1,cryptolocker_addr_2}} \\ \cline{1-2}
			4           & 1KP72fBmh3XBRfuJDMn53APaqM6iMRspCh    &                    \\ \hline
		\end{tabular}
		\caption{CryptoLocker}
		\label{cryptolocker_address}
	\end{table}

	\begin{table}[H]
		\centering
		\begin{tabular}{|c|l|c|}
			\hline
			\textbf{\#} & \multicolumn{1}{c|}{\textbf{Address}} & \textbf{Source(s)} \\ \hline
			1           & 19DyWHtgLgDKgEeoKjfpCJJ9WU8SQ3gr27    & \multirow{2}{*}{\cite{cryptodefense_addr_1}} \\ \cline{1-2}
			2           & 1EmLLj8peW292zR2VvumYPPa9wLcK4CPK1    &                    \\ \hline
		\end{tabular}
		\caption{CryptoDefense}
		\label{cryptodefense_address}
	\end{table}
	
	\begin{table}[H]
		\centering
		\begin{tabular}{|c|l|c|}
			\hline
			\textbf{\#} & \multicolumn{1}{c|}{\textbf{Address}} & \textbf{Source(s)}      \\ \hline
			1           & 1PoebUjR5pdH88tc9ECQ1PCLaCrtPnG9fm    & \cite{cryptowall_addr_1}                       \\ \hline
			2           & 128pJdREzcR6xorYPQAPzGf8RwMQjRBzDt    & \cite{cryptowall_addr_5}                       \\ \hline
			3           & 15WUYqKerTtxi4rUEmnakw5gRMkr3nZCQd    & \cite{cryptowall_addr_6}                       \\ \hline
			4           & 1L66AcnbuZkYjs8eE6uVbTUxmorHYGKxFJ    & \cite{cryptowall_addr_7}                       \\ \hline
			5           & 16REtGSobiQZoprFnXZBR2mSWvRyUSJ3ag    & \cite{cryptowall_addr_8}                       \\ \hline
			6           & 16Z6sidfLrfNoxJNu4qM5zhRttJEUD3XoB    & \cite{cryptowall_addr_9}                       \\ \hline
			7          & 12LE1yNak3ZuNTLa95KYR2CQSKb6rZnELb    & \cite{cryptowall_addr_10}                      \\ \hline
			8          & 1JYYzNHDaGC7noiE4eKatuYA4AThqVocDd    & \cite{cryptowall_addr_11}                      \\ \hline
			9           & 1BhLzCZGY6dwQYgX4B6NR5sjDebBPNapvv    & \cite{secureworks_cryptowall,cryptowall_addr_2}                   \\ \hline
			10           & 16yd1Wj2NZa2uLZ6W4UDCDJ2Ttw92uFaT7    & \cite{secureworks_cryptowall,cryptowall_addr_3}                   \\ \hline
			11           & 1LGnuv6KX9SXB8eM72dnBAcECeaC8Z2zje    & \cite{secureworks_cryptowall,cryptowall_addr_4}                   \\ \hline
			12          & 1L7SLmazbbcy614zsDSLwz4bxz1nnJvDeV    & \multirow{4}{*}{\cite{secureworks_cryptowall,cryptowall_addr_12}} \\ \cline{1-2}
			13          & 19yqWit95eFGmUTYDLr3memcDoJiYgUppc    &                         \\ \cline{1-2}
			14          & 16N3jvnF7UhRh74TMmtwxpLX6zPQKPbEbh    &                         \\ \cline{1-2}
			15          & 1ApF4XayPo7Mtpe326o3xMnSgrkZo7TCWD    &                         \\ \hline
			\multicolumn{3}{|c|}{And, 27 other distinct addresses that are listed on {\cite{secureworks_cryptowall}}}     \\ \hline
		\end{tabular}
		\caption{CryptoWall}
		\label{cryptowall_address}
	\end{table}

	\begin{table}[H]
		\centering
		\begin{tabular}{|c|l|c|}
			\hline
			\textbf{\#} & \multicolumn{1}{c|}{\textbf{Address}} & \textbf{Source(s)} \\ \hline
			1           & 1MrKJhiECV3RufrY1dSybSXRCwSw11Co6i    & \cite{dmalocker_addr_2}                  \\ \hline
			2           & 1C8yA7wJuKD4D2giTEpUNcdd7UNExEJ45r    & \cite{dmalocker_addr_3}                  \\ \hline
			3           & 166vHLnGB1pCQGxdBkRiMkHW5WGQDbsw6s    & \cite{dmalocker_addr_4}                  \\ \hline
			4           & 1BA48s9Eeh77vwWiEgh5Vt29G3YJN1PRoR    & \cite{dmalocker_addr_5}                  \\ \hline
			5           & 18mfoGHSfe9h145e8djHK5rChDTnGfPDU9    & \cite{dmalocker_addr_6}                  \\ \hline
			6           & 16hHkyuzCDRFzoejVuqajqrnbmKHSmEfQM    & \cite{dmalocker_addr_7}                  \\ \hline
			7           & 1382JAg5xbQv7QNwq1svDeyw6ELtNCmujG    & \multirow{2}{*}{\cite{dmalocker_addr_1}} \\ \cline{1-2}
			8           & 1KXw7aJR4THWAxtnxZYzmysdLXVhLfa97n    & 							                  \\ \hline
		\end{tabular}
		\caption{DMA Locker}
		\label{dmalocker_address}
	\end{table}
	
	\begin{table}[H]
		\centering
		\begin{tabular}{|c|l|c|}
			\hline
			\textbf{\#} & \multicolumn{1}{c|}{\textbf{Address}} & \textbf{Source(s)} \\ \hline
			1           & 13dN96pRTQDhpWRqKyLTbgRxeTN52p2CqY    & \cite{mischa_addr_1}                  \\ \hline
		\end{tabular}
		\caption{Mischa}
		\label{mischa_address}
	\end{table}

	\begin{table}[H]
		\centering
		\begin{tabular}{|c|c|c|}
			\hline
			\textbf{\#} & \textbf{Address}                   & \textbf{Source(s)} \\ \hline
			1           & 1BAdEKq6zE1JDL8g2pA1MDRHbW1wvYCWhT & \cite{goldeneye_addr_1}                  \\ \hline
			2           & 1MGnopAa6MAGjUpCEmRiSAcVKZNB6n8gnR & \cite{goldeneye_addr_2}                  \\ \hline
			3           & 17xV74Hp2zNR74yG3AJvPpNMchPJHm2iUo & \cite{goldeneye_addr_3}                  \\ \hline
		\end{tabular}
		\caption{GoldenEye}
		\label{goldeneye_address}
	\end{table}
	
	\begin{table}[H]
		\centering
		\begin{tabular}{|c|l|c|}
			\hline
			\textbf{\#} & \multicolumn{1}{c|}{\textbf{Address}} & \textbf{Source(s)} \\ \hline
			1           & 1Mz7153HMuxXTuR2R1t78mGSdzaAtNbBWX    & \cite{petya_symantec,NotPetya_addr_1}                  \\ \hline
		\end{tabular}
		\caption{NotPetya}
		\label{NotPetya_address}
	\end{table}

	\begin{table}[H]
		\centering
		\begin{tabular}{|c|l|c|}
			\hline
			\textbf{\#} & \multicolumn{1}{c|}{\textbf{Address}} & \textbf{Source(s)} \\ \hline
			1           & 1PGAUBqHNcwSHYKnpHgzCrPkyxNxvsmEof    & \cite{keranger_addr_1}                  \\ \hline
			2           & 1Lhgda4K77rFMTkgBKqmsdinDNYYVbLDJN    & \cite{keranger_addr_2}                  \\ \hline
			3           & 1KGusS7xB9hnqZQdCZ1G8Tno16RfTS95ey    & \cite{keranger_addr_3}                  \\ \hline
			4           & 1KPPqHpd8Z9S6pQH1qVovzyejyfDMghp4u    & \cite{keranger_addr_4}                  \\ \hline
			5           & 1J9PMCpbrnicZoBUdyuNBwi4QvXwq6Korq    & \cite{keranger_addr_5}                  \\ \hline
			6           & 16hhyeg7WMh4Go7JqNKRwmD95bRd4aenwz    & \cite{keranger_addr_6}                  \\ \hline
		\end{tabular}
		\caption{KeRanger}
		\label{keranger_address}
	\end{table}
	
	\begin{table}[H]
		\centering
		\begin{tabular}{|c|l|c|}
			\hline
			\textbf{\#} & \multicolumn{1}{c|}{\textbf{Address}} & \textbf{Source(s)} \\ \hline
			1           & 13AM4VW2dhxYgXeQepoHkHSQuy6NgaEb94    & \multirow{3}{*}{\cite{secureworks_wannacry,wannacry_addr_2}} \\ \cline{1-2}
			2           & 12t9YDPgwueZ9NyMgw519p7AA8isjr6SMw    &                    \\ \cline{1-2}
			3           & 115p7UMMngoj1pMvkpHijcRdfJNXj6LrLn    &                    \\ \hline
		\end{tabular}
		\caption{WannaCry}
		\label{wannacry_address}
	\end{table}


	\begin{table}[H]
		\centering
		\begin{tabular}{|c|l|c|}
			\hline
			\textbf{\#} & \multicolumn{1}{c|}{\textbf{Address}} & \textbf{Source(s)} \\ \hline
			1           & 1EfuwPcYeCTes24X8CVGMUCR1H4yZ4CyoE    & \cite{ctblocker_addr_1}                  \\ \hline
			2           & 1EhJcMYwQKKWQcLFBjjYaMGTVncpQMJbbv    & \cite{ctblocker_addr_2}                  \\ \hline
			3           & 1Bj2z4j3weU1g9jwu4oHQQA6x8x2G2FRRm    & \cite{ctblocker_addr_3}                  \\ \hline
			4           & 1MScgv8kvbVLwGbciuw44gvy23rocaNCc8    & \cite{ctblocker_addr_4}                  \\ \hline
			5           & 1JXMiCkbrPiDWxoZ8oJ9yQZutHoaGQtXCF    & \multirow{4}{*}{\cite{ctblocker_addr_6}} \\ \cline{1-2}
			6           & 12UrsknT8hqYGpi8NToS2GWCWaLKtR2UXn    &                    \\ \cline{1-2}
			7           & 1PAVxqYtWD1RBAjE5voSDnUSefGGUvCwpm    &                    \\ \cline{1-2}
			8           & 1N3qTaZsUqU2owUVjmijVyHB4uiid2JoXd    &                    \\ \hline
			9           & 1PWLk2FP6r3FzKcqq9UgsYVZ9Ev6gufCsJ    & \multirow{4}{*}{\cite{ctblocker_addr_5}} \\ \cline{1-2}
			10           & 1BLeMsrSLB8H1fDDLRhQbLHScoC58ncf4x    &                    \\ \cline{1-2}
			11           & 1A6GJMhpPhCcM557o62scEtuVXNAFe74fa    &                    \\ \cline{1-2}
			12           & 1BGDTqDZyD446Q71eGhdmWLzyCHVPZUJxv    &                    \\ \hline
			
		\end{tabular}
		\caption{CTB-Locker}
		\label{ctblocker_address}
	\end{table}

	\begin{table}[H]
		\centering
		\begin{tabular}{|c|l|c|}
			\hline
			\textbf{\#} & \multicolumn{1}{c|}{\textbf{Address}} & \textbf{Source(s)} \\ \hline
			1           & 1KpP1YGGxPHKTLgET82JBngcsBuifp3noW    & \cite{cryptotorlocker2015_addr_1}                  \\ \hline
		\end{tabular}
		\caption{CryptoTorLocker2015}
		\label{cryptotorlocker2015_address}
	\end{table}

	\begin{table}[H]
		\centering
		\begin{tabular}{|c|l|c|}
			\hline
			\textbf{\#} & \multicolumn{1}{c|}{\textbf{Address}} & \textbf{Source(s)} \\ \hline
			1           & 1NRn15kJnVRrptTSQJJnMD9KJcWkVFh1Gv    & \cite{secureworks_teslacrypt}             \\ \hline
			2           & 15Y2TmHrxjmRFxfNUttwb9aU4DifvDpWKM    & \cite{teslacrypt_addr_1}                  \\ \hline
			3           & 1JthvnK8aoieXpx8YCAEtQwhfZSjSkdNox    & \cite{teslacrypt_addr_3}                  \\ \hline
			4           & 1L2jriaKw39jZysdH7nhe6eMSLSPNHvvHx    & \cite{teslacrypt_addr_4}                  \\ \hline
			5           & 1GQf1kEFK3SmVw8AMjRcn7jX1mvrGSDTkK    & \cite{teslacrypt_addr_5}                  \\ \hline
		\end{tabular}
		\caption{TeslaCrypt}
		\label{teslacrypt_address}
	\end{table}

	\begin{table}[H]
		\centering
		\begin{tabular}{|c|l|c|}
			\hline
			\textbf{\#} & \multicolumn{1}{c|}{\textbf{Address}} & \textbf{Source(s)} \\ \hline
			1           & 1HqoNfpAJFMy9E36DBSk1ktPQ9o9fn2RxX    & \cite{chimera_addr_1}                  \\ \hline
			2           & 15QzHEbNZWp2w1i2mfZSx7pV5YNM4ahszB    & \cite{chimera_addr_2}                  \\ \hline
			3           & 1GaVKrVT17DN4dnWbTqGB9qG3rQrk1JBe9    & \cite{chimera_addr_3}                  \\ \hline
			4           & 1MZsTFUNMGxQxz38wWm8CtBoycW7VD5z7v    & \cite{chimera_addr_4}                  \\ \hline
			5           & 1DGqEKZJdCd4YftWPuK5Z1HFBdeyz9RNDU    & \cite{chimera_addr_5}                  \\ \hline
		\end{tabular}
		\caption{Chimera}
		\label{chimera_address}
	\end{table}

	\begin{table}[H]
		\centering
		\begin{tabular}{|c|l|c|}
			\hline
			\textbf{\#} & \multicolumn{1}{c|}{\textbf{Address}} & \textbf{Source(s)} \\ \hline
			1           & 1AoNMLZfhw7cbMCKAhaKHiveMdwFyVUGeA    & \cite{hibuddy_addr_1}                  \\ \hline
		\end{tabular}
		\caption{Hi Buddy!}
		\label{hibuddy_address}
	\end{table}

	\begin{table}[H]
		\centering
		\begin{tabular}{|c|l|c|}
			\hline
			\textbf{\#} & \multicolumn{1}{c|}{\textbf{Address}} & \textbf{Source(s)}  \\ \hline
			1           & 15fbyNgDnqYQR5vSHJ8PTAEJbKy4dwNBCZ    & \cite{jigsaw_addr_1}                   \\ \hline
			2           & 12YHmaLEAbWx3o3p6BvegG9WH47EYs8t1V    & \cite{jigsaw_addr_2}                   \\ \hline
			3           & 15MHczWfcYxf3P3NwYqCthaNiieGP8RY9d    & \cite{jigsaw_addr_3}                   \\ \hline
			4           & 3NQoq5MVPfEMw12gB4a2c1G61mRZyMymsB    & \cite{jigsaw_addr_4}                   \\ \hline
			5           & 12vfQqmMxiDvZdzYHndfURupmcjjs8uSpY    & \cite{jigsaw_addr_5}                   \\ \hline
			6           & 1FLjcTFpz9MhwLdZ4xm9onpAnUGfRbGdXg    & \cite{jigsaw_addr_6}                   \\ \hline
			7           & 1Cj37Tw5uHwfye6Srd1zHzSMhUekp3jM63    & \cite{jigsaw_addr_7}                   \\ \hline
			8           & 1Q5B5udzDLpNJbpedGpyGMLVU5DR5dTqx6    & \cite{jigsaw_addr_8}                   \\ \hline
			9           & 13VEVaJUMdJyQ7ttPfBaVNKjj2dS9ahU1z    & \cite{jigsaw_addr_9}                   \\ \hline
			10          & 1HxkJ3vz2tvpcHgdt9yyY4XivdY9jKkcZH    & \multirow{3}{*}{\cite{jigsaw_addr_10}} \\ \cline{1-2}
			11          & 1LBhCecBmT23hybSUYyFW1YYqtTJcvFui2    &                     \\ \cline{1-2}
			12          & 1H8BXLJsLk9YCoNeBahYbgWo5ZqEn752ey    &                     \\ \hline
			13          & 1L9GdBW65Rt6e8UY69bnWNWomsppFFFR2X    & \multirow{7}{*}{\cite{jigsaw_addr_11}} \\ \cline{1-2}
			14          & 1ESe1nekuFJcEWycb1JjCz9KneNEm8yjg3    &                     \\ \cline{1-2}
			15          & 1EVNFaX7HktW1ud6fPueoMJ2Xw4UfYGY5Y    &                     \\ \cline{1-2}
			16          & 1CcAYfsKNNFPq7AKkbKQzRKw2kqjrqUeN9    &                     \\ \cline{1-2}
			17          & 18jCCAR2QZf6uZTnu4769ZknPfXjbmh1mw    &                     \\ \cline{1-2}
			18          & 1EH3yoQciVcWUufa4NWJvftyvvFxjbFLtQ    &                     \\ \cline{1-2}
			19          & 1F5RJzWN1g38wD9XbcspcxaYDU5hKpdvm8    &                     \\ \hline
		\end{tabular}
		\caption{Jigsaw}
		\label{jigsaw_address}
	\end{table}

	\begin{table}[H]
		\centering
		\begin{tabular}{|c|l|c|}
			\hline
			\textbf{\#} & \multicolumn{1}{c|}{\textbf{Address}} & \textbf{Source(s)} \\ \hline
			1           & 17XajwHHeWbfKfNwn57sHRMAEXxvQUUGNd    & \cite{zcryptor_addr_1,zcryptor_addr_2}                  \\ \hline
		\end{tabular}
		\caption{ZCryptor}
		\label{zcryptor_address}
	\end{table}

	\begin{table}[H]
		\centering
		\begin{tabular}{|c|l|c|}
			\hline
			\textbf{\#} & \multicolumn{1}{c|}{\textbf{Address}} & \textbf{Source(s)} \\ \hline
			1           & 16jvWspVfvhjRgJhGCDETf29cjQAyNmx9G    & \cite{venuslocker_addr_2}                  \\ \hline
			2           & 1JKVwmeokitMHAFxCUeC4yrd8pdWxDAjZW    & \cite{venuslocker_addr_3}                  \\ \hline
			3           & 1Dj9YnMiciNgaKuyzKynygu7nB21tvV6QD    & \cite{venuslocker_addr_1,venuslocker_addr_4}                  \\ \hline
		\end{tabular}
		\caption{VenusLocker}
		\label{venuslocker_address}
	\end{table}

	\begin{table}[H]
		\centering
		\begin{tabular}{|c|l|c|}
			\hline
			\textbf{\#} & \multicolumn{1}{c|}{\textbf{Address}} & \textbf{Source(s)} \\ \hline
			1           & 1N82pq3XovKoJYqUmTrRiXftpNHZyu4jyv    & \cite{trumplocker_addr_1}                  \\ \hline
		\end{tabular}
		\caption{The Trump Locker}
		\label{trumplocker_address}
	\end{table}

	\begin{table}[H]
		\centering
		\begin{tabular}{|c|l|c|}
			\hline
			\textbf{\#} & \multicolumn{1}{c|}{\textbf{Address}} & \textbf{Source(s)} \\ \hline
			1           & 19fhNi9L2aYXTaTFWueRhJYGsGDaN6WGcP    & \cite{lltp_addr_1}                  \\ \hline
		\end{tabular}
		\caption{The LLTP Locker}
		\label{lltp_address}
	\end{table}

	\begin{table}[H]
		\centering
		\begin{tabular}{|c|l|c|}
			\hline
			\textbf{\#} & \multicolumn{1}{c|}{\textbf{Address}} & \textbf{Source(s)} \\ \hline
			1           & 1Q94RXqr5WzyNh9Jn3YLDGeBoJhxJBigcF    & \cite{killdisk}                  \\ \hline
		\end{tabular}
		\caption{KillDisk}
		\label{killdisk_address}
	\end{table}

	\begin{table}[H]
		\centering
		\begin{tabular}{|c|l|c|}
			\hline
			\textbf{\#} & \multicolumn{1}{c|}{\textbf{Address}} & \textbf{Source(s)} \\ \hline
			1           &  1EZrvz1kL7SqfemkH3P1VMtomYZbfhznkb   & \cite{findzip}                  \\ \hline
		\end{tabular}
		\caption{FindZip}
		\label{findzip_address}
	\end{table}

	\begin{table}[H]
		\centering
		\begin{tabular}{|c|l|c|}
			\hline
			\textbf{\#} & \multicolumn{1}{c|}{\textbf{Address}} & \textbf{Source(s)} \\ \hline
			1           & 18yfx86BwNK5xYKw71uaHwAxPgCGRJaqgg    & \cite{thundercrypt_addr_1}                  \\ \hline
			2           & 1HFY12o56xbHer3oeNxC99A7SGyXaR64hs    & \cite{thundercrypt_addr_2}                  \\ \hline
			3           & 18KfMJBTDWUUa1h4tm58swbkvsgHNZ6d2g    & \cite{thundercrypt_addr_3}                  \\ \hline
		\end{tabular}
		\caption{ThunderCrypt}
		\label{thundercrypt_address}
	\end{table}

	\begin{table}[H]
		\centering
		\begin{tabular}{|c|l|c|}
			\hline
			\textbf{\#} & \multicolumn{1}{c|}{\textbf{Address}} & \textbf{Source(s)} \\ \hline
			1           & 1CvcvetHZ81V8itkDtF8iRpLfPp7Zz8UER    & \cite{doublelocker_addr_1}                  \\ \hline
			2           & 1HxKouDDK9WbkizMEnf23tftHSefWhUyXR    & \cite{doublelocker_addr_2}                  \\ \hline
		\end{tabular}
		\caption{DoubleLocker}
		\label{doublelocker_address}
	\end{table}

	\begin{table}[H]
		\centering
		\begin{tabular}{|c|l|c|}
			\hline
			\textbf{\#} & \multicolumn{1}{c|}{\textbf{Address}} & \textbf{Source(s)} \\ \hline
			1           & 1GxXGMoz7HAVwRDZd7ezkKipY4DHLUqzmM    & \multirow{2}{*}{\cite{badrabbit_addr_1}} \\ \cline{1-2}
			2           & 17GhezAiRhgB8DGArZXBkrZBFTGCC9SQ2Z    &                    \\ \hline
		\end{tabular}
		\caption{Bad Rabbit}
		\label{badrabbit_address}
	\end{table}
	
	\ifCLASSOPTIONcaptionsoff
	\newpage
	\fi
	
	\bibliographystyle{IEEEtran}
	\bibliography{bib}
	\balance

	%

	\begin{IEEEbiography}[{\includegraphics[width=1in,height=1.25in,clip
			]{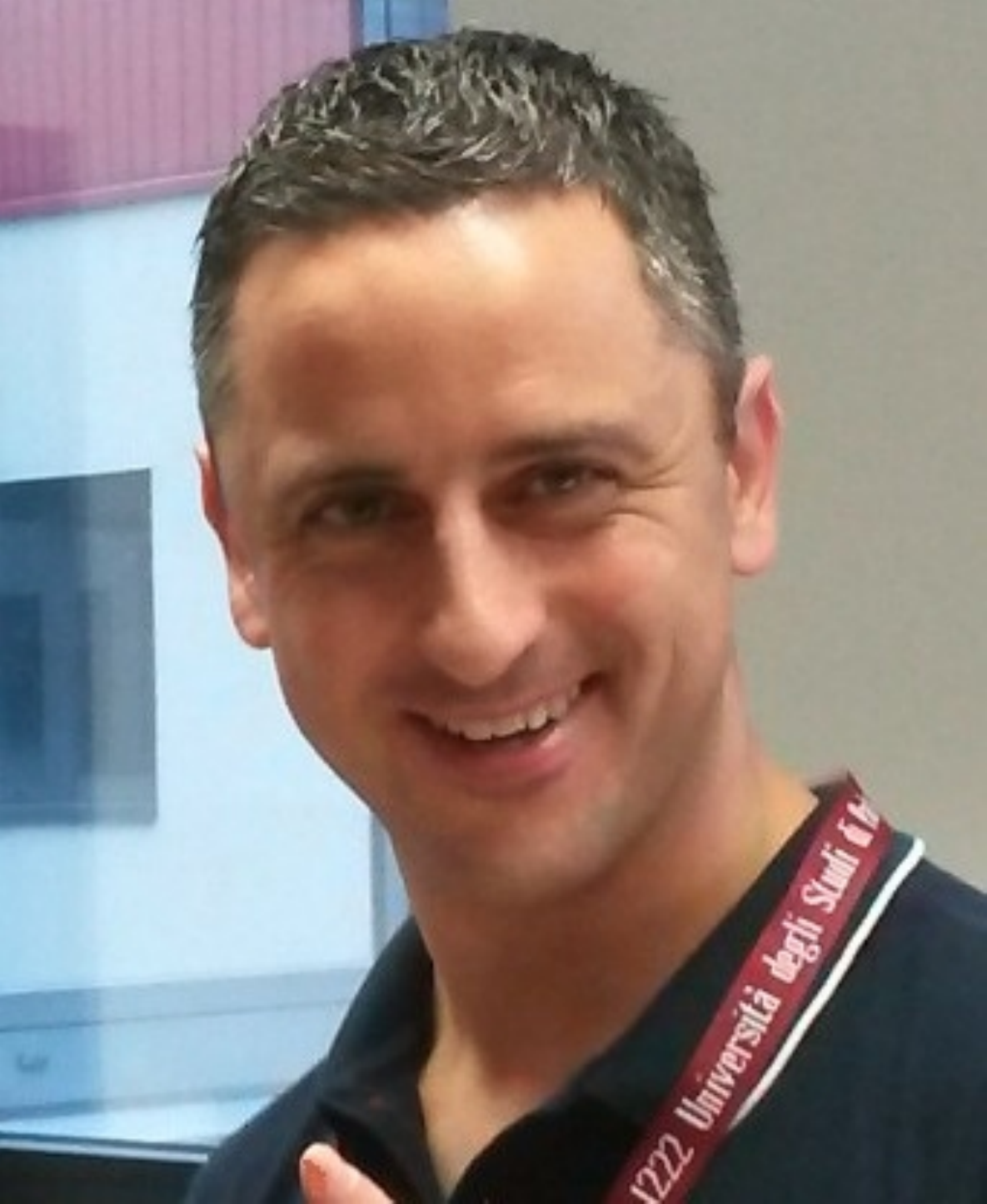}}]{Mauro Conti}
		received the Ph.D. degree from Sapienza University of Rome, Italy, in 2009. He is an Associate Professor with the University of Padua, Italy. He was a Post-Doctoral Researcher with Vrije Universiteit Amsterdam, The Netherlands. He was a recipient of the Marie Curie Fellowship (2012) by the European Commission, and a Fellowship by the German DAAD (2013). His main research interest is in the area of security and privacy. In this area, he published over 170 papers in topmost international peer-reviewed journals and conference. He is an Associate Editor for several journals, including the IEEE COMMUNICATIONS SURVEYS \& TUTORIALS and the IEEE TRANSACTIONS ON INFORMATION FORENSICS AND SECURITY. He was a Program Chair for TRUST 2015, ICISS 2016, WiSec 2017, and the General Chair for SecureComm 2012 and ACM SACMAT 2013.
	\end{IEEEbiography}
	
	\begin{IEEEbiography}[{\includegraphics[width=1in,height=1.25in,clip]{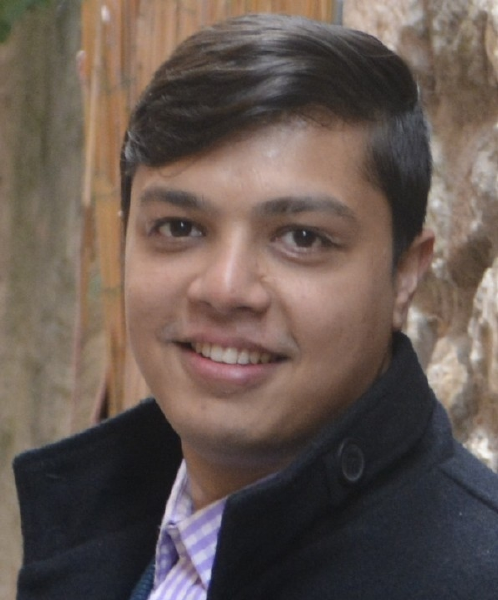}}]{Ankit Gangwal} received the B.Tech. degree in Information Technology from RTU, Kota, India in 2011 and the M.Tech. degree in Computer Engineering from Malaviya National Institute of Technology, Jaipur, India in 2016. Currently, he is a Ph.D. student in the Department of Mathematics, University of Padua, Padua, Italy with a fellowship for international students funded by Fondazione Cassa di Risparmio di Padova e Rovigo (CARIPARO). His~current research interest is in the area of security and privacy of blockchain technology and novel networking architectures, in particular, software-defined networking.
	\end{IEEEbiography}
	
	\begin{IEEEbiography}[{\includegraphics[width=1in,height=1.25in,clip]{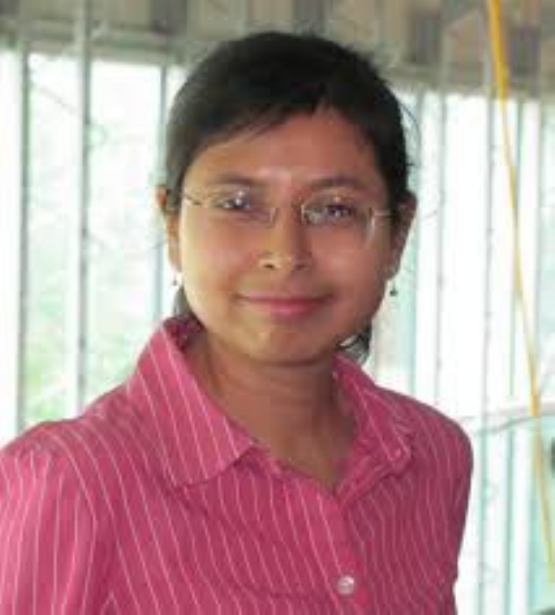}}]{Sushmita Ruj} received her B.E. in Computer Science from Bengal Engineering and Science University, Shibpur, India in 2004, and Masters and Ph.D. in Computer Science from Indian Statistical Institute, India in 2006 and 2010, respectively. She~was a Erasmus Mundus Post Doctoral Fellow at Lund University, Sweden and Post Doctoral Fellow at University of Ottawa, Canada. She is currently an Assistant Professor at Indian Statistical Institute, Kolkata, India. Prior to this, she was an Assistant Professor at IIT, Indore. She was a visiting researcher at INRIA, France, University of Wollongong, Australia, Kyushu University, Japan and Microsoft Research Labs, India. Her research interests~are in applied cryptography, security, combinatorics and complex network analysis. She works in mobile ad hoc networks, vehicular networks, cloud security, security in smart grids. She has served as program co-chair of IEEE ICCC (P\&STrack), IEEE ICDCS, IEEE ICC, etc and served on many TPCs. She won a Samsung GRO award in 2014. Sushmita is a Senior Member of IEEE.
	\end{IEEEbiography}
	\balance
\end{document}